\documentclass[a4paper, 12pt]{book}
\pdfoutput=1

\usepackage{fancyheadings}
\usepackage{amsfonts}
\usepackage{amsthm}
\usepackage{amsmath}
\usepackage{graphicx}
\usepackage{psfrag}
\usepackage{cite}
\usepackage[arrow, matrix, curve]{xy}
\usepackage{graphicx}
\usepackage{psfrag}
\usepackage{pdfpages}

\newtheorem{lemma}{Lemma}[section]
\newtheorem{proposition}{Proposition}[section]
\newtheorem{theorem}{Theorem}[section]
\newtheorem{corollary}{Corollary}[section]
\theoremstyle{remark}
\newtheorem{remark}{Remark}[section]
\newtheorem{example}{Example}[section]
\theoremstyle{definition}
\newtheorem{definition}{Definition}[section]

\newcommand{\Dom}[1]{\ensuremath{\mathcal{D}(#1)}}
\newcommand{\B}[1]{\ensuremath{\mathcal{B}(#1)}}
\newcommand{\NVS}[1]{\ensuremath{\left (#1, \|\cdot\|_{#1} \right )}}
\renewcommand{\Re}{\ensuremath{\mathrm{Re}}}

\newcommand{\im}{\ensuremath{\mathrm{im}}}

\begin{document}

\includepdf{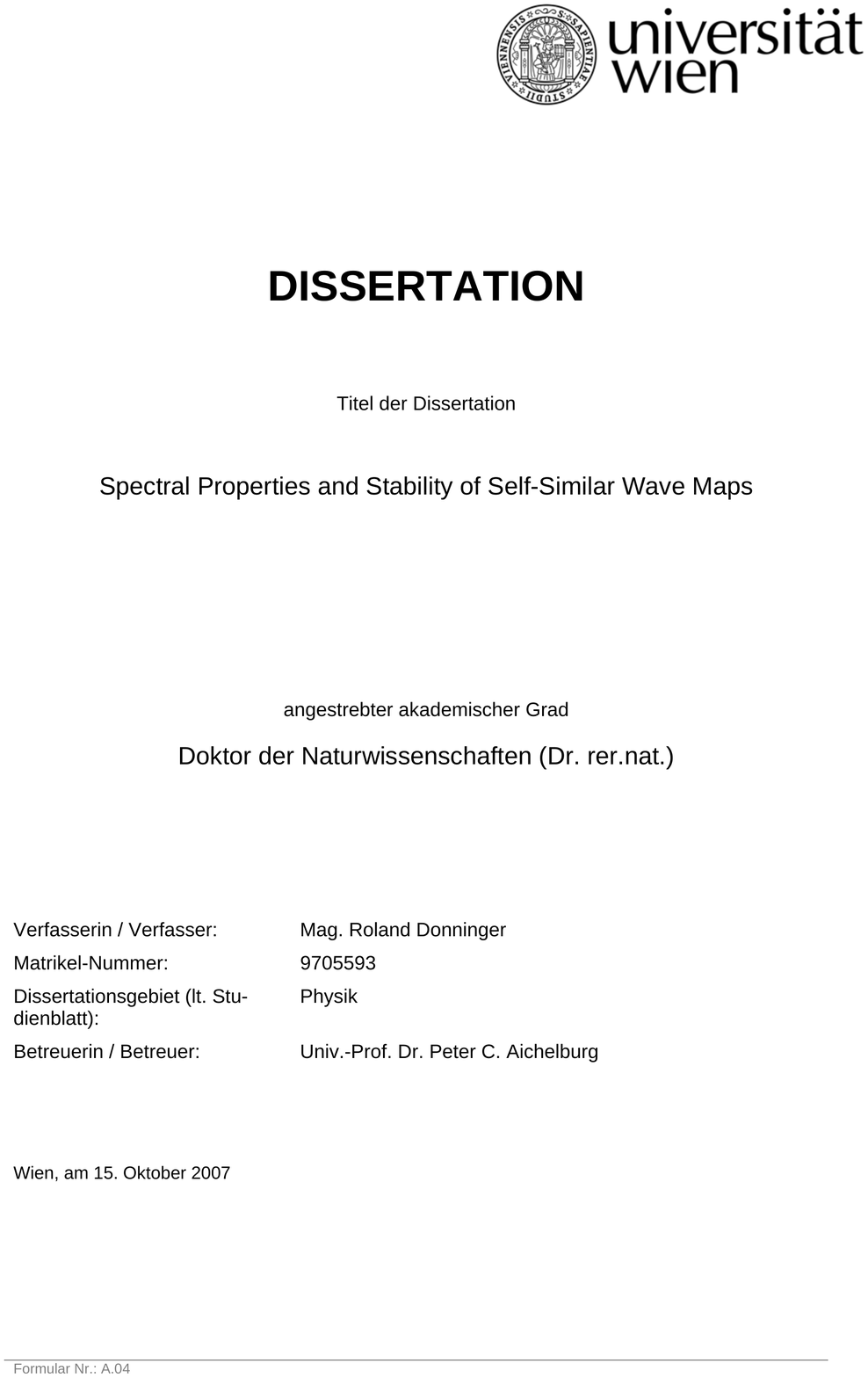}
\mbox{}
\thispagestyle{empty}
\newpage
\setcounter{page}{1}
\begin{center}
{\bf \large Acknowledgments}
\end{center}

First of all I want to thank my advisor Prof. Peter C. Aichelburg and Prof. 
Piotr Bizo\'n for helpful discussions and their support 
during the development of this thesis.
Furthermore, I am very grateful to Prof. Horst Beyer for taking the time to
answer my e--mail questions  
which helped me to overcome some conceptual difficulties.
Finally, I want to thank Prof. Robert Beig for many interesting discussions 
during lunchtime in our library (yes, we eat in the library).
This work has been supported by the Austrian Fond zur F\"orderung der
wissenschaftlichen Forschung FWF Projects P15738 and P19126.

\begin{center}
{\bf \large Danksagungen}
\end{center}

Von privater Seite m\"ochte ich mich zuerst bei meinen Freunden und Kollegen
Michael, Mark und Flo bedanken f\"ur viele interessante Diskussionen \"uber
wichtige und weniger wichtige Themen, sowie diverse Freizeitaktivit\"aten die
ich nicht missen m\"ochte.
Weiters bedanke ich mich bei meinen Eltern, die mich immer in allem
unterst\"utzt haben.
Am meisten Dank geb\"uhrt jedoch meiner Freundin Lisa f\"ur ihre Geduld, ohne
die sie mich wohl nicht ertragen w\"urde.

\thispagestyle{empty}
\newpage
\thispagestyle{empty}

\tableofcontents
\newpage 

\chapter{Introduction}
\thispagestyle{empty}

The huge field of partial differential equations (PDEs) is a rapidly 
developing branch
of mathematics which has a countless number of applications in 
natural science, engineering, economics, etc.
The formulation of physical theories would be entirely impossible
without the language of partial differential equations.
A mathematically rigorous investigation of PDEs started in the 
$20^\mathrm{th}$ century
although many corresponding physical theories are much older.
While in the old days the studies were mainly confined to
the construction of explicit solutions (e.g. in terms of integrals),
a paradigm shift
took place with the introduction of functional analytic methods 
in the $20^\mathrm{th}$ century.
It has been realized that for most equations it is hopeless to try to construct 
explicit solutions and even if this is possible, the arising formulas are 
completely useless due to their complexity.
Thus, the new methods
focused on studying properties of solutions, 
function spaces they belong to and
estimates they satisfy without knowing them explicitly.
This approach turned out to be very fruitful and the highly successful 
modern PDE theory is entirely based on it.

A special class of PDEs are so--called evolution equations
where, roughly speaking, one of the independent variables is interpreted as
time.
For that kind of equations one considers the Cauchy problem, i.e. one 
prescribes initial data at a certain instance of time and studies the future 
development.
Typical results are related to the existence and uniqueness of solutions 
as well as
their dependence on the data.
In this respect one has to distinguish carefully between
local (for small times) and global (for all times) existence.
It may well be possible that the Cauchy problem is locally well--posed 
but it does not admit global
solutions, at least not for arbitrary data.
This fact leads to the problem of formation of singularities, i.e. the
breakdown of solutions in finite time.
Once one has a local existence result one can ask the question: 
Given regular data, 
is it possible that the
corresponding solution
ceases to exist after a finite time in the future and if so, 
how does this happen?
Closely related to this is the question under what circumstances the solution 
extends
globally in time.
It is worth mentioning that the possible breakdown of solutions 
(singularity formation) is a feature which
is common to many nonlinear evolution equations arising from various branches of
physics, chemistry, biology, etc.

This thesis is devoted to the investigation of some aspects of the Cauchy
problem for wave maps from Minkowski space to the three--sphere $S^3$.
The wave maps system, a nonlinear generalization of the wave equation,
arises from a geometric action principle.
Nowadays, local well--posedness for the Cauchy problem of the wave maps system
is well--understood.
Furthermore, some major progress concerning global
aspects has been made quite recently.
Some of these results are reviewed.
However, in this particular model of wave maps with $S^3$ target, singularity
formation is possible.
One can explicitly construct self--similar solutions that "blow up" at a
prescribed finite time.
Imposing additional symmetry assumptions, this wave maps system reduces to a 
single semilinear wave equation which exhibits a rich phenomenology despite its
simplicity.
In particular, it shares some properties with Einstein's equations 
in connection with
gravitational collapse.
Thus, this semilinear wave equation is used as a toy model for the much more 
involved evolution problem in general relativity
since it is simple enough to be accessible for analytic techniques.
The main goal of this thesis is to obtain a better understanding of 
singularity formation for wave maps from Minkowski space to the three--sphere.
To this end, 
spectral theoretic aspects of
self--similar solutions and their linear stability are studied.
The mathematical machinery required for this consists of functional analysis,
operator theory, Sturm--Liouville theory and semigroups.
Since this thesis is intended to be essentially self--contained, all the
mathematical requirements are carefully introduced although not all results are
proved since this would go beyond the scope of this work.
However, if the proof of a certain result is omitted then there is given an easy
accessible reference or at least the idea is sketched.
The organization of the thesis is as follows.

In ch. \ref{wm_ch} we define the wave maps model. First, we recall some
basic notions from differential geometry required to state the action
functional.
Then, we derive the associated Euler--Lagrange equations which constitute the
wave maps system.
This system reduces to a
single semilinear wave equation under the assumption of co--rotationality.
Finally, we give some historical remarks and applications of wave maps.

In ch. \ref{math_ch} we collect some basic mathematical concepts. We introduce
Banach, Hilbert and Lebesgue spaces and note some elementary properties.
Then, we briefly recall the basics of linear operator theory on Banach spaces.
Next, we turn to some aspects of spectral theory for closed operators and 
in the last part of this chapter we give a short overview of basic 
semigroup theory.
 
In ch. \ref{sa_ch} we focus on Sturm--Liouville operators.
First, we state some general properties of self--adjoint operators.
Then, we turn to the study of Sturm--Liouville operators.
We discuss endpoint classification, calculate adjoints for regular 
and singular operators and state the Weyl alternative.
    
In ch. \ref{wp_ch} we apply semigroup techniques to abstract wave equations.
By this we obtain a well--posedness result for the linear homogeneous case.
Then, we turn to the inhomogeneous problem whose solution relies on a Bochner
integral version of the variation of constants formula.
In order to state this result we recall some basic aspects of measure theory.
Finally, a nonlinear equation is investigated to illustrate the application of
the Banach fixed point theorem to such problems.

In ch. \ref{wpwm_ch} we give a short overview of some known results concerning
the Cauchy problem for wave maps.
We state a local well--posedness theorem for general semilinear wave equations
which is applicable to the wave maps system.
Then, we mention some global results and give historical remarks.

Ch. \ref{css_ch} is devoted to the study of self--similar solutions.
We discuss blow up, criticality class and finite speed of propagation for the
wave maps model.
Then, we state a result concerning the existence of a countable set of 
self--similar solutions
which provide explicit examples of blow up solutions.
Furthermore, we give some numerical evidence for the universality of the blow up
profile.
We introduce a new coordinate system adapted to self--similarity and linearize
the evolution equation.
Using Sturm--Liouville theory we construct a self--adjoint operator which
governs the linearized flow around a self--similar solution.
Applying the previously developed operator techniques we prove 
well--posedness of
the corresponding Cauchy problem.

In ch. \ref{specl0_ch} we study the spectrum of the operator $A_0$ which drives
the linearized flow around the first self--similar solution.
Using estimates for solutions around the singular endpoints of the eigenvalue
equation and an oscillation argument we show that the point spectrum is empty.
Furthermore, we identify a continuous spectrum.
These results lead to the best possible growth estimate for solutions of the
linearized equation around the first self--similar solution.
Finally, we give an intuitive explanation why no further improvement of
this result can be expected. 

In ch. \ref{funcalc_ch} we introduce the functional calculus, a powerful method
to define functions of self--adjoint operators.
This yields an alternative approach to the study of
abstract wave equations.
We apply this method to the problem under investigation and reproduce the
result previously obtained by semigroup techniques.

In ch. \ref{specln_ch} we study the spectra of the operators $A_n$ which drive
the linearized flow around the $n$--th self--similar wave map for $n \geq 1$.
We show that $A_n$ has exactly $n$ negative eigenvalues and derive a rough
lower bound.
Then, we numerically calculate the point spectra of $A_n$ and
reveal a certain convergence of the eigenvalues.
Finally, we construct a "limiting" operator $A_\infty$.
A numerical investigation strongly suggests that the spectrum
of $A_\infty$ is the limit of the spectra of $A_n$.

In the last chapter we give an outlook, sketch some ideas how
one could proceed and mention further results. 

\chapter{Wave Maps}
\thispagestyle{empty}
\label{wm_ch}

In this chapter we define the model to be studied and note some of its basic
properties.

\section{The Wave Maps Equation}
The wave maps equation is a system of semilinear wave equations where the 
involved nonlinearities
have a geometric nature.
We define an action functional for wave maps from Minkowski space to a
Riemannian manifold.

\subsection{Basic Definitions and Notation}
We collect some basic notions from differential geometry required to define a
wave map. 
The main purpose of this section is to fix notation in order to avoid confusion
due to different conventions.
For more information see e.g. \cite{Spivak1999}, \cite{Cap2004} or any other textbook on
differential geometry.

\paragraph{Smooth manifolds}
Let $M$ be a set and $d: M \times M \to \mathbb{R}$ a \emph{distance function}
on $M$, i.e. $d(x,y) \geq 0$, $d(x,y)=0 \Leftrightarrow x=y$, $d(x,y)=d(y,x)$
and $d(x,z) \leq d(x,y)+d(y,z)$ for all $x,y,z \in M$.
Then $(M,d)$ is said to be a \emph{metric space}.
We define a \emph{manifold} as a 
metric space $M$ which is locally homeomorphic to $\mathbb{R}^m$, i.e. for every
$x \in M$ there exists a neighbourhood $U \subset M$ of $x$, an integer $m$ 
and a continuous bijective mapping $\phi: U \to \mathbb{R}^m$ such that
$\phi^{-1}$ is continuous as well.
If the number $m$ is constant for all $x \in M$ (which we will always assume), 
the manifold $M$ is said to be
\emph{$m$-dimensional} and we write $\dim(M)=m$.
An open subset $U \subset M$ together with a homeomorphism $u: U \to u(U)
\subset \mathbb{R}^m$ is said to be a \emph{chart}, denoted by $(U,u)$.
Two charts $(U,u)$ and $(V,v)$ are said to be \emph{compatible} if the mappings
$u \circ v^{-1}: v(U \cap V) \to u(U \cap V)$ and $v \circ u^{-1}: u(U \cap V)
\to v(U \cap V)$ are smooth.
A family $\mathcal{A}=\{(U_i, u_i): i \in I\}$ of mutually compatible charts
such that $M=\cup_{i \in I}U_i$ is called an \emph{atlas}.
Two atlases $\mathcal{A}_1$ and $\mathcal{A}_2$ 
are said to be \emph{equivalent} if any chart in $\mathcal{A}_1$ is compatible
with any chart in $\mathcal{A}_2$.
A manifold together with an equivalence class of atlases is called a \emph{smooth
manifold}.

\paragraph{Derivations, tangent space}
Let $M, N$ be smooth manifolds. 
A continuous function $f: M \to N$ is said to be \emph{smooth} if for every $x
\in M$ there exists a chart $(U,u)$ on $M$ with $x \in U$ and a chart $(V,v)$ on $N$ with $f(x) \in V$ such that the mapping 
$v \circ f \circ u^{-1}: u(U \cap f^{-1}(V)) \to v(V)$ is smooth.
The set of smooth functions from $M$ to $N$ is denoted by $C^{\infty}(M,N)$.  
A \emph{derivation} $\xi_x$ at $x \in M$ is a linear mapping $\xi_x:
C^\infty(M,\mathbb{R}) \to \mathbb{R}$ such that
$\xi_x(fg)=\xi_x(f)g(x)+f(x)\xi_x(g)$ for all $f,g \in C^\infty(M,\mathbb{R})$.
It turns out that the set of derivations at $x \in M$ form an 
$\dim(M)$--dimensional
vector space.
This vector space is referred to as the \emph{tangent space} $T_xM$ of $M$ at
$x$.

\paragraph{Tangent map, tangent bundle}
Let $M$, $N$ be smooth manifolds, $\dim(M)=m$ and $f \in C^\infty(M,N)$.
For $x \in M$ we define the \emph{tangent map at $x$} $T_xf: T_xM \to T_{f(x)}N$ by
$\xi_x \mapsto T_xf \cdot \xi_x$ and $T_xf\cdot \xi_x(g):=\xi_x(g \circ f)$ for
$\xi_x \in T_xM$ and $g \in C^\infty(N, \mathbb{R})$. 
The \emph{tangent bundle} $TM$ is defined as the disjoint union of all tangent
spaces, i.e. $TM:=\bigcup_{x \in M}\{x\} \times T_xM$.
There is a canonical projection $p: TM \to M$ given by $(x, \xi_x) \mapsto x$. 
One defines the \emph{tangent map} 
$Tf: TM \to TN$ as $(x, \xi_x) \mapsto (f(x), T_xf \cdot \xi_x)$.
Given an atlas $\{(U_i,u_i): i \in I\}$ on $M$ there exists a canonical 
atlas on
$TM$, namely $\{(p^{-1}(U_i), Tu_i): i \in I\}$.
This leads to the observation that $TM$ is a $2m$--dimensional smooth manifold.

\paragraph{Vector fields, metric, pullback}
Let $M$ be a smooth manifold. 
A \emph{vector field} $\xi$ on $M$ is a smooth mapping $\xi: M \to TM$ such that
$p \circ \xi=\mathrm{id}_M$.
The set of smooth vector fields on $M$ is denoted by $\mathfrak{X}(M)$.
A \emph{metric} $g$ on $M$ is a mapping which associates to every $x \in M$ a
scalar product \footnote{By a \emph{scalar product} on a real vector space $X$
we mean a symmetric bilinear mapping $b: X \times X \to \mathbb{R}$ which is
non-degenerate, i.e. $b(x,y)=0$ for all $y \in X$ is equivalent to $x=0$.} 
on $p^{-1}(\{x\})$ (or, equivalently, on $T_xM$) such that 
the mapping $x \mapsto g(x)(\xi(x), 
\zeta(x)): M
\to \mathbb{R}$ is smooth for all $\xi, \zeta \in \mathfrak{X}(M)$.
To simplify notation one usually writes $g(\xi, \zeta)$ instead of
$g(\cdot)(\xi(\cdot), \zeta(\cdot))$.
If the scalar product is positive definite, $g$ is said to be a \emph{Riemannian
metric}.
If it is indefinite, $g$ is called a \emph{pseudo--Riemannian metric}.
$(M, g)$ is called a \emph{(pseudo--) Riemannian manifold}.

Given a (pseudo--) Riemannian manifold $(N, g)$ and a smooth
mapping $f \in C^\infty(M,N)$ one can "pull back" the metric $g$ on $N$ to $M$.
The pullback metric $f^*g$ is defined by $f^*g(x)(\xi(x),\zeta(x)):=g(f(x))(Tf\cdot
\xi(x), Tf\cdot \zeta(x))$ for $x \in M$ and $\xi, \zeta \in \mathfrak{X}(M)$.

\paragraph{Cotangent bundle}
Given the tangent space $T_xM$ at $x \in M$ of a smooth manifold $M$ with
$\dim(M)=m$, one denotes
the \emph{dual space} or \emph{cotangent space} 
(in the sense of linear algebra) of $T_xM$ by $T^*_xM$.
$T^*_xM$ consists of all linear functionals on $T_xM$ and is again an
$m$--dimensional vector space.
Using a similar construction as in the case of tangent spaces, the disjoint
union of all cotangent spaces can be viewed as a $2m$--dimensional manifold, the
\emph{cotangent bundle} $T^*M$.

Similarly to vector fields, one defines a \emph{one--form} $\omega$ as a smooth mapping
$\omega: M \to T^*M$ such that $p \circ \omega=\mathrm{id}_M$ where $p: T^*M \to
M$ denotes the canonical projection.

\paragraph{Identification of $TM$ with $T^*M$}
Let $(M, g)$ be a (pseudo--) Riemannian manifold.
Then there is a natural way of
identifying elements of $T_xM$ with elements of $T^*_xM$: At each $x \in M$ the
metric $g$ induces a scalar product $g_x$ on $T_xM$.
Let $\xi_x \in T_xM$ and consider the mapping $\xi_x \mapsto g_x(\xi_x, \cdot):
T_xM \to T^*_xM$.
Due to the properties of a metric, this mapping is linear and bijective and
therefore provides a natural way of identifying $TM$ with $T^*M$. 

\subsection{Wave Maps on Minkowski Space} 

\paragraph{Minkowski space}
Consider the set $\mathbb{R}^4$. 
The identity $\mathrm{id}: \mathbb{R}^4 \to \mathbb{R}^4$ yields a global 
chart on the 
metric space $M:=\mathbb{R}^4$ and therefore
$M$ can be viewed as a 4--dimensional smooth manifold.
Consider the $i$--th partial derivative for functions in $C^\infty(M,
\mathbb{R})$ which induces a derivation at $x \in M$, i.e. an element of 
$T_xM$.
We identify this element with the $i$--th unit vector in $M$.
Therefore, the tangent space $T_xM$ at $x \in M$ is naturally isomorphic to
$M$ itself.
We define a pseudo--Riemannian metric $\eta$ on $M$ which induces a scalar 
product $\eta_x$ on $T_xM$ given by 
$$ \eta_x(v,w):=-v^0 w^0 + \sum_{i=1}^3 v^i w^i $$
for $v=(v^0,v^1,v^2,v^3), w=(w^0,w^1,w^2,w^3) \in \mathbb{R}^4$.
The pseudo--Riemannian manifold $(M, \eta)$ is called \emph{Minkowski space}.

\paragraph{The trace of the pullback metric}
Let $(N,g)$ be a Riemannian manifold and $(M,\eta)$ be Minkowski space.
For a smooth mapping $f: M \to N$ consider the pullback metric $f^* g$.
At each $x \in M$ it induces a bilinear mapping $(f^* g)_x: T_xM \times T_xM
\to \mathbb{R}$.
This defines a linear mapping 
$\xi_x \mapsto (f^* g)_x(\xi_x, \cdot): T_xM \to 
T^*_xM \cong T_xM$, i.e. a linear operator on the 4--dimensional vector space
$T_xM$
(we have implicitly identified $T_xM$ with $T^*_xM$ using $\eta$).
We denote the trace (in the sense of linear
algebra) of this operator by $\mathrm{tr}_\eta f^* g(x)$.
Then, $\mathrm{tr}_\eta f^* g$ is a smooth real--valued function on $M$.

\paragraph{Variational formulation}
We define a functional $S$ on the set $C^\infty(M,N)$ by
$$ S(f):=\int_M \mathrm{tr}_\eta f^* g $$
for $f \in C^\infty(M,N)$.
The integral is understood with respect to the ordinary Lebesgue measure on
$\mathbb{R}^4$.
We are interested in critical points of that functional, i.e. we intend to study
compactly supported variations.
However, for a function attaining values in a manifold it is not clear 
what it
means to have compact support.
Moreover, in the standard approach one considers an expression like
$S(f+\varepsilon \varphi)$ which makes absolutely no sense if the functions
$f$ and $\varphi$ have values in a manifold.

There are essentially two ways to go around this difficulty.
The first possibility is to restrict oneself to a single chart on $N$, 
i.e. one
actually solves the problem for functions $f$ having values in some
$\mathbb{R}^n$.
Clearly, this is a local notion and one does not obtain a global statement 
in this case.
However, this approach is sufficient if the whole target manifold 
can be covered "almost completely" by a single chart.

A second possibility is to embed $N$ into some $\mathbb{R}^n$.
Of course, in this case one has to ensure that all variations
$f+\varepsilon \varphi$ have values in $N$ which restricts the set of
admissible test functions $\varphi$.
Hence, one has to make sure that the set of test functions is 
"large enough" (cf.
\cite{Shatah1998}).
 
\paragraph{Definition of wave maps}
For our purposes the first approach is sufficient, i.e. we restrict ourselves to
a fixed single chart $(V,v)$ on $N$. 
Via $v^{-1}: v(V) \to V \subset N$ we pull back the metric $g$ to
$v(V) \subset \mathbb{R}^n$ where $n=\dim(N)$.
The open set $v(V)$ is a smooth manifold itself and
$(v^{-1})^*g$ is a Riemannian metric on $v(V)$. Hence, 
$(v(V), (v^{-1})^*g)$ is a Riemannian manifold.
$$ 
\begin{xy} \xymatrix{
(M,\eta) \ar[r]^f \ar[dr]_{v \circ f} & (N,g) \ar[d]^v \\
& (v(V),(v^{-1})^*g)
}
\end{xy}
$$
Fix $\Phi \in C^\infty(M,v(V))$ and consider $\Phi+\varepsilon \varphi$ for
$\varepsilon \in \mathbb{R}$ and smooth $\varphi: M \to \mathbb{R}^n$
having compact support.
In general the function $\Phi + \varepsilon \varphi$ does not have values in
$v(V)$ but for $\varepsilon$ sufficiently close to $0$ it does.
Having these considerations in mind we can define an action functional
for wave maps.

\begin{definition}
\label{wm_def_wm}
Let $(M,\eta)$ be Minkowski space, $(N,g)$ an $n$--dimensional Riemannian 
manifold and $(V,v)$ a chart on $N$.
Define a functional $S$ on $C^\infty(M,v(V))$ by
$$ S(\Phi):=\int_M \mathrm{tr}_\eta \Phi^* (v^{-1})^* g $$
for $\Phi \in C^\infty(M,v(V))$.
A function $\Phi \in C^\infty(M,v(V))$ is said to be a \emph{critical point} of the
functional $S$
if
$$ \left. \frac{d}{d \varepsilon}S(\Phi+
\varepsilon \varphi) \right |_{\varepsilon=0}=0 $$
for all $\varphi \in C^\infty_c(M,\mathbb{R}^n)$.
A \emph{wave map on Minkowski space with target
N} is a mapping $v^{-1} \circ \Phi: M \to N$ such that $\Phi$ is a critical
point of $S$.
\end{definition}

\paragraph{Generalizations}
It is obvious that this definition can be extended to the case where $M$ is
a more general pseudo--Riemannian manifold.
Such a generalization is essential to study 
self--gravitating wave maps, 
i.e. to use wave maps as a matter model in general relativity.
This has been done by various authors (e.g. \cite{Aichelburg2005},
\cite{Szybka2004}, \cite{Bizon2002}) although 
most of the work in
this direction is confined to numerical studies or heuristic arguments due to
the complexity of such systems.

\subsection{Local Expressions}
We intend to derive a local expression for the action functional which defines
wave maps.
As before, let $(M,g)$ be Minkowski space, $(N,g)$ an $n$--dimensional
Riemannian manifold and $(V,v)$ a chart on $N$.

\paragraph{Lagrangian in local coordinates}
The $\mu$--th partial derivative for functions defined on $\mathbb{R}^4$ 
yields a smooth vector field
on $M$ which is denoted by $\partial_\mu$ ($\mu=0,\dots,4$).
We set $\eta_{\mu \nu}:=\eta(\partial_\mu, \partial_\nu)$ and thus we have
$(\eta_{\mu \nu})=\mathrm{diag}(-1,1,1,1)$.
Similarly, the $A$--th partial derivative on $\mathbb{R}^n$ defines a smooth
vector field $\partial_A$ on $v(V) \subset \mathbb{R}^n$ ($A=1,\dots,n$).
Consider a smooth function $\Phi \in C^\infty(M,v(V))$.
The $A$--th component of $\Phi$ is denoted by $\Phi^A$.
By carefully inserting the definitions one readily calculates $T_x\Phi \cdot
\partial_\mu|_x=\frac{\partial \Phi^A}{\partial
x^\mu}(x)\partial_A|_{\Phi(x)}$ for $x \in M$ where Einstein's
summation convention is assumed in the sequel.
Thus, we have 
$\Phi^*(v^{-1})^*g(\partial_\mu, \partial_\nu)(x)=[(v^{-1})^*
g(\Phi(x))](T_x\Phi\cdot \partial_\mu|_x, T_x\Phi \cdot \partial_\nu|_x)
=\frac{\partial \Phi^A}{\partial x^\mu}(x)\frac{\partial \Phi^B}{\partial
x^\nu}(x)g_{AB}(\Phi(x))$ with $g_{AB}:=(v^{-1})^*g(\partial_A,
\partial_B)$.

\paragraph{Vectors and one--forms}
We define the one--forms $dx^\mu$ by
$dx^\mu(\partial_\nu)=\delta^\mu{}_\nu$.
Consider a vector field $\xi^\mu \partial_\mu$.
We assign to it the one--form $\omega_\mu dx^\mu$ where $\omega_\mu:=
\eta_{\mu \nu} \xi^\nu$.
Since this is the usual multiplication of a matrix with a vector, 
the inverse operation is given by $\xi^\mu=\eta^{\mu \nu} \omega_\nu$ where
$\eta^{\mu \nu}$ are the components of the inverse of the matrix $(\eta_{\mu
\nu})$. 
Thus, the identification of vectors with their duals takes place via
multiplication with the matrix $(\eta_{\mu \nu})$ and its inverse $(\eta^{\mu
\nu})$.

\paragraph{The wave maps system}
Taking the above considerations into account we arrive at
$$\mathrm{tr}_\eta \Phi^* (v^{-1})^* g(x)=\eta^{\mu
\nu}(\partial_\mu \Phi^A)(x) (\partial_\nu \Phi^B)(x) g_{AB}(\Phi(x)). $$
Hence, the action functional is given by
$$ S(\Phi)=\int_{\mathbb{R}^4} \eta^{\mu
\nu}(\partial_\mu \Phi^A) (\partial_\nu \Phi^B) (g_{AB} \circ \Phi). $$
Following the standard approach (cf. \cite{Shatah1998}, \cite{Evans1998}) one
interprets $\Phi^A$ and $\partial_\mu \Phi^A$ as independent variables and
defines the Lagrangian $\mathcal{L}$ by $\mathcal{L}(\Phi, \partial \Phi):=\eta^{\mu
\nu}(\partial_\mu \Phi^A) (\partial_\nu \Phi^B) (g_{AB} \circ \Phi)$.
Then, critical points (with respect to compactly supported variations) 
of $S$ satisfy the Euler--Lagrange equations 
$$ \partial_\mu \frac{\partial \mathcal{L}}{\partial (\partial_\mu
\Phi^A)}-\frac{\partial \mathcal{L}}{\partial \Phi^A}=0. $$
A straight--forward calculation yields the equation
\begin{equation}
\label{wmeq_eq_wmsys} 
\Box \Phi^A + \eta^{\mu \nu} \Gamma^A{}_{BC}(\Phi)(\partial_\mu
\Phi^B)(\partial_\nu \Phi^C)=0 
\end{equation}
where $\Box \Phi^A:=\eta^{\mu \nu}\partial_\mu \partial_\nu \Phi^A$ and
$$ \Gamma^A{}_{BC}:=\frac{1}{2}g^{AD}\left (\partial_B g_{CD}+\partial_C
g_{BD}-\partial_D g_{BC} \right ) $$
is a \emph{Christoffel symbol} of the second kind associated to the metric $g$.
Eq. (\ref{wmeq_eq_wmsys}) is known as the \emph{wave maps system}.

We remark that one can also study wave maps on Minkowski
spaces $\mathbb{R}^{m+1}$ with different spatial dimension ($m \not=3$).
It is obvious how the above considerations are generalized to this case.

\section{Wave Maps from Minkowski Space to the Three--Sphere}
Now we choose the three--sphere as a target manifold and derive an explicit
expression for the action under certain symmetry assumptions.

\subsection{Basic Definitions}

\paragraph{The three--sphere}
The three--sphere 
$ S^3 \subset \mathbb{R}^4$ is defined as 
$$ S^3:=\left \{(x^1, x^2,x^3,x^4) \in \mathbb{R}^4: \sum_{k=1}^4 (x^k)^2=1
\right \}. $$
As a subset of the metric space $\mathbb{R}^4$ it is a metric space itself and 
via stereographic projections it is locally homeomorphic to $\mathbb{R}^3$.
Thus, $S^3$ is a 3--dimensional manifold.

To define a chart $(V,v)$ on $S^3$ we give an 
explicit expression for $v^{-1}$.
We define the mapping $v^{-1}: (0,\pi) \times (0,\pi) \times (0,2 \pi) 
\to S^3 $ 
by
$$ (\psi, \Theta, \phi) \mapsto (\sin \psi \sin \Theta \cos \phi, 
\sin \psi \sin \Theta \sin \phi, \sin \psi \cos \Theta, \cos \psi). $$
The function $v$ is a homeomorphism onto its image and therefore, 
$(V,v)$  is a chart on $S^3$ where $V:=\mathrm{im}( v^{-1})$.
Similarly one can construct a whole atlas for $S^3$ and 
therefore, $S^3$ is a smooth manifold.
 
Let $\delta$ be the standard metric on $\mathbb{R}^4$ which induces 
the 
Euclidean scalar product on every $T_x\mathbb{R}^4 \cong \mathbb{R}^4$. 
We consider the natural embedding $i: S^3 \to \mathbb{R}^4$, $i(x)=x$ for 
$x \in S^3$ and pull back the metric $\delta$ via $i$.
The pullback metric $g:=i^*\delta$ is a Riemannian metric on $S^3$ and
therefore, 
$(S^3, g)$ is a Riemannian manifold. 

\paragraph{Co--rotational maps}
We formally define the usual spherical coordinates on Minkowski space 
$(M,\eta)$.
Let $u^{-1}: \mathbb{R} \times (0, \infty) \times (0, \pi) \times (0,2 \pi) \to 
M$ be given by
$$ (t,r,\theta,\varphi) \mapsto (t, r \sin \theta \cos \varphi, r \sin \theta
\sin \varphi, r \cos \theta). $$
Then, $u^{-1}$ is a homeomorphism onto its image and therefore,
$(U,u)$ is a chart on $M$ where $U:=\mathrm{im}(u^{-1})$.

Consider the coordinate representation $v \circ f \circ u^{-1}$ of a smooth 
mapping $f: M \to S^3$.
The function $v \circ f \circ u^{-1}$ assigns to a 4--tuple 
$(t,r,\theta,\varphi)$ a 3--tuple 
$(\psi, \Theta, \phi)$.
Therefore, $\psi, \Theta$ and $\phi$ are functions of $t,r,\theta$
and $\varphi$.
In what follows we restrict ourselves to mappings $f$ such that $\psi$ is a 
function of $t$ and $r$ only and $\Theta \equiv \theta$, $\phi \equiv \varphi$.
Such maps are called \emph{co--rotational}.

\subsection{Explicit Local Expressions}

\paragraph{Coordinate representations}
We define $\Phi:=v \circ f$ and calculate the local expression 
$h:=(\mathrm{tr}_\eta \Phi^* (v^{-1})^*g) \circ u^{-1}$. 
However, in order to restrict ourselves to co--rotational maps it is
desireable to work with the mapping $\Phi \circ u^{-1}: u(U) \to v(V)$ rather
than with $\Phi$ itself.
To do so we use the fact that the following
diagram commutes. 

$$
\begin{xy} 
\xymatrix{
 & 
(M,\eta) \ar@/_ 0.7cm/[ldd]_{\mathrm{tr}_\eta \Phi^* (v^{-1})^*g}
\ar[r]^f \ar[dr]_\Phi \ar[d]_u & 
(S^3,g) \ar[d]^v \ar[r]^i & (\mathbb{R}^4, \delta) \\
& (u(U), (u^{-1})^* \eta) 
\ar@/^/[ld]^{h} \ar[r]_{\Phi \circ u^{-1}}
& (v(V),(v^{-1})^*g) & \\
\mathbb{R}
}
\end{xy}
$$
Hence, we read off the useful identity 
$$ (\mathrm{tr}_\eta \Phi^* (v^{-1})^*g) \circ u^{-1} 
=\mathrm{tr}_{(u^{-1})^* \eta} (\Phi \circ u^{-1})^*(v^{-1})^*g. $$

\paragraph{The metric on the three--sphere}
The $i$--th partial derivative ($i=1,2,3$) for functions in $C^\infty(v(V),\mathbb{R})$ 
defines
a smooth vector field on $v(V)$ which will be denoted by $\partial_i$. 
We define
$g_{ij}:=(v^{-1})^*g(\partial_i, \partial_j)$, $i,j=1,2,3$.
A short calculation yields 
$$ (g_{ij}(\psi, \Theta, \phi))=\left (
\begin{array}{ccc}
1 & 0 & 0 \\
0 & \sin^2 \psi & 0 \\
0 & 0 & \sin^2 \psi \sin^2 \Theta 
\end{array}
\right ).
$$

\paragraph{Minkowski metric in spherical coordinates}
Similarly, one defines the vector fields $\partial_\mu$, $\mu=0, \dots, 3$
induced by the partial derivatives on $u(U)$.
The components 
$\eta_{\mu \nu}:=(u^{-1})^*\eta(\partial_\mu, \partial_\nu)$ of the Minkowski 
metric in spherical 
coordinates are 
given by
$$ (\eta_{\mu \nu}(t,r,\theta,\varphi))=\left ( 
\begin{array}{cccc}
-1 & 0 & 0 & 0 \\
0 & 1 & 0 & 0 \\
0 & 0 & r^2 & 0 \\
0 & 0 & 0 & r^2 \sin^2 \theta
\end{array} \right ).
$$

\paragraph{Calculation of the Lagrangian}
We make the ansatz $(\Phi \circ u^{-1})(t,r, \theta,
\varphi)=(\psi(t,r),\theta, \phi)$, i.e. we restrict ourselves to co--rotational
maps.
Calculating the local expression $h_{\mu \nu}:=(\Phi \circ
u^{-1})^*(v^{-1})^*g(\partial_\mu, \partial_\nu)$ for the pullback of the target
metric we obtain
$$ (h_{\mu \nu})=\left ( \begin{array}{cccc}
\psi_t^2 & \psi_t \psi_r & 0 & 0 \\
\psi_t \psi_r & \psi_r^2 & 0 & 0 \\
0 & 0 & \sin^2 \psi & 0 \\
0 & 0 & 0 & \sin^2 \psi \sin^2 \Theta
\end{array} \right ). $$
Then $h$ is given by $h=\mathrm{tr}
(\eta^{\mu \lambda} h_{\lambda \nu})$,
explicitly
$$ h(t, r)=-\psi_t^2(t,r)+\psi_r^2(t,r)
+\frac{2}{r^2}\sin^2 \psi(t,r). $$ 

\paragraph{The action functional}
Since the chart $(U,u)$ covers the whole Minkowski space apart from a set of 
Lebesgue
measure 0, integration over $M$ is equivalent to integration over $U$.
Applying the standard substitution rule we obtain
$$ S(\Phi)=\int_U \mathrm{tr}_\eta \Phi^* (v^{-1})^* g = \int_{u(U)} (\mathrm{tr}_\eta
\Phi^* (v^{-1})^* g) \circ u^{-1} |\det J_{u^{-1}}| $$
where $J_{u^{-1}}$ denotes the Jacobian matrix of the function $u^{-1}$ between 
the two open
sets $u(U)$ and $U$ of $\mathbb{R}^4$. 
Thus, critical points of $S$ can be obtained by calculating critical points of
the functional
\begin{equation}
\label{wm_eq_action}
\psi \mapsto 4 \pi \int_{- \infty}^\infty \int_0^\infty 
\left (\psi_t^2(t,r)-\psi_r^2(t,r)-\frac{2}{r^2}\sin^2 \psi(t,r) \right ) r^2 
dr dt 
\end{equation}
for $\psi: u(U) \to \mathbb{R}$ and setting
$\Phi \circ u^{-1}(t, r, \theta, \varphi):=(\psi(t,r), \theta, \varphi)$.
Then, $f:=v^{-1} \circ \Phi: M \to S^3$ is a co--rotational wave map.
$$ 
\begin{xy} 
\xymatrix{
M \ar[r]^{v^{-1} \circ \Phi} \ar[dr]_\Phi \ar[d]_u & 
S^3 \ar[d]^v \\
u(U) \ar[r]_{\Phi \circ u^{-1}} \ar[d]_\psi
& v(V) & \\
\mathbb{R}
}
\end{xy}
$$

Compactly supported variations of the functional (\ref{wm_eq_action}) together 
with integration by 
parts yields the equation
\begin{equation}
\label{wm_eq_wm}
\psi_{tt} - \psi_{rr} -\frac{2}{r} \psi_r + \frac{\sin(2 \psi)}{r^2}=0
\end{equation}
for critical points $\psi$.
Thus, in the co--rotational case the wave map problem reduces to the single
semilinear wave equation (\ref{wm_eq_wm}) for the function $\psi$.

\section{Additional Remarks and References}
To be precise one distinguishes between wave maps on a Lorentzian manifold and
the analogous mappings defined on a Riemannian manifold.
The latter ones are usually called \emph{harmonic maps}
and were introduced by Fuller \cite{Fuller1954}.
In particle physics wave maps first appeared in a paper by Gell--Mann and 
L\'evy \cite{Gell-Mann1960} as
\emph{nonlinear sigma models}.
For further applications in physics and historical remarks see the surveys 
\cite{Misner1978}, \cite{Misner1982} and references therein.
Misner \cite{Misner1978} points out that the wave maps equation contains
nonlinearities in a very natural geometric way. 
Due to the lack of a vector space structure on the target manifold, 
the resulting
field equation has to be nonlinear.
Therefore, wave maps provide a rich source for nonlinear field 
theories.
Moreover, they include well--known equations of
mathematical physics as special cases.
For instance, harmonic maps from Euclidean $\mathbb{R}^3$ to $\mathbb{R}$
satisfy the Laplace equation
while wave maps from Minkowski space to $\mathbb{R}$ are solutions of the wave
equation.
The simplest nonlinear example is a harmonic map from an open interval $(a,b)
\subset \mathbb{R}$ to a Riemannian manifold $M$.
The resulting field equation is the geodesic equation on $M$.
Therefore, the wave map functional provides an elegant and unified way for the
derivation of many interesting equations (see \cite{Misner1982} for more
examples).

For our purposes the most important feature is the geometric nature 
of the involved nonlinearities which classifies
wave maps as "naturally nonlinear
fields" \cite{Misner1982}.
Another example of such a "naturally nonlinear" field theory is general 
relativity and
therefore, wave maps are promising candidates for modelling the 
more involved nonlinearities of Einstein's equations.
Hence, the study of wave maps can provide insights into mechanisms which might
be common to many field theories of that kind, in particular to general
relativity which is the main physical motivation.
From the mathematical point of view eq. (\ref{wm_eq_wm}) provides a simple
example of a semilinear wave equation.
The study of nonlinear wave equations is a large field which is rapidly
developing and many difficult questions are still unanswered, even for examples
as simple as eq. (\ref{wm_eq_wm}).  
Therefore, the wave map equation (\ref{wm_eq_wm}) provides an interesting object
for a rigorous mathematical analysis.
For a recent survey on open mathematical questions concerning general wave 
maps see
\cite{Tataru2004}.
An introduction to semilinear wave equations and wave maps
as partial differential equations is given in \cite{Shatah1998}.

\chapter{Mathematical Preparation}
\thispagestyle{empty}
\label{math_ch}
Before continuing the analysis of wave maps, we have to introduce the
mathematical machinery required later on.
In this chapter we give an overview of function spaces, linear operators, 
spectral theory and semigroup theory.
These tools are required for the treatment of evolution
equations in an elegant abstract way.

\section{Function Spaces}
We define the standard Lebesgue and Sobolev spaces required for the analysis
of partial differential equations.
However, we restrict ourselves to the special case of spaces of functions defined on
(open subsets of) the
real line which is sufficient for our purposes.

\subsection{Banach Spaces}
\paragraph{Norms, completeness}
We start with the basic definition of a normed vector space. Let $X$ be a (not
necessarily finite--dimensional) nontrivial vector space
over the field $\mathbb{K}$ where
$\mathbb{K}$ is either $\mathbb{R}$ or $\mathbb{C}$.
A \emph{norm} on $X$ is a mapping $\|\cdot\|: X \to \mathbb{R}$ that satisfies
\begin{itemize}
\item $\|x\|\geq 0$ and $\|x\|=0 \Leftrightarrow x=0$
\item $\|\lambda x\|=|\lambda| \|x\|$
\item $\|x+y\| \leq \|x\|+\|y\|$ "triangle inequality" 
\end{itemize}
for all $\lambda \in \mathbb{K}$ and $x, y \in X$.
A vector space $X$ together with a norm $\|\cdot\|$ on $X$ (denoted by $(X,
\|\cdot\|)$) is said to be a \emph{normed vector space}.
For brevity we will write $X$ instead of $(X,\|\cdot\|)$ if the norm we refer to
follows from the context.
A normed vector space is a special case of a metric space (in the sense of point
set topology) since $d(x,y):=\|x-y\|$ defines a \emph{distance} on $X$.
On a metric space there exists a distinguished topology defined by the distance
function $d$ which is called the \emph{metric topology}.
The \emph{open $\varepsilon$ balls} $B_\varepsilon(x):=\{y \in X:
\|x-y\|<\varepsilon\}$ form a basis of this topology.
Therefore it follows by definition that $x \mapsto \|x\|$ is a continuous 
mapping from $X$ to $\mathbb{R}$ with respect to the metric topology.

A \emph{Cauchy sequence} in $X$ is a sequence $(x_k) \subset X$ with the
property that for
every $\varepsilon>0$ there exists a $N \in \mathbb{N}$ such that
$\|x_k-x_j\|<\varepsilon$ for all $k,j > N$.
If every Cauchy sequence in $X$ has a limit in $X$, the normed vector space
$X$ is said to be $\emph{complete}$.
A complete normed vector space is called a $\emph{Banach space}$.
Not every normed vector space is a Banach space.
Heuristically spoken it is possible that a Cauchy sequence 
"converges to a hole".
We give several elementary examples.

\begin{example}
The vector space $\mathbb{R}^n$ together with the Euclidean norm
$\|x\|^2:=\sum_{k=1}^n x_k^2$ for $x=(x_1,\dots,x_n) \in \mathbb{R}^n$ is a Banach
space which follows from the completeness of the real numbers.
Similarly, every finite--dimensional normed vector space is complete. 
\end{example}

\begin{example}
\label{banach_ex_c}
The space $C[a,b]$ of continuous functions from an interval $[a,b]$ to
$\mathbb{C}$ together with the norm $\|u\|_{C[a,b]}:=\sup_{x \in [a,b]}|u(x)|$, 
$u \in C[a,b]$ is a Banach
space.
Similarly, the space $C^k[a,b]$ of $k$--times continuously differentiable
functions from $[a,b]$ to $\mathbb{C}$ together with $\|u\|_{C^k[a,b]}:=
\sum_{j=0}^k
\sup_{x \in [a,b]}|u^{(j)}(x)|$, $u \in C^k[a,b]$ is a Banach space.
However, the space $C^1[a,b]$ equipped with the norm $\|u\|_{C[a,b]}$, 
$u \in C^1[a,b]$ provides an example of a normed vector space
which is not complete. 
\end{example} 

\paragraph{Dense subsets, completion}
A subset $A \subset X$ is said to be \emph{dense} if for every $x \in X$ there
exists a sequence $(x_k) \subset A$ such that $x_k \to x$ in $X$.
A Banach space is called \emph{separable} if there exists a countable dense
subset.

\begin{example}
Weierstra\ss' approximation theorem states that polynomials are dense in 
$C[a,b]$
and therefore, $C[a,b]$ is separable since the set of all polynomials is
countable.
\end{example} 

For every normed vector space $(X,\|\cdot\|_X)$ there exists a \emph{completion},
i.e. a Banach space $(\overline{X}, \|\cdot\|)$ such that $X$ is dense in
$\overline{X}$ and $\|x\|_X=\|x\|$ for all $x \in X$.
A completion is unique up to isomorphisms. \footnote{To be more precise one 
requires the existence of a Banach space $(Y, \|\cdot\|_Y)$ and a
norm--preserving linear mapping $i: X \to Y$ (an \emph{isometry}) such that
$i(X)$ is dense in $Y$. 
Then $Y$ is called the completion of $X$. 
In our definition we have implicitly identified $X$ with $i(X)$ which is
possible since $\|i(x)\|_Y=\|x\|_X$ for all $x \in X$  
implies injectivity of $i$.
}  

\begin{example}
\label{banach_ex_Lp}
Consider the space $C^\infty_c(a,b)$ of infinitely often differentiable 
functions from
$(a,b)$ to $\mathbb{C}$ having compact support.
For $p \geq 1$ we define 
$$ \|u\|_{L^p(a,b)}:=\left ( \int_a^b |u(x)|^p dx \right )^{1/p} $$ for $u \in
C^\infty_c(a,b)$.
It is a consequence of Minkowski's inequality that $\|\cdot\|_{L^p(a,b)}$ 
satisfies the
triangle inequality and therefore, $\|\cdot\|_{L^p(a,b)}$ is a norm on 
$C^\infty_c(a,b)$.
The normed vector space $(C^\infty_c(a,b), \|\cdot\|_{L^p(a,b)})$ is not
complete.
Its completion is denoted by $L^p(a,b)$.
\end{example}

\paragraph{Dual space, reflexivity}
A \emph{bounded linear functional} on a normed vector space $(X, \|\cdot\|_X)$
is a linear mapping $f: X \to \mathbb{K}$ for which there exists a constant
$c>0$ such that $|f(x)|\leq c \|x\|_X$ for all $x \in X$.
We define 
the \emph{(topological)
dual space} $X^*$ of $X$ as the set of all bounded linear functionals on $X$.
One can define a norm on $X^*$ by $\|f\|_{X^*}:=\sup\{|f(x)|: x \in X, \|x\|_X 
\leq 1\}$.
It is easy to prove that $(X^*, \|\cdot\|_{X^*})$ is a Banach space (even if $X$
is not complete).
There exists a natural inclusion map $i: X \to X^{**}$ from $X$ to the
\emph{bidual space} $X^{**}$ given by
$(i(x))(f):=f(x)$ ($x \in X, f \in X^*$).
The linear map $i$ is an \emph{isometry}, i.e. it is norm--preserving. 
However, in general $i$ is not surjective and therefore one cannot identify $X$
with $X^{**}$.
A Banach space $X$ which can be identified with its bidual space $X^{**}$ (i.e.
the inclusion $i$ is an isometric isomorphism) is
said to be \emph{reflexive}.

\begin{example}
Consider the Banach space $X:=C[0,1]$ (ex. \ref{banach_ex_c}).
The linear mapping $f: C[0,1] \to \mathbb{C}$ given by $f(u):=u(0)$ for $u \in
C[0,1]$ is bounded since $\|u\|_{C[0,1]} \leq 1$ implies $|u(0)| \leq 1$ and 
therefore
$\|f\|_{X^*} \leq 1$ (recall the definition of $\|\cdot\|_{X^*}$). 
Hence, $f \in X^*$.
\end{example}

\begin{example}
Let $X:=C[0,1]$ and define a norm on $X$ by 
$$ \|u\|_X:=\int_0^1 |u(x)|dx $$
for $u \in X$.
Consider the mapping $f: X \to \mathbb{K}$ given by $f(u):=u(0)$. Then $f$ is
linear but it is not bounded: It is easy to construct a sequence $(u_k)$
of continuous
functions with $\|u_k\|_X=1$ for all $k$ and $u_k(0) \to \infty$ for $k \to
\infty$. Therefore, $\|f\|_{X^*}$ does not exist and $f \notin X^*$.  
\end{example}

\subsection{Hilbert Spaces}
A Hilbert space is a special case of a Banach space defined by the existence of
an inner product.
Many issues are dramatically simplified in that case and therefore Hilbert 
spaces deserve special attention.

\paragraph{Inner products}
Let $H$ be a vector space over $\mathbb{K}$. An \emph{inner product}
$(\cdot | \cdot)$ is a mapping from $H \times H$ to $\mathbb{K}$ that satisfies
\begin{itemize}
\item $(\lambda x + \mu y | z)=\lambda (x | z) + \mu (y | z)$
\item $(x|y)=\overline{(y|x)}$
\item $(x|x)>0$ if $x \not= 0$
\end{itemize}
for all $x,y,z \in H$ and $\lambda, \mu \in \mathbb{K}$. 
It is easily seen that $\|x\|:=\sqrt{(x|x)}$, $x \in H$ defines a norm on $H$.
$H$ is said to be a \emph{Hilbert space} if $(H, \|\cdot\|)$ is a Banach space.

\begin{example}
The vector space $\mathbb{R}^n$ together with the Euclidean inner product
$(x|y):=\sum_{k=1}^n x_k y_k$ for $x:=(x_1, \dots, x_n), y:=(y_1, \dots, y_n)
\in \mathbb{R}^n$ is a (finite--dimensional) Hilbert space.
\end{example}

\begin{example}
The space $L^2(a,b)$ (cf. ex. \ref{banach_ex_Lp}) equipped with the inner
product 
$$ (u|v)_{L^2(a,b)}:=\int_a^b u(x)\overline{v(x)}dx $$
is a prominent example of a Hilbert space.
\end{example}

\paragraph{Dual space}
The characterization of the dual space of a Hilbert space is drastically
simplified by the so--called \emph{Riesz representation theorem}.

\begin{theorem}
\label{hilbert_thm_Riesz}
Let $H$ be a Hilbert space over $\mathbb{K}$ and $f: H \to
\mathbb{K}$ a bounded linear functional on $H$.
Then there exists a unique $y \in H$ such that $f(x)=(x|y)$ for all $x \in H$
where $(\cdot | \cdot)$ denotes the inner product on $H$.
\end{theorem}

The proof is similar to the finite--dimensional case.
Due to the Riesz representation theorem it is possible to identify the dual
of a Hilbert space with the Hilbert space itself.
In particular it follows that every Hilbert space is reflexive.

\subsection{Lebesgue-- and Sobolev Spaces}
Now we switch from abstract Banach spaces to concrete function spaces.
However, as mentioned in the beginning we will restrict ourselves to functions
defined on one--dimensional real intervals.
For the general definition of Sobolev spaces on subsets of $\mathbb{R}^n$ 
see e.g. \cite{Evans1998}.
For an extensive treatment of weighted Sobolev spaces as well as more general
function spaces we refer to
\cite{Triebel1995}.

\paragraph{Lebesgue spaces}
Let $(a,b) \subset \mathbb{R}$ be an open interval and $p \geq 1$.
In ex. \ref{banach_ex_Lp} we have already defined the \emph{Lebesgue space} 
$L^p(a,b)$ as the completion of compactly supported smooth functions with
respect to the $L^p$--norm.
We note the important fact that in general a function $u \in L^p(a,b)$ cannot 
be evaluated at a point, i.e. $u(x)$ for $x \in (a,b)$ is not well--defined!
The reason for this lies in the process of completion.
By construction the completion is in fact a quotient space and therefore it
consists of equivalence classes of functions rather than functions themselves.
Two functions belong to the same class if they coincide almost everywhere, i.e.
up to a set of Lebesgue measure 0.
A single point $x \in (a,b)$ has Lebesgue measure $0$ and therefore, one can
assign arbitrary values to $u(x)$ without changing the function class.
Thus, it makes no sense to speak of the value of a function $u \in L^p(a,b)$ at
a point $x \in (a,b)$.
An important exception exists if a function class contains a continuous element.
In this case one can define the value at a point by the
evaluation of the unique continuous function in the class at that point.
In the sequel we will do that implicitly.

Additionally we remark that one often encounters statements like "$u_j \to u$
in $L^p(a,b)$ implies $u_j(x) \to u(x)$ for almost all $x \in (a,b)$" although,
strictly speaking, this makes no sense.
For convenience we will adopt this sloppy formulation having in mind what is 
actually meant by this, namely that $u_j \to u$ in $L^p(a,b)$ implies that
$v_j(x) \to v(x)$ for almost all $x \in (a,b)$ and for 
\emph{all representatives $v_j$
and $v$ of the function classes $u_j$ and $u$, respectively}. 

\paragraph{Weighted Lebesgue spaces}
A \emph{weight function} is a continuous, positive function $w: (a,b) \to
\mathbb{R}$.
Define the weighted norm 
$$ \|u\|_{L^p_w(a,b)}:=\left ( \int_a^b |u(x)|^p w(x)dx \right )^{1/p} $$
for $u \in C^\infty_c(a,b)$.
The completion of $C^\infty_c(a,b)$ with respect to the norm
$\|\cdot\|_{L^p_w(a,b)}$ is referred to as the \emph{weighted Lebesgue space
$L^p_w(a,b)$ with weight $w$}.
Again, the space $L^2_w(a,b)$ is a Hilbert space.

\paragraph{Weak derivatives}
In order to apply functional analytic methods to partial differential equations
it is necessary to generalize the notion of differentiability.
The space $L^p_{\mathrm{loc}}(a,b)$ of \emph{locally Lebesgue integrable functions} 
is defined by
$$ L^p_{\mathrm{loc}}(a,b):=\{u: (a,b) \to \mathbb{C}: u \in L^p(c,d) \mbox{ for all }
[c,d] \subset (a,b)\}. $$
Inspired by the integration--by--parts--formula one says a function $u \in
L^p(a,b)$ is \emph{weakly differentiable} if there exists a $v \in
L^1_{\mathrm{loc}}(a,b)$ such that
$$ \int_a^b u(x) \varphi'(x)dx=-\int_a^b v(x) \varphi(x) dx $$
for all $\varphi \in C^\infty_c(a,b)$.
In this case one sets $u':=v$ and $u'$ is said to be the 
\emph{weak derivative} of $u$.  
A weak derivative is unique (as a function class).
As usual we adopt the notation $u^{(j)}$ for the $j$--th weak derivative and set
$u^{(0)}:=u$.

\paragraph{Sobolev spaces}
The Sobolev space $W^{k,p}(a,b)$, $k \in \mathbb{N}$ is defined to be the 
vector space of functions $u \in L^p(a,b)$ such that weak derivatives
$u^{(j)}$ for all $j=1, \dots , k$ exist and belong to $L^p(a,b)$.
The normed vector space $(W^{k,p}(a,b), \|\cdot\|_{W^{k,p}(a,b)})$, where
$$ \|u\|_{W^{k,p}(a,b)}:=\left ( \sum_{j=0}^k \|u^{(j)}\|_{L^p(a,b)}^p \right
)^{1/p}, $$
is a Banach space.
The special cases $H^k(a,b):=W^{k,2}(a,b)$ are of particular interest since they
are Hilbert spaces with respect to the inner product
$$ (u|v)_{H^k(a,b)}:=\sum_{j=0}^k \int_a^b u^{(j)}(x)\overline{v^{(j)}(x)}dx. $$
Note the set inclusions $C^\infty_c(a,b) \subset W^{k,p}(a,b) \subset L^p(a,b)$.
It follows that $W^{k,p}(a,b)$ is dense in $L^p(a,b)$ for all $k \in
\mathbb{N}$.

\paragraph{Weighted Sobolev spaces}
For a $(k+1)$--tuple $w=(w_0, w_1, \dots, w_k)$ of weight functions we define 
the weighted
Sobolev space $W^{k,p}_w(a,b)$ as the vector space of functions 
$u \in L^p_{w_0}(a,b)$ such that weak derivatives $u^{(j)}$ for all $j=1, \dots,
k$ exist and $u^{(j)} \in L^p_{w_j}(a,b)$.
The \emph{weighted Sobolev norm} $\|\cdot\|_{W^{k,p}_w(a,b)}$ is defined by
$$ \|u\|_{W^{k,p}_w(a,b)}:=\left ( \sum_{j=0}^k \|u^{(j)}\|_{L^p_{w_j}}^p \right
)^{1/p}. $$
Again, $\NVS{W^{k,p}_w(a,b)}$ is a Banach space.
In particular, $H^k_w(a,b):=W^{k,p}_w(a,b)$ is a Hilbert space with respect to
the inner product
$$ (u|v)_{H^k_w(a,b)}:=\sum_{j=0}^k \int_a^b w_j(x)u^{(j)}(x)\overline{v^{(j)}(x)}dx. $$

\section{Operator Theory}
In this section we give a brief review of well--known facts about basic operator
theory. All of the material presented here can be found in standard textbooks, e.g.
Kato's classic \cite{Kato1980}.

\subsection{Linear Operators}
\paragraph{Linear operators, extensions, restrictions}
A linear operator $A$ on a Banach space $(X, \|\cdot\|_X)$ is a 
linear mapping from a subspace $\mathcal{D}(A) \subset X$ (the \emph{domain}) to
$X$.
We write $A: \mathcal{D}(A) \subset X \to X$ to emphasize that
topological properties of $A$ are considered with respect to the
topology on $X$ and not to a possible different topology on $\mathcal{D}(A)$. 
It is essential for the definition of a linear operator to specify the domain
$\mathcal{D}(A)$.
Let $A$ and $B$ be two linear operators on a Banach space $(X, \|\cdot\|_X)$.
$A$ and $B$ are considered to be \emph{equal} if 
both their 
domains and 
their mapping rules coincide, i.e. $A=B$ if and only if
$\mathcal{D}(A)=\mathcal{D}(B)$ and $Ax=Bx$ for all $x \in \mathcal{D}(A)$.
$B$ is said to be an \emph{extension} (a \emph{restriction}) of $A$ if 
$\mathcal{D}(A) \subset
\mathcal{D}(B)$ ($\mathcal{D}(A) \supset
\mathcal{D}(B)$) and $Ax=Bx$ for all $x \in \mathcal{D}(A) \cap \mathcal{D}(B)$.  

\begin{example}
Let $X:=C[0,1]$ (cf. ex. \ref{banach_ex_c}), $\mathcal{D}(A):=C^1[0,1]
\subset X$ and $Au:=u'$, $u \in \mathcal{D}(A)$.
Then $A$ is a linear operator on $X$.
Define $\mathcal{D}(B):=C^2[0,1]$, $Bu:=u'$ for $u \in \mathcal{D}(B)$. 
Then $B$ is a restriction of $A$.
\end{example} 

\paragraph{Sums and products}
The sum $A+B$ of two linear operators on $X$ is defined by
$\mathcal{D}(A+B):=\mathcal{D}(A) \cap \mathcal{D}(B)$ and 
$(A+B)x:=Ax+Bx$ for $x \in \mathcal{D}(A+B)$. 
Note that $A+B$ might be trivial, i.e. $\mathcal{D}(A+B)=\{0\}$.
The \emph{product} of $A$ and $B$ is defined by 
$$\mathcal{D}(AB):=\{x \in \mathcal{D}(B): Bx \in \mathcal{D}(A)\}$$
and $ABx:=A(Bx)$ for $x \in \Dom{AB}$.
Again, $AB$ might be trivial although $A$ and $B$ are not.

\subsection{Bounded and Closed Operators}
\paragraph{Bounded operators}
A linear operator $A: \Dom{A} \subset X \to X$ is said to be \emph{bounded} if 
$\mathcal{D}(A)=X$ and there exists a
constant $c>0$ such that $\|Ax\|_X \leq c\|x\|_X$ for all $x \in X$.
Since $A$ is linear, boundedness is equivalent to continuity.
The vector space of bounded linear operators on $X$ is denoted by $\mathcal{B}(X)$.
It is possible to define a norm on $\B{X}$ by
$$ \|A\|_{\B{X}}:=\sup\{\|Ax\|_X: x \in X, \|x\|_X \leq 1\}, $$
the so--called $\emph{operator norm}$.
It turns out that $\NVS{\B{X}}$ is a Banach space (provided that $X$ is
complete).

\begin{example}
Let $X=C[0,1]$ (cf. ex. \ref{banach_ex_c}) and define $\Dom{A}:=X$, 
$$ (Au)(t):=\int_0^t u(s)ds \mbox{, } u \in X. $$
Then
$$ |Au(t)|=\left |\int_0^t u(s)ds \right | \leq \int_0^t |u(s)|ds \leq \int_0^1
|u(s)|ds \leq \|u\|_X $$
for all $t \in [0,1]$ and therefore $\|Au\|_X \leq \|u\|_X$ for all $u \in X$.
Since $A$ is linear this shows $A \in \B{X}$. 
\end{example}

\paragraph{Closed operators}
Without requiring additional analytical properties, the study of unbounded
linear operators reduces to linear algebra.
An important property which is often present in applications
is closedness which is a certain generalization of
continuity.
A linear operator $A: \Dom{A} \subset X \to X$ on a Banach space $\NVS{X}$ is
said to be \emph{closed} if for any sequence $(x_n) \subset \Dom{A}$ with $x_n
\to x$ and $Ax_n \to y$ it follows that $x \in \Dom{A}$ and $Ax=y$.
If $A$ is bounded then $\Dom{A}=X$ and $x_n \to x$ already implies $Ax_n \to Ax$.
Hence, any bounded operator is closed.
The converse, of course, is not true.

\begin{example}
Consider again the Banach space $X:=C[0,1]$ (cf. ex. \ref{banach_ex_c}). Define
$$ \Dom{A}:=\left \{u \in X: t \mapsto \frac{u(t)}{t} \in X \right \} $$
and 
$$ (Au)(t):=\frac{u(t)}{t} \mbox{, } u \in \Dom{A}. $$
Then, $A$ is a linear operator on $X$.
It is easy to show that $A$ is closed but it is not bounded.
\end{example} 

The \emph{closed graph theorem} states an important result concerning closed
operators.

\begin{theorem}
Let $A: \Dom{A} \subset X \to X$ be a closed linear operator on a Banach space
$X$. If $\Dom{A}=X$ then $A$ is bounded.
\end{theorem}

This theorem has many useful applications.
If $A: \Dom{A} \subset X \to X$ is bijective we can define the 
\emph{inverse operator} $A^{-1}: X \to X$ ($A^{-1}(Ax)=x$ for all $x \in
\Dom{A}$ and $A(A^{-1}x)=x$ for all $x \in X$).
One easily shows that the inverse of a bijective closed operator is closed.
Since $\Dom{A^{-1}}=X$ it follows by the closed graph theorem that $A^{-1}$ is
bounded.
This observation is important for the spectral theory of closed operators. 

\paragraph{Closeable operators, closure, graph norm}
A linear operator $A: \Dom{A} \subset X \to X$ on a Banach space $\NVS{X}$ is
said to be \emph{closeable} if it admits a closed extension.
If $A$ is closeable then the smallest closed extension of $A$ is called the
\emph{closure} of $A$ and denoted by $\overline{A}$.
There exists an easy criterion to decide whether $A$ is closeable or not.
It turns out that $A$ is closeable if and only if $x_n \to 0$ for $x_n \in
\Dom{A}$ and $Ax_n \to y$ imply $y=0$.

Let $A$ be closed. 
A subspace $Y \subset \Dom{A}$ is said to be a \emph{core} of $A$ if there exists a
closeable operator $B: \Dom{B} \subset X \to X$ with $\Dom{B}=Y$ and
$\overline{B}=A$. 

It is possible to define a new norm on the domain of $A$, the \emph{graph norm}
$\|\cdot\|_A$ given by 
$$ \|u\|_A^2:=\|Au\|_X^2+\|u\|_X^2 \mbox{, } u \in \Dom{A}. $$
$\Dom{A}$ equipped with $\|\cdot\|_A$ is a normed vector space which is complete 
if and only if $A$ is closed.

\subsection{Self--Adjointness}

\paragraph{The adjoint}
Let $A: \Dom{A} \subset H \to H$ and $B: \Dom{B} \subset H \to H$ be linear 
operators on a Hilbert space $(H, (\cdot | \cdot)_H)$.
$B$ is said to be \emph{adjoint} to $A$ if $(Ax|y)_H=(x|By)_H$ for all $x \in
\Dom{A}$ and $y \in \Dom{B}$.

\begin{example}
Let $H:=L^2(0,1)$, $\Dom{A}:=C^1[0,1]$ and $Au:=u'$ for $u \in \Dom{A}$. Then
$A$ is a linear operator on the Hilbert space $H$.
Define $\Dom{B}:=\{u \in C^1[0,1]: u(0)=u(1)=0\}$ and $Bu:=-u'$ for $u \in
\Dom{B}$. Then $B$ is a linear operator on $H$ which is adjoint to $A$
(integration by parts).
\end{example}

In general there are many operators which are adjoint to a given operator $A$. 
However, if $A$ is densely defined (i.e. $\Dom{A}$ is dense in $H$) then there
exists a unique maximal operator $A^*$ adjoint to $A$ (i.e. $B \subset A^*$ for
all $B$ adjoint to $A$). 
$A^*$ is called \emph{the adjoint} and is constructed as follows.
$$ \Dom{A^*}:=\{y \in H: \exists z \in H \mbox{ such that } (Ax|y)_H=(x|z)_H
\mbox{ for all } x \in \Dom{A}\} $$
and whenever $(Ax|y)_H=(x|z)_H$ holds for all $x \in \Dom{A}$ we write $A^*y=z$.
Since $\Dom{A}$ is dense in $H$ this construction yields a well--defined 
operator $A^*$ on $H$.
It turns out that $A^*$ is always closed and therefore $A^{**}$ is a closed
extension of $A$.

\paragraph{Symmetric operators, self-adjointness}
A densely defined linear operator $A: \Dom{A} \subset H \to H$ on a 
Hilbert space $(H, (\cdot|\cdot)_H)$ is said to
be \emph{symmetric} (or \emph{hermitian}) if $(Ax|y)_H=(x|Ay)_H$ for all 
$x,y \in \Dom{A}$.
It is said to be \emph{self--adjoint} if $A=A^*$ (i.e. $A$ is symmetric 
\emph{and} $\Dom{A}=\Dom{A^*}$).
Obviously, self--adjoint operators are closed.
A symmetric operator $A$ is said to be \emph{essentially self--adjoint} if its
closure $\overline{A}$ is self--adjoint.

\begin{example}
Let $H:=L^2(0,1)$, $\Dom{A}:=\{u \in C^1[0,1]: u(0)=u(1)=0\}$, $Au:=-iu'$.
Then $A: \Dom{A} \subset H \to H$ is symmetric but it is not self-adjoint: 
Let $v \in C^1[0,1]$ with $v(0) \not=0$.
Then we have $-iv' \in H$ and $(Au|v)_H=(u|-iv')_H$ for all $u \in \Dom{A}$. 
This shows that
$v \in \Dom{A^*}$ but $v \notin \Dom{A}$ and hence, $\Dom{A} \not=
\Dom{A^*}$.
\end{example}

\section{Spectral Theory}
We give an outline of the spectral theory of closed operators.
Again, we refer to \cite{Kato1980} for a detailed treatment of this subject.
In what follows we assume $A: \Dom{A} \subset X \to X$ to be a closed linear
operator on a Banach space $\NVS{X}$ over $\mathbb{C}$.

\subsection{The Resolvent}

\paragraph{Resolvent set, resolvent}
For an operator on a finite--dimensional vector space the spectrum is defined as
the set of all eigenvalues.
In principle this definition could be carried over to the infinite--dimensional
case without any modifications since the notions "eigenvalue"
and "eigenvector" still
make sense.
However, if one intends to generalize results from finite--dimensional
spectral theory it turns out that this definition is too restrictive.
It is more fruitful to consider the resolvent instead.
One defines the \emph{resolvent set} $\rho(A)$ of the operator $A$ as the set 
of all $\lambda \in \mathbb{C}$ such that the inverse of the operator
$\lambda-A$ exists and is bounded. 
The resolvent set is always open.
For $\lambda \in \rho(A)$ we define the \emph{resolvent} $R_A(\lambda):=(\lambda
- A)^{-1}$.

\paragraph{Properties of the resolvent}
Since $A$ is closed it follows that $\lambda - A$ is closed as well.
Therefore, $(\lambda-A)^{-1}: X \to X$ is automatically bounded (if it exists)
because it is the inverse of a 
closed operator (closed graph theorem).
This means that 
$\lambda \in \rho(A)$ is equivalent to $\lambda - A$ being bijective.
The resolvent $R_A$ is an operator--valued function on $\rho(A)$:
It assigns to
every complex number $\lambda \in \rho(A)$ a bounded linear operator 
$R_A(\lambda): X \to X$.
In abstract notation this means that $R_A$ is a mapping from $\rho(A)$
to $\B{X}$.
One can show that $R_A: \rho(A) \to \B{X}$ is piecewise holomorphic
(piecewise because $\rho(A)$ is not necessarily connected).
Holomorphy in this context is defined as follows: Let $f: U \subset \mathbb{C}
\to \B{X}$ be a mapping from an open subset $U$ of 
$\mathbb{C}$ to the Banach
space of bounded operators on $X$.
Then $f$ is said to be complex differentiable at $z_0 \in U$ if 
$$ \lim_{z \to z_0}\frac{\|f(z)-f(z_0)\|_{\B{X}}}{z-z_0} $$
exists.
$f$ is said to be \emph{holomorphic} on $U$ if it is
complex differentiable at all $z \in U$.   

\subsection{Spectra}

\paragraph{The spectrum}
The \emph{spectrum} $\sigma(A)$ of $A$ is defined as the complement of 
$\rho(A)$ in $\mathbb{C}$, i.e. $\sigma(A):=\mathbb{C} \backslash \rho(A)$.
Therefore, the spectrum is the set of all $\lambda \in \mathbb{C}$ such that the
operator $\lambda-A$ fails to be bijective.
In the finite--dimensional case the failure of $\lambda-A$ to be
bijective is equivalent to $\lambda-A$ having a nontrivial kernel.
In infinite dimensions there are more possibilities.
Hence, there are different types of spectra.
\begin{itemize}
\item The point spectrum $\sigma_p(A)$: 
We say that $\lambda$ is in the \emph{point spectrum} $\sigma_p(A)$ of $A$ if
there exists an $x \in \Dom{A}$, $x \not= 0$ such that $Ax=\lambda x$, 
i.e. $\lambda-A$ fails to be injective.
Like in finite dimensions, $x$ is called an \emph{eigenvector} and
elements of $\sigma_p(A)$ are called \emph{eigenvalues}.
\item The continuous spectrum $\sigma_c(A)$:
$\lambda \in \mathbb{C}$ belongs to the \emph{continuous spectrum} $\sigma_c(A)$ of $A$
if $\lambda - A$ is injective but 
fails to be surjective and the range of $\lambda-A$ is dense in
$X$. 
\item The residual spectrum $\sigma_r(A)$:
$\lambda \in \mathbb{C}$ is said to be an element of the \emph{residual
spectrum} $\sigma_r(A)$ of $A$ if $\lambda-A$ is injective but 
fails to be surjective and the
range of $\lambda-A$ is not dense in $X$.
\end{itemize}
These three cases cover all possibilities.
Therefore we have $\sigma(A)=\sigma_p(A) \cup \sigma_c(A) \cup \sigma_r(A)$.
Note that $\sigma_p(A)$, $\sigma_c(A)$ and $\sigma_r(A)$ are disjoint.
In the finite--dimensional case we have $\sigma(A)=\sigma_p(A)$.
We remark that there is no overall standard for these definitions.
Other classifications of 
spectra are possible and may be found in the literature. 

\subsection{Structure of the Spectrum}
\label{strucspec_sec}

\paragraph{The spectra of closed operators}
In a finite--dimensional vector space over $\mathbb{C}$ the spectrum of a linear
operator is bounded and nonempty due to the fundamental theorem of algebra.
This is not true anymore when one considers unbounded operators in 
infinite--dimensional Banach spaces.
We give two very simple examples to illustrate different spectral behaviour
which one has to expect.

\begin{example}
\label{spec_ex_whole}
Let $X:=C[0,1]$ (cf. ex. \ref{banach_ex_c}), $\Dom{A}:=C^1[0,1]$ and define
$Au:=u'$ for
$u \in \Dom{A}$. Then $A: \Dom{A} \subset X \to X$ is an unbounded linear 
operator on $X$.
We show that $A$ is closed. 
Consider a sequence $(u_k) \in \Dom{A}$ with $u_k \to u$ and $Au_k \to f$.
Note that
$$ \lim_{k \to \infty}u_k(t)=\lim_{k \to \infty} \left ( 
u_k(0)+\int_0^t u'_k(s)ds \right ). $$
Since $Au_k=u'_k \to f$ uniformly, we can interchange the limit with
the integral sign and obtain
$$ u(t)=u(0)+\int_0^t f(s)ds. $$
Hence, $u \in \Dom{A}$ and $Au=f$ which proves closedness of $A$. 

Now we calculate the spectrum of $A$. For every $\lambda \in \mathbb{C}$ the
equation $(\lambda - A)u=0$ has a nontrivial solution $u(t):=e^{\lambda t}$ in
$\Dom{A}$.
This shows that $\sigma(A)=\sigma_p(A)=\mathbb{C}$ and $\rho(A)=\emptyset$.
We remark that the term "point spectrum" might be misleading as this
example shows. It is possible that the point spectrum consists of the whole 
complex plane 
and therefore it is not discrete in general. 
\end{example}

\begin{example}
Let $X:=C[0,1]$, $\Dom{A}:=\{u \in C^1[0,1]: u(0)=0\}$ and $Au=u'$ for $u \in
\Dom{A}$.
Then $A: \Dom{A} \subset X \to X$ is an unbounded linear operator on $X$.
A similar reasoning as in ex. \ref{spec_ex_whole} shows that $A$ is closed.
To calculate the point spectrum of $A$ we consider the equation
$(\lambda-A)u=0$.
Formally, this equation has the solution $u(t):=ce^{\lambda t}$, $c \in
\mathbb{C}$ but $u \notin
\Dom{A}$ as long as $c \not= 0$.
Therefore, the only solution which is in $\Dom{A}$ is the trivial one.
This shows that $\sigma_p(A)=\emptyset$.
Next we show surjectivity of $\lambda - A$.
Let $f \in X$ and consider the equation $(\lambda - A)u=f$.
A formal solution is given by 
$$ u(t)=-e^{\lambda t} \int_0^t e^{- \lambda s}f(s)ds $$
as is easily checked by direct computation.
The so--defined $u$ is continuously differentiable, satisfies $u(0)=0$ and
hence,
is an element of $\Dom{A}$. 
Therefore, $\lambda - A$ is surjective for all $\lambda \in \mathbb{C}$ which,
together with $\sigma_p(A)=\emptyset$, shows $\sigma(A)=\emptyset$ and
$\rho(A)=\mathbb{C}$.
The resolvent $R_A(\lambda): X \to X$ is given by
$$ (R_A(\lambda)f)(t)=-e^{\lambda t} \int_0^t e^{-\lambda s}f(s)ds. $$
\end{example}



These two examples show that the spectrum of a closed operator can fill the 
whole complex plane or it
may be empty.
Roughly speaking, everything in between is also possible.
Therefore, the spectra of general closed operators can be very complicated.
However, for many differential operators arising from 
mathematical physics one rediscovers the familiar situation of having a pure 
point spectrum which is discrete. 
The analysis is additionally simplified if the problems can be formulated in a
self--adjoint way.
In this case one has a good knowledge of the structure of the spectrum.

\paragraph{The Spectrum of a self--adjoint operator}
Like a symmetric operator in finite dimensions, 
a self--adjoint operator can only have real
eigenvalues. This statement is already true for symmetric operators and the 
proof is the same as in finite dimensions.
However, in the case of a self--adjoint operator a much stronger result holds.
It turns out that the whole spectrum has to be a subset of the real axis and 
not only the point spectrum.
Furthermore, self--adjoint operators do not have residual spectra as the
following argument shows.
Consider a self--adjoint operator $A: \Dom{A} \subset H \to H$ on a Hilbert
space $H$.
Suppose $\lambda \in \sigma_r(A)$.
It follows that $\lambda$ is real and
by definition of
$\sigma_r(A)$, $\lambda$ is not an eigenvalue and the range of $\lambda-A$ is
not dense in $H$.
Hence, there exists a $y \in H$, $y \not= 0$ such that $((\lambda-A)x|y)_H=0$
for all $x \in \Dom{A}$.
This implies $(Ax|y)_H=(x|\lambda y)_H$ for all $x \in \Dom{A}$ which shows that
$y \in \Dom{A^*}=\Dom{A}$.
Therefore, we have $(x|Ay)_H=(x|\lambda y)_H$ for all $x \in \Dom{A}$ and, since
$\Dom{A}$ is dense in $H$, we conclude that $Ay=\lambda y$ which shows that
$\lambda$ is an eigenvalue of $A$, a contradiction.

Moreover, due to their importance for quantum mechanics, self--adjoint operators
have been investigated systematically for a long time and there exists a 
well--developed general theory (cf. \cite{Reed1975}, \cite{Reed1978}).
For instance, there is a spectral theorem and it is possible to define
functions of self--adjoint operators with the help of the so--called functional
calculus.
These properties simplify the analysis tremendously 
and therefore it is desireable to formulate the problem to be studied in a
self--adjoint way whenever this is possible.

\section{Semigroup Theory}
A thorough introduction to the theory of semigroups of linear operators can be
found in \cite{Engel2000}. We also mention the classic \cite{Pazy1983}.
For a treatment of semigroup theory with a focus on physical applications (in
particular relativity and astrophysics) we refer to \cite{Beyer2007}.

\subsection{Motivation}
Let $X$ be a Banach space, $A: \Dom{A} \subset X \to X$ a linear operator and
$u: [0,\infty) \to X$ a function.
We are interested in solving the evolution equation
\begin{equation}
\label{semigroup_eq_odebanach}
\frac{d}{dt}u(t)=A u(t) 
\end{equation}
with initial data $u(0)=u_0$.
Formally, the solution is given by $u(t)=\exp(tA)u_0$.
Therefore, our aim is to give precise meaning to the exponential of a linear
operator. 
There is no problem if the operator $A$ is bounded. 
In this case the exponential series converges absolutely and therefore,
$\exp(tA)$ can be defined as usual via a power series.
This situation is completely analogous to the case of a linear operator on a
finite--dimensional vector space.
However, as soon as the operator $A$ is unbounded, serious problems arise.
Unfortunately, this is the most important case for applications.
For instance, if we intend to write a partial differential equation in the 
form
of eq. (\ref{semigroup_eq_odebanach}) then $A$ is a 
differential operator and hence, it is unbounded.
Semigroup theory provides a framework for tackling these problems and yields a
method for the treatment of ordinary differential
equations on Banach spaces.

\subsection{Generation of Semigroups}
\paragraph{Strongly continuous semigroups}
Let $X$ be a Banach space and $S$ a family of bounded operators on
$X$ depending on a real nonnegative parameter, i.e. $S(t) \in \mathcal{B}(X)$
for $t \geq 0$.
$S$ is called a \emph{one--parameter semigroup} if $S(0)=\mathrm{id}_X$
and $S(t+s)=S(t)S(s)$ for all $t,s \geq 0$.
The semigroup $S$ is said to be \emph{strongly continuous} if the mapping $t
\mapsto S(t)x: [0, \infty) \to X$ is continuous for all $x \in X$.
We remark that this is a weaker notion of continuity than the more 
natural--looking
requirement of 
$S: [0, \infty) \to \mathcal{B}(X)$ being continuous.
The latter is referred to as $\emph{uniform continuity}$ and will not play a
role in our further considerations since it is too restrictive.

\paragraph{Generators}
Let $S: [0,\infty) \to \mathcal{B}(X)$ be a strongly continuous one--parameter 
semigroup of linear operators on a Banach space $X$.
We define 
$$ \Dom{A}:=\left \{x \in X: \lim_{t \to 0+} \frac{S(t)x-x}{t} \mbox{ exists}
\right \}. $$
For $x \in \Dom{A}$ define $Ax:=\lim_{t \to 0+} \frac{1}{t}(S(t)x-x)$.
One easily shows that $A: \Dom{A} \subset X \to X$ is a linear operator which is
called the \emph{generator} of the semigroup $S$.
It turns out that the generator determines the semigroup uniquely, i.e.
different semigroups have different generators.
Another useful property of semigroups is the fact that $x \in \Dom{A}$ implies
$S(t)x \in \Dom{A}$ for all $t>0$ which will be important for the application to
partial differential equations. 

\paragraph{Generation results}
An obviously important problem is to determine whether a given linear operator
generates a semigroup or not.
It turns out that generators can be completely characterized by spectral
properties.
This result is known as the \emph{Hille-Yosida theorem}.

\begin{theorem}[Hille--Yosida]
Let $X$ be a Banach space and $A: \Dom{A} \subset X \to X$ a linear operator on
$X$.
Then, $A$ generates a strongly continuous one-parameter semigroup $S$ on $X$
satisfying $\|S(t)\|_{\mathcal{B}(X)} \leq 1$ for all $t>0$, if and
only if
\begin{itemize}
\item $\Dom{A}$ is dense in $X$
\item $A$ is closed
\item $(0,\infty) \subset \rho(A)$
\item $\|R_A(\lambda)\|_{\mathcal{B}(X)} \leq \frac{1}{\lambda}$ for all
$\lambda > 0$.
\end{itemize}
\end{theorem}

A semigroup $S: [0,\infty) \to \mathcal{B}(X)$ which satisfies
$\|S(t)\|_{\mathcal{B}(X)} \leq 1$ for all $t>0$ is called a \emph{contraction semigroup}.
We remark that semigroups which satisfy a more general growth bound given by
$\|S(t)\|_{\mathcal{B}(X)} \leq e^{\omega t}$ for $\omega \in \mathbb{R}$ can 
be reduced 
to the Hille--Yosida case by rescaling, i.e. considering the semigroup $t \mapsto
e^{-\omega t}S(t)$ instead of $S$.
For even more general bounds given by $\|S(t)\|_{\mathcal{B}(X)} \leq M
e^{\omega t}$, $M>0$, there is a similar generation theorem but one has to take 
powers of the resolvent into account (cf. \cite{Engel2000}).
 
\subsection{The Abstract Cauchy Problem}
\paragraph{Statement of the problem}
Let $X$ be a Banach space and $A: \Dom{A} \subset X \to X$ a linear operator.
We consider the abstract evolution problem
\begin{equation}
\label{semigroups_eq_cauchy}
\left \{ \begin{array}{l} \frac{d}{dt}u(t)=Au(t) \mbox{ for } t>0 \\
u(0)=u_0 \end{array}
\right.
\end{equation}
for a function $u: [0,\infty) \to X$ and initial data $u_0 \in X$.

\begin{definition}
A function $u: [0,\infty) \to \Dom{A} \subset X$ is called a \emph{classical
solution} of eq. (\ref{semigroups_eq_cauchy}) if it is continuously 
differentiable and satisfies eq. (\ref{semigroups_eq_cauchy}) 
(in particular we must have $u_0 \in \Dom{A}$). 
\end{definition}

\paragraph{Well--posedness}
In order to obtain a reasonable Cauchy problem we require eq.
(\ref{semigroups_eq_cauchy}) to possess the following property: 
For given initial
data there exists a unique solution which depends continuously on the data.
This assumption is usually referred to as well--posedness.
Now we give the precise definition. 

\begin{definition}
The abstract evolution problem (\ref{semigroups_eq_cauchy}) is said to be 
\emph{well--posed} if
for every $u_0 \in \Dom{A}$ there exists a unique classical solution $u:
[0,\infty) \to \Dom{A}$ of eq.
(\ref{semigroups_eq_cauchy}) such that for each fixed $t>0$ the mapping 
$u_0 \mapsto u(t): \Dom{A} \subset X \to X$ is uniformly continuous in $t$ on
compact intervals $[0,T]$, $T>0$. 
\end{definition}
We remark that there is no overall standard and different definitions of 
well--posedness might be found in the literature, e.g. 
one sometimes includes growth bounds for the solution.
In this sense our requirements are weaker.

\paragraph{Well--posedness and semigroups}
Suppose $A$ generates a strongly continuous semigroup $S$ on $X$ satisfying
$\|S(t)\|_{\mathcal{B}(X)} \leq C(t)$ where $C$ is a continuous positive
function.
Let $u_0 \in \Dom{A}$. Then the unique classical solution $u$ of eq. 
(\ref{semigroups_eq_cauchy}) is given by $u(t)=S(t)u_0$ (recall that $S(t)u_0
\in \Dom{A}$ for all $t>0$ if $u_0 \in \Dom{A}$).
Moreover, we have $\|u(t)\|_X = \|S(t)u_0\|_X \leq C(t)\|u_0\|_X \leq \sup_{t
\in [0,T]} C(t)\|u_0\|_X$ and therefore, the
mapping $u_0 \mapsto u(t): \Dom{A} \subset X \to X$ is uniformly continuous 
in $t$ on each $[0,T]$, $T>0$. 
Hence, if $A$ generates a strongly continuous semigroup $S$ satisfying 
$\|S(t)\|_{\mathcal{B}(X)}
\leq C(t)$ then the abstract Cauchy problem eq. (\ref{semigroups_eq_cauchy}) is
well--posed.

Furthermore, the existence of a semigroup extends the notion of a solution of eq.
(\ref{semigroups_eq_cauchy}).
Since $S(t)u_0$ is defined for general $u_0 \in X$ and not only for $u_0 \in
\Dom{A}$ we can define \emph{(generalized) solutions} $u: [0,\infty) \to X$ of 
eq. (\ref{semigroups_eq_cauchy}) by $u(t):=S(t)u_0$ for $u_0 \in X$.

\subsection{Second Order Cauchy Problems}
\paragraph{Reduction to first order system}
We are concerned with wave equations and hence it is desireable to have a 
semigroup formulation for second order Cauchy problems.
This can be achieved by the usual reduction of a second order equation to a
first order system.
Consider informally the Cauchy problem
$$ 
\left \{ \begin{array}{l} \frac{d^2}{dt^2}u(t)=Au(t) \mbox{ for } t>0 \\ 
u(0)=u_0, u_t(0)=u_1 \end{array}
\right. . $$
This second order equation can be rewritten as the first order system
$$ \frac{d}{dt} \left ( \begin{array}{c}u(t) \\ u_t(t) \end{array} \right )
=\left ( \begin{array}{cc}  0 & 1 \\ A & 0 \end{array}\right ) \left ( 
\begin{array}{c}u(t) \\ u_t(t) \end{array} \right ) $$
with initial data $(u(0),u_t(0))=(u_0,u_1)$. 

\paragraph{Function spaces}
We have avoided a precise consideration of the involved function spaces 
because this issue can be very subtle.
Instead we will prove a generation result for the case when $A$ satisfies
certain additional conditions which will be enough for our purposes.
However, higher order Cauchy problems on Banach spaces have been systematically
investigated and we refer to \cite{Fattorini1985} and \cite{Xiao1998} 
for more information.

\chapter{Self--Adjoint Operators}
\thispagestyle{empty}
\label{sa_ch}

We continue collecting mathematical prerequisites.
In the first part of this chapter we note some important properties of 
self--adjoint operators.
Then we turn to Sturm--Liouville theory which deals with
self--adjoint operators that are generated by symmetric ordinary 
differential expressions of second order.

\section{Properties of Self--Adjoint Operators}
\label{propsa_sec}
We recall some well--known properties of general self--adjoint operators.
Everything in this section can be found in standard textbooks, e.g.
\cite{Kato1980}, \cite{Yosida1980}.

\subsection{The Square Root}
Let $A: \Dom{A} \subset H \to H$ be a self--adjoint operator on a Hilbert space
$H$.
Then, for $u \in \Dom{A}$, we have 
$(Au|u)_H=\overline{(u|Au)_H}=\overline{(Au|u)_H}$ and hence, 
$(Au|u)_H \in \mathbb{R}$ for all $u \in \Dom{A}$.
$A$ is said to be \emph{nonnegative} if $(Au|u)_H \geq 0$ for all $u \in
\Dom{A}$.
For a nonnegative operator $A$ there exists a square root $A^{1/2}$ with nice
properties.

\begin{theorem}
\label{growth_thm_sqrt}
Let $H$ be a Hilbert space and $A: \Dom{A} \subset H \to H$ a self--adjoint
operator which satisfies $(Au|u)_H \geq 0$ for all $u \in \Dom{A}$.
Then, there exists a unique self--adjoint operator $A^{1/2}$ such that
$\Dom{A}$ is a core of $A^{1/2}$, $A=\left (A^{1/2} \right)^2$ and
$(A^{1/2}u|u)_H \geq 0$ for all $u \in \Dom{A^{1/2}}$.
Moreover, $BA^{1/2} \subset A^{1/2}B$ whenever $BA \subset AB$ for $B \in
\mathcal{B}(H)$, i.e. $A^{1/2}$ commutes with any bounded operator that commutes
with $A$.
\end{theorem}

\begin{proof}
The claim follows from \cite{Kato1980}, p. 281 Theorem 3.35 together with 
\cite{Kato1980}, p. 279, Problem 3.32.
\end{proof}

In connection with the square root we note the following triviality which will
be useful later on.

\begin{lemma}
Let $A: \Dom{A} \subset H \to H$ be a nonnegative self--adjoint operator on a
Hilbert space $H$. If $(Au|u)_H \geq \gamma (u|u)_H$ for some $\gamma >0$ and
all $u \in \Dom{A}$ then $(A^{1/2}u|A^{1/2}u)_H \geq \gamma (u|u)_H$ for all $u
\in \Dom{A^{1/2}}$.
\end{lemma}

\begin{proof}
From $(Au|u)_H \geq \gamma (u|u)_H$ we conclude that $(A^{1/2}u|A^{1/2}u)_H \geq
\gamma (u|u)_H$ for all $u \in \Dom{A}$.
Let $u \in \Dom{A^{1/2}}$.
Since $\Dom{A}$ is a core for $A^{1/2}$ by Theorem \ref{growth_thm_sqrt}, there
exists a sequence $(u_j) \subset \Dom{A}$ such that $u_j \to u$ in $H$ and
$A^{1/2}u_j \to A^{1/2}u$ in $H$.
We have $(A^{1/2}u_j|A^{1/2}u_j)_H \geq \gamma (u_j|u_j)_H$ for all $j \in
\mathbb{N}$ and hence, this inequality remains valid in the limit $j \to
\infty$.
\end{proof}

\begin{remark}
A self--adjoint operator $A$ satisfying $(Au|u)_H \geq \gamma (u|u)_H$ for all
$u \in \Dom{A}$ and some $\gamma \in \mathbb{R}$ is called \emph{semibounded
from below}.
\end{remark}

\subsection{Boundedness of the Spectrum}
For self--adjoint operators there exists an important connection between 
semiboundedness and
boundedness of the spectrum similar to the finite dimensional case.
The following result makes it possible to deduce inequalities by studying 
spectra.

\begin{proposition}
\label{propsa_prop_semibound}
Let $A: \Dom{A} \subset H \to H$ be a self--adjoint operator on a Hilbert space
$H$ and $\gamma \in \mathbb{R}$.
Then, $(Au|u)_H \geq \gamma (u|u)_H$ for all $u \in \Dom{A}$ 
if and only if $\inf \sigma (A) \geq
\gamma$.
\end{proposition}

\begin{proof}
See \cite{Kato1980}, p. 278.
\end{proof}

\section{Sturm--Liouville Operators}
\label{sl_sec}
In this section we review some aspects of Sturm--Liouville theory.
We will also give proofs for most of the results stated below 
since they are very instructive.
For more detailed expositions see e.g. 
\cite{Naimark1968},
\cite{Teschl2007}.

\subsection{Basic Definitions}

\paragraph{Absolutely continuous functions}
We introduce a new function space which turns out to be useful in connection
with ordinary differential operators.
Let $u: [a,b] \to \mathbb{C}$.
Then, $u \in AC[a,b]$ if there exists a function $v \in L^1(a,b)$ and a 
$c \in (a,b)$ such that $u(x)-u(c)=\int_c^x v(s)ds$ for all $x \in [a,b]$.
In particular, it follows that $u$ is continuous and possesses a
weak derivative given by $v$. 
Clearly, $AC[a,b]$ is a vector space and it is called the space of
\emph{absolutely continuous functions}.
As usual, the local version $AC_{\mathrm{loc}}(a,b)$ is defined as
$AC_{\mathrm{loc}}(a,b):=\{u \in AC[c,d]: [c,d] \subset (a,b)\}$.

\paragraph{Sturm--Liouville operators}
A Sturm--Liouville operator is generated by a formal differential 
expression $\alpha$
of the form
$$ \alpha u:=\frac{1}{w}\left (-(pu')'+qu \right ) $$
where $p,q: (a,b) \to \mathbb{R}$, $1/p,q \in L^1_{\mathrm{loc}}(a,b)$ and $w$
is a weight function.

Formal integration by parts yields the \emph{Green's formula}
\begin{equation}
\label{Ainf_eq_Green} \int_c^d (\alpha u)(x)\overline{v(x)}w(x)dx=[u,v]_p(d)-[u,v]_p(c)+\int_c^d
u(x)\overline{(\alpha v)(x)}w(x)dx 
\end{equation}
for any $[c,d] \subset (a,b)$ where
$[u,v]_p(x):=u(x)\overline{(pv')(x)}-(pu')(x)\overline{v(x)}$. 

\paragraph{The initial value problem}
Consider the equation 
$$ -(pu')'+qu=\lambda w u $$
on $(a,b) \subset \mathbb{R}$ for a complex parameter $\lambda$.
The assumptions on $p,q$ are sufficient to guarantee 
existence and uniqueness of
the initial value problem.
More precise, we have the following theorem.

\begin{theorem}
\label{opth_thm_ivp}
Let $c \in (a,b)$ and $\xi,\eta \in \mathbb{C}$.
Then, there exists a unique function $u(\cdot, \lambda) \in
AC_\mathrm{loc}(a,b)$ with $pu'(\cdot,\lambda) \in AC_\mathrm{loc}(a,b)$ 
such that
$-(pu')'+qu=\lambda w u$ and $u(c, \lambda)=\xi$, $(pu')(c, \lambda)=\eta$.
Moreover, the function $u(x, \cdot)$ is holomorphic on $\mathbb{C}$ for any $x
\in (a,b)$.
\end{theorem}

The proof is similar to the classic Picard--Lindel\"of theorem and can be
found e.g. in \cite{Naimark1968}.

\paragraph{The maximal operator}
We set $H:=L^2_w(a,b)$ and define the \emph{maximal operator} $A_1: \Dom{A_1}
\subset H \to H$ associated
to $\alpha$ by 
$$ \Dom{A_1}:=\{u \in H: u,pu' \in AC_\mathrm{loc}(a,b), \alpha u \in
H\} $$
and $A_1u:=\alpha u$ for $u \in \Dom{A_1}$.
Note that functions $u$ in $\Dom{A_1}$ satisfy the minimal requirements to give
sense to $\alpha u$ and to guarantee $u \in H$ as well as $\alpha u \in H$.
That is why $A_1$ is called the maximal operator.
Note further that $u,v \in \Dom{A_1}$ implies the existence of 
$$ [u,v]_p(a):=\lim_{x \to a+}[u,v]_p(x) \mbox{ and }[u,v]_p(b):=\lim_{x \to
b-}[u,v]_p(x) $$
as follows from the Green's formula eq. (\ref{Ainf_eq_Green}).

\paragraph{Endpoint classification}
The endpoint $a$ of the interval $(a,b)$ is classified as follows.
\begin{itemize}
\item The endpoint $a$ is said to be \emph{regular} if $a>-\infty$ and 
there exists a $c \in
(a,b)$ such that $1/p,q,w \in L^1(a,c)$.
\item $a$ is said to be in the \emph{limit--circle} case if there exist $u,v \in
\Dom{A_1}$ such that $[u,v]_p(a) \not= 0$.
\item Finally, $a$ is said to be in the \emph{limit--point} case if
$[u,v]_p(a)=0$ for all $u,v \in \Dom{A_1}$.
\end{itemize}
An analogous classification is applied to $b$.

There exists a close connection
between endpoint classification and integrability of solutions of the equation
$\alpha u=0$ known as the \emph{Weyl alternative}.
This fact will be discussed later on.

\paragraph{The minimal operator}
The \emph{minimal operator} $A_0: \Dom{A_0} \subset H \to H$ 
associated to $\alpha$ is defined by
$$ \Dom{A_0}:=\{u \in \Dom{A_1}: [u,v]_p(a)=[u,v]_p(b)=0 \mbox{ for all
} v \in \Dom{A_1}\} $$
and $A_0u:=\alpha u$ for $u \in \Dom{A_0}$.
Obviously, we have $A_0 \subset A_1$.
Furthermore, $(A_0u|v)_H=(u|A_1v)_H$ for all $u \in \Dom{A_0}$ and $v \in
\Dom{A_1}$, i.e. the operators $A_0$ and $A_1$ are adjoint to each other.
This observation leads to the first easy relationship between $A_0$ and $A_1$.

\begin{lemma}
\label{opth_lem_imsubker}
We have the inclusion $\im{A_0} \subset (\ker{A_1})^\perp$.
\end{lemma}

\begin{proof}
Let $f \in \im{A_0}$, i.e. there exists a $u \in \Dom{A_0}$ such that $A_0u=f$.
Choose any $v \in \ker{A_1}$.
Then, since $A_0$ and $A_1$ are adjoint to each other, we have 
$(f|v)_H=(A_0u|v)_H=(u|A_1v)_H=0$.
\end{proof}

\subsection{Regular Sturm--Liouville Operators}
We prove some properties of $A_0$ and $A_1$.
However, it is easier to consider the regular case first.
Thus, we assume that
both endpoints $a$ and $b$ are regular, i.e. $1/p,q,w \in L^1(a,b)$.
The regular case allows major simplifications.
First of all, any $u \in C[a,b]$ belongs to $H=L^2_w(a,b)$ thanks to the
inequality $\int_a^b |u(x)|^2w(x)dx \leq \||u|^2\|_{C[a,b]}
\|w\|_{L^1(a,b)}<\infty$
and, secondly, the initial value problem can be uniquely solved at the
endpoints, i.e. the point $c$ in Theorem \ref{opth_thm_ivp} can be chosen to be
$a$ or $b$.
Moreover, any solution $u$ of $A_1 u=0$ satisfies $u,pu' \in AC[a,b]$.

\paragraph{Relation between $\im{A_0}$ and $\ker{A_1}$}

\begin{lemma}
\label{opth_lem_A1surj}
Let $f \in H$. Then, we can find a function $u \in \Dom{A_1}$ satisfying
$u(a)=(pu')(a)=0$ and $A_1u=f$. In particular, $A_1$ is surjective.
\end{lemma}

\begin{proof}
Let $f \in H$.
Since $A_1$ is regular we can find two linearly independent functions $u_1, u_2
\in \Dom{A_1}$ satisfying $A_1u_j=0$ and $u_j,pu_j' \in AC[a,b]$ for $j=1,2$.
Moreover, $[u_1,u_2]_p(x)$ is a constant $\not= 0$ 
as follows from the Green's formula eq.
(\ref{Ainf_eq_Green}) and thus, by normalization we can assume that
$[u_1,u_2]_p(x)=1$. 
We set 
$$ u(x):=u_1(x)\int_a^x u_2(s) f(s) w(s)ds-u_2(x)\int_a^x u_1(s) f(s) w(s)ds. $$
The involved integrals exist thanks to the inequality 
\begin{multline*}
\int_a^b u_j(x) f(x) w(x)dx \leq 
\left ( \int_a^b |u_j(x)|^2 w(x)dx \right )^{1/2}
\|f\|_H \\
\leq \left ( \||u_j|^2\|_{C[a,b]}\|w\|_{L^1(a,b)} \right )^{1/2} \|f\|_H<\infty. 
\end{multline*}
A direct computation reveals $u,pu' \in AC[a,b]$, $u(a)=(pu')(a)=0$
and $A_1u=f$.
\end{proof}

Now we are ready to prove the converse statement to Lemma
\ref{opth_lem_imsubker}.

\begin{lemma}
\label{opth_lem_kersubim}
Let $f \in (\ker{A_1})^\perp$. Then, $f \in \im{A_0}$.
\end{lemma}

\begin{proof}
Let $f \in (\ker{A_1})^\perp$. According to Lemma \ref{opth_lem_A1surj} we can
find a $u \in \Dom{A_1}$ such that $u(a)=(pu')(a)=0$ and $A_1 u =f$.
Let $u_1,u_2 \in \ker{A_1}$ satisfy $u_1(b)=(pu'_2)(b)=0$ and
$u_2(b)=(pu'_1)(b)=1$.
Such functions exist thanks to Theorem \ref{opth_thm_ivp} and the regularity
of $A_1$. 
Invoking Green's formula eq. (\ref{Ainf_eq_Green}) we calculate
$$ 0=(f|u_j)_H=(A_1u|u_j)_H=[u,u_j]_p(b)-[u,u_j]_p(a)
+(u|A_1u_j)_H=[u,u_j]_p(b) $$
for $j=1,2$ which yields $u(b)=(pu')(b)=0$.
Thus, $u \in \Dom{A_0}$ and we have $A_0u=A_1u=f$ which shows that $f \in
\im{A_0}$.
\end{proof}

\begin{remark}
\label{opth_rem_imker}
Taking together Lemmas \ref{opth_lem_imsubker} and \ref{opth_lem_kersubim} we
have proved that $\im{A_0}=(\ker{A_1})^\perp$. 
Note that the existence and uniqueness Theorem \ref{opth_thm_ivp} implies that
$\ker{A_1}$ is two--dimensional and in particular, as a finite--dimensional
normed vector space, it is closed.
Thus, by taking the orthogonal complement, the relation
$\im{A_0}=(\ker{A_1})^\perp$ implies $(\im{A_0})^\perp=\ker{A_1}$.
\end{remark} 

\paragraph{Density of $\Dom{A_0}$}
Next we prove that $\Dom{A_0}$ is dense in $H$ which shows that $A_0$ and $A_1$
are densely defined.

\begin{lemma}
\label{opth_lem_A0dense}
The domain $\Dom{A_0}$ is dense in $H$.
\end{lemma}

\begin{proof}
It suffices to show that any element which is orthogonal to $\Dom{A_0}$ is zero.
Let $f \in \Dom{A_0}^\perp$, i.e. $(u|f)_H=0$ for all $u \in \Dom{A_0}$.
Invoking Lemma \ref{opth_lem_A1surj} we see that 
there exists a $v \in \Dom{A_1}$ such
that $A_1 v=f$.
Thus, we have
$0=(u|f)_H=(u|A_1v)_H=(A_0u|v)_H$
for all $u \in \Dom{A_0}$ which shows that $v \in (\im{A_0})^\perp$.
However, according to Remark \ref{opth_rem_imker} we have
$(\im{A_0})^\perp=\ker{A_1}$ and thus, $v \in \ker{A_1}$ which yields
$0=A_1v=f$.
\end{proof}

\paragraph{The adjoints}
We calculate the adjoints of $A_0$ and $A_1$ (which are now known to exist since
$A_0$ and $A_1$ are densely defined).

\begin{lemma}
\label{opth_lem_adjA0}
The operator $A_1$ is the adjoint of $A_0$, i.e. $A_0^*=A_1$.
\end{lemma}

\begin{proof}
According to the Green's formula eq. (\ref{Ainf_eq_Green}) we know that
$(A_0u|v)_H=(u|A_1v)_H$ for all $u \in \Dom{A_0}$ and $v \in \Dom{A_1}$ which
means that $A_1 \subset A_0^*$.
Let $v \in \Dom{A_0^*}$, i.e. there exists an $f \in H$ such that
$(u|f)_H=(A_0u|v)_H$ for all $u \in \Dom{A_0}$.
However, since $A_1$ is surjective by Lemma \ref{opth_lem_A1surj} there exists a
$\tilde{v} \in \Dom{A_1}$ such that $A_1 \tilde{v}=f$.
Hence, we have
$$ (A_0u|v)_H=(u|f)_H=(u|A_1 \tilde{v})_H=(A_0u|\tilde{v})_H $$
which shows that $(A_0u|v-\tilde{v})_H=0$.
We conclude that 
$v-\tilde{v} \in (\im{A_0})^\perp=\ker{A_1} \subset \Dom{A_1}$ (Remark
\ref{opth_rem_imker}) and therefore, since $\tilde{v} \in \Dom{A_1}$,
we infer that $v \in \Dom{A_1}$.
Thus, we have shown that $A_0^* \subset A_1$ which finishes the proof.
\end{proof}

\begin{remark}
In particular it follows that $A_1$ is closed since it coincides with the
adjoint of a densely defined operator which is always closed.
\end{remark}

\begin{lemma}
\label{opth_lem_adjA1}
The operator $A_0$ is the adjoint of $A_1$, i.e. $A_1^*=A_0$.
\end{lemma}

\begin{proof}
Applying "$^*$" to the relation $A_1=A_0^*$ (Lemma \ref{opth_lem_adjA0}) 
yields $A_1^*=A_0^{**} \supset A_0$.
Thus, it remains to show that $A_1^* \subset A_0$.
The relation $A_1 \supset A_0$ implies $A_1^* \subset A_0^*$ and we infer $A_1^*
\subset A_1$.
Let $v \in \Dom{A_1^*} \subset \Dom{A_1}$.
By definition of the adjoint we have $(A_1u|v)_H=(u|A_1^*v)_H=(u|A_1v)_H$ for
all $u \in \Dom{A_1}$.
On the other hand, by the Green's formula eq. (\ref{Ainf_eq_Green}), we have
$(A_1u|v)_H=[u,v]_p(b)-[u,v]_p(a)+(u|A_1v)_H$ for all $u \in \Dom{A_1}$ 
and thus, $[u,v]_p(b)-[u,v]_p(a)=0$ for all $u \in \Dom{A_1}$.
However, since $A_1$ is regular we can choose $u(a),(pu')(a),u(b),(pu')(b)$
arbitrarily and we conclude that $v(a)=(pv')(a)=v(b)=(pv')(b)=0$ 
which shows that $v
\in \Dom{A_0}$ and therefore, $A_1^* \subset A_0$.
\end{proof}

\begin{remark}
Again, it follows that $A_0$ is closed.
\end{remark}

\paragraph{A special case}
The operator $A_0$ is symmetric but it is not self--ajoint since $A_0^*=A_1
\not=A_0$.
Thus, possible self--adjoint extensions of $A_0$ 
lie between $A_0$ and $A_0^*=A_1$.
We give a simple example.
Let $\Dom{A}:=\{u \in \Dom{A_1}: u(a)=u(b)=0\}$ and $Au:=\alpha u$ for $u \in
\Dom{A}$.
Then, $A_0 \subset A \subset A_1$.
We claim that $A$ is self--adjoint. 
 
\begin{lemma}
The operator $A$ is self--adjoint.
\end{lemma}

\begin{proof}
First of all we note that $(Au|v)_H=(u|Av)_H$ for all $u,v \in \Dom{A}$ by the
Green's formula eq. (\ref{Ainf_eq_Green}) and therefore, since $A$ is densely
defined by Lemma \ref{opth_lem_A0dense}, $A$ is symmetric.
Hence, it remains to show that $A^* \subset A$.

The relation $A_0 \subset A \subset A_1$ together with Lemmas
\ref{opth_lem_adjA0} and \ref{opth_lem_adjA1} imply 
$A_1=A_0^* \supset A^* \supset A_1^*=A_0$.
Let $v \in \Dom{A^*} \subset \Dom{A_1}$.
By definition we have $(Au|v)_H=(u|A^*v)_H=(u|A_1v)_H$ for all $u \in \Dom{A}$.
On the other hand, according to the Green's formula eq. (\ref{Ainf_eq_Green}), 
we observe that $(Au|v)_H=[u,v]_p(b)-[u,v]_p(a)+(u|A_1v)_H$ for all $u \in
\Dom{A}$ which yields $[u,v]_p(b)-[u,v]_p(a)=0$ for all $u \in \Dom{A}$.
However, since $A$ is regular we can choose $(pu')(a)$ and $(pu')(b)$
arbitrarily which yields $v(a)=v(b)=0$.
Thus, $v \in \Dom{A}$ and therefore we have $A^* \subset A$.
\end{proof}

\subsection{Singular Sturm--Liouville Operators}
The results of the previous section are not very useful for concrete 
applications since most of the interesting Sturm--Liouville problems are
singular.
Thus, we have to generalize the theory to the singular case, i.e. in the sequel
we merely
assume $1/p,q,w \in L^1_\mathrm{loc}(a,b)$.

\paragraph{An auxiliary operator}
We define an auxiliary operator $\tilde{A}_0: \Dom{\tilde{A}_0}
\subset H \to H$ by 
$$ \Dom{\tilde{A}_0}:=\{u \in \Dom{A_1}: u \mbox{ has compact support}\}$$ 
and $\tilde{A}_0 u:=\alpha u$
for $u \in \Dom{\tilde{A}_0}$.

\paragraph{Regularized operators}
The main idea is to restrict the problem to a fixed interval 
$I:=[c,d] \subset (a,b)$ where everything is regular and then use the fact that
$I$ is arbitrary.
We define $H^I:=L^2_w(c,d)$.
Consider the mapping $i: H^I \to H$, $u \mapsto i(u)$ defined by
$$ (i(u))(x):=\left \{ \begin{array}{l} u(x) \mbox{ if }x \in [c,d] \\
0 \mbox{ if } x \in (a,b)\backslash [c,d] \end{array} \right. $$
Since $i: H^I \to H$ is injective we can identify $u$ with $i(u)$
and in this sense we have the inclusion $H^I \subset H$.
We define the regular operators $A^I_0$ and $A^I_1$ on $H^I$ analogous to 
$A_0$ and $A_1$ where $(a,b)$ is substituted by $(c,d)$.
According to the results of the previous section we have $(A_0^I)^*=A_1^I$ and
$(A_1^I)^*=A_0^I$.
Note further that $u \in \Dom{A_0^I}$ implies $i(u) \in \Dom{\tilde{A}_0}$ and
hence, we have $\Dom{A_0^I} \subset \Dom{\tilde{A}_0}$ in the same sense as $H^I
\subset H$.

\paragraph{Density of $\tilde{A}_0$}
We show that $\Dom{\tilde{A}_0}$ is dense in $H$.

\begin{lemma}
\label{opth_lem_tildeA0dense}
The operator $\tilde{A}_0$ is densely defined.
\end{lemma}

\begin{proof}
Let $f \in H$ with $(u|f)_H=0$ for all $u \in \Dom{\tilde{A}_0}$.
Since $\Dom{A_0^I} \subset \Dom{\tilde{A}_0}$ it follows that $(u|f)_H=0$ for
all $u \in \Dom{A_0^I}$ and we can substitute the inner product on $H$ by the inner
product on $H^I$ which yields $(u|f|_I)_{H^I}=0$ for all $u \in \Dom{A_0^I}$.
Since $A_0^I: \Dom{A_0^I} \subset H^I \to H^I$ is densely defined by
Lemma \ref{opth_lem_A0dense} we conclude that $f|_I=0$.
However, $I$ was arbitrary and therefore we infer that $f=0$ almost
everywhere.
\end{proof}

Lemma \ref{opth_lem_tildeA0dense} together with $(\tilde{A}_0
u|v)_H=(u|\tilde{A}_0v)_H=0$ for all $u,v \in \Dom{\tilde{A}_0}$ imply that
$\tilde{A}_0$ is symmetric.
Furthermore, since $\tilde{A}_0 \subset A_0 \subset A_1$, we observe that $A_0$
and $A_1$ are densely defined as well.

\paragraph{The equation $\alpha u=f$}
We consider the inhomogeneous equation $\alpha u=f$.

\begin{lemma}
\label{opth_lem_auf}
Let $f \in H$. Then, there exists a $u \in AC_\mathrm{loc}(a,b)$ with $pu' \in
AC_\mathrm{loc}(a,b)$ satisfying
$\alpha u=f$.
\end{lemma}

\begin{proof}
According to Theorem \ref{opth_thm_ivp} there exist two linearly independent
functions $u_1,u_2 \in AC_\mathrm{loc}(a,b)$ with 
$pu_j' \in AC_\mathrm{loc}(a,b)$ satisfying $\alpha u_j=0$ ($j=1,2$).
Without loss of generality we can assume $[u_1,u_2]_p(x)=1$ (Green's formula eq.
(\ref{Ainf_eq_Green})).
We define $u$ by
\begin{multline*}
u(x):=c_1u_1(x)+c_2u_2(x)+u_1(x) \int_{x_1}^x u_2(s)f(s)w(s)ds
\\ -u_2(x) \int_{x_2}^x
u_1(s)f(s)w(s)ds 
\end{multline*}
where the constants $c_1,c_2 \in \mathbb{C}$ and $x_1,x_2 \in (a,b)$ can be
chosen arbitrarily.
A direct computation shows that $u,pu' \in AC_\mathrm{loc}(a,b)$ and 
$\alpha u=f$.
\end{proof}   

\paragraph{The adjoint operator}
Since $\tilde{A}_0$ is densely defined by Lemma \ref{opth_lem_tildeA0dense}, the
adjoint $\tilde{A}_0^*$ is defined.

\begin{lemma}
\label{opth_lem_adjtildeA0}
The operator $A_1$ is the adjoint of $\tilde{A_0}$, i.e. $\tilde{A}_0^*=A_1$.
\end{lemma}

\begin{proof}
According to the Green's formula eq. (\ref{Ainf_eq_Green}) we have
$(\tilde{A}_0u|v)_H=(u|A_1v)_H$ for all $u \in \Dom{\tilde{A}_0}$ and all $v \in
\Dom{A_1}$ which means that $A_1$ is adjoint to $\tilde{A}_0$, i.e. $A_1 \subset
\tilde{A}_0^*$.
Thus, it suffices to show that $A_1 \supset \tilde{A}_0^*$.
Let $v \in \Dom{\tilde{A}_0^*}$, i.e. there exists an $f \in H$ such that 
$(\tilde{A}_0u|v)_H=(u|f)_H$ for all $u \in \Dom{\tilde{A}_0}$.
In particular it follows that $(\tilde{A}_0u|v)_H=(u|f)_H$ for all $u \in
\Dom{A_0^I}$ since $\Dom{A_0^I} \subset \Dom{\tilde{A}_0}$.
Invoking Lemma \ref{opth_lem_auf} we find a $\tilde{v} \in AC_\mathrm{loc}(a,b)$
satisfying $\alpha \tilde{v}=f$.
Hence, we have
$(\tilde{A}_0u|v)_H=(u|f)_H=(u|\alpha \tilde{v})_H
=(u|\alpha \tilde{v}|_I)_{H^I}=(u|A_1^I
\tilde{v}|_I)_{H^I}=(A_0^Iu|\tilde{v}|_I)_{H^I}$ for all $u \in \Dom{A_0^I}$.
On the other hand we can write $(\tilde{A}_0u|v)_H=(A_0^I u|v|_I)_{H^I}$ 
for all $u
\in \Dom{A_0^I}$.
This yields $(A_0^Iu|(v-\tilde{v})|_I)_{H^I}=0$ for all $u \in \Dom{A_0^I}$
which shows that $(v-\tilde{v})|_I \in (\im{A_0^I})^\perp=\ker{A_1^I}$.
Thus, $\alpha (v-\tilde{v})|_I=0$ which implies $\alpha v|_I=f|_I$.
Since this relation holds for all $I=[c,d] \subset (a,b)$ we infer 
$\alpha v=f \in H$.
Hence, we have $v,pv' \in AC_\mathrm{loc}(a,b)$ and $v, \alpha v \in H$ 
which shows
that $v \in \Dom{A_1}$.
Therefore, we have shown that $\tilde{A}_0^* \subset A_1$ and we are done.
\end{proof} 

\paragraph{The special case limit--point, limit--point}
We consider the special case of a singular Sturm--Liouville operator 
where both endpoints are limit--point.

\begin{lemma}
\label{sl_lem_lplp}
Suppose that both endpoints $a$ and $b$ are in the limit--point case.
Then, the maximal operator $A_1$ is self--adjoint.
\end{lemma}

\begin{proof}
According to Green's formula eq. (\ref{Ainf_eq_Green}), the operator $A_1$ is
symmetric and thus, $A_1 \subset A_1^*$.
However, we have $\tilde{A}_0 \subset A_1$ which implies $A_1^* \subset
\tilde{A}_0^*=A_1$ by Lemma \ref{opth_lem_adjtildeA0}.
\end{proof}

\begin{corollary}
\label{sl_cor_closAtilde}
Let both endpoints $a$ and $b$ be in the limit--point case. Then, the maximal
operator $A_1$ is the closure of $\tilde{A}_0$.
\end{corollary}

\begin{proof}
Lemma \ref{opth_lem_adjtildeA0} tells us that $\tilde{A}_0^*=A_1$.
This implies $\tilde{A}_0^{**}=A_1^*=A_1$ by Lemma \ref{sl_lem_lplp} and we are
done since $\tilde{A}_0^{**}$ is the closure of $\tilde{A}_0$.
\end{proof}

\paragraph{The special case limit--circle, limit--point}
We study the special case of a singular Sturm--Liouville operator on $(a,b)$ 
where $a$ is limit--circle and $b$ is limit--point.
The question is what kind of boundary condition one can choose 
in order to construct a self--adjoint
operator.

\begin{lemma}
\label{opth_lem_Aselfadj}
Suppose $a$ is in the limit--circle case and $b$ in the limit--point case.
Fix a $\chi \in \Dom{A_1}$ such that there exists a $v \in \Dom{A_1}$ with
$[v,\chi]_p(a) \not=0$ (such a $\chi$ exists since $a$ is limit--circle) 
and define $\Dom{A}:=\{u \in \Dom{A_1}:
[u,\chi]_p(a)=0\}$, $Au:=\alpha u$ for $u \in \Dom{A}$.
Then, the operator $A: \Dom{A} \subset H \to H$ is self--adjoint.
\end{lemma}

\begin{proof}
For $u_1,u_2,u_3,u_4 \in \Dom{A_1}$ we have the so--called \emph{Pl\"ucker 
identity}
$$
[u_1,u_2]_p(x)[u_3,u_4]_p(x)+[u_1,u_3]_p(x)[u_4,u_2]_p(x)
+[u_1,u_4]_p(x)[u_2,u_3]_p(x)=0. $$
Thus, choosing $u_1,u_2 \in \Dom{A}$, $u_3:=\chi$ and $u_4$ such that
$[\chi,u_4]_p(a)\not=0$ we obtain $[u_1,u_2]_p(a)=0$ for all $u_1,u_2 \in
\Dom{A}$.
Hence, the operator $A$ is symmetric ($b$ is limit--point) 
and it remains to show that $A^* \subset A$.

We have $\tilde{A}_0 \subset A$ which implies $A_1=\tilde{A}_0^*
\supset A^*$ by Lemma \ref{opth_lem_adjtildeA0}.
Let $v \in \Dom{A^*} \subset \Dom{A_1}$.
By definition we have $(Au|v)_H=(u|A^*v)_H=(u|A_1v)_H$ for all $u \in \Dom{A}$.
Thus, the Green's formula yields $[u,v]_p(a)=0$ for all $u \in \Dom{A}$.
Note that $\chi \in \Dom{A}$ by definition and hence, the above formula with
$u=\chi$ yields $[\chi,v]_p(a)=0$ which shows that $v \in \Dom{A}$.
Therefore, we have $A^* \subset A$.
\end{proof}

\subsection{The Weyl Alternative}
Finally, we come back to the already mentioned relationship between
integrability of solutions of $\alpha u=0$ and endpoint classification.
The following theorem, known as the \emph{Weyl alternative}, 
is very important for
applications since it simplifies the endpoint classification for concrete
Sturm--Liouville operators a lot.

\begin{theorem}[Weyl alternative]
\label{sl_thm_Weyl}
The endpoint $a$ is in the limit--circle case if and only if there exist two
linearly independent functions $u_1,u_2 \in AC_\mathrm{loc}(a,b)$ with 
$pu_j' \in
AC_\mathrm{loc}(a,b)$ which belong to $H$ near
$a$ \footnote{One says that the function $u$ belongs to $L^2_w(a,b)$ near $a$ 
if there exists a $c \in
(a,b)$ such that $u|_{(a,c)} \in L^2_w(a,c)$.} and satisfy 
$(i-\alpha) u_j=0$ ($j=1,2$).
\end{theorem}

\begin{proof}[Sketch of Proof]
Given two functions $u_1,u_2$ satisfying the properties stated in the theorem
one can easily construct $u,v \in \Dom{A_1}$ such that $u=u_1$ and $v=v_1$
near $a$ (use appropriate cut--off functions).
Since $u_1,u_2$ are linearly independent it follows that $u,v$ have the same
property which implies $[u,v]_p(a)\not=0$ and hence, $a$ is limit--circle.

Conversely, let $a$ be limit--circle, i.e. there exist 
$\chi,\eta \in \Dom{A_1}$ such that $[\chi,\eta]_p(a)\not=0$.
By specifying an appropriate boundary condition at $b$ (or none, if $b$ is
limit--point) we can construct two self--adjoint operators $A_\chi$ and $A_\eta$
where $u \in \Dom{A_\chi}$ implies $[u,\chi]_p(a)=0$ and $u \in \Dom{A_\eta}$
satisfies $[u,\eta]_p(a)=0$ (cf. Lemma \ref{opth_lem_Aselfadj}).
We choose an $f \in H$ with compact support and define
$u_\chi:=(i-A_\chi)^{-1}f$, $u_\eta:=(i-A_\eta)^{-1}f$.
Since $f|_{(a,c)} \equiv 0$ for some $c \in (a,b)$, it follows that
$(i-\alpha)u_\chi|_{(a,c)}=(i-\alpha)u_\eta|_{(a,c)}=0$.
By playing around with the variation of constants formula one can show that $f$
can be chosen in such a way that neither $u_\chi|_{(a,c)}$ nor $u_\eta|_{(a,c)}$ 
are identically zero.
Moreover, from $[u_\chi,\chi]_p(a)=[u_\eta,\eta]_p(a)=0$ it follows easily that
$u_\chi$ and $u_\eta$ are linearly independent.
Hence, by solving an initial value problem at $c$ we can extend
$u_\chi|_{(a,c)}$ and $u_\eta|_{(a,c)}$
to $(a,b)$.
By construction the resulting functions belong to $H$ near $a$ and 
solve $(i-\alpha)u=0$. 
\end{proof}

\begin{remark}
Clearly, Theorem \ref{sl_thm_Weyl} is equally valid for the endpoint $b$.
Moreover, one can show that $i$ can be substituted by any $\lambda \in
\mathbb{C}$.
\end{remark}

\chapter{Semigroups and Abstract Wave Equations}
\thispagestyle{empty}
\label{wp_ch}

In this chapter we prove a generation result for an abstract second order Cauchy
problem on a Banach space. 
Then we consider the inhomogeneous abstract Cauchy problem and prove
its well--posedness under certain assumptions.
Next, using a fixed point argument, we show existence and uniqueness of 
solutions
of a nonlinear abstract wave equation.

\section{The Abstract Wave Equation}
\label{cauchy_sec}
We prove well--posedness of an abstract second order Cauchy problem on a Banach
space where the involved operator satisfies certain conditions.

\subsection{Reduction to a First Order Equation}

\paragraph{Statement of the problem}
In what follows we will assume that $B: \Dom{B} \subset H \to H$ is a  
densely defined closed linear operator on a Hilbert space $H$ satisfying 
$(Bu|Bu)_H \geq \gamma (u|u)_H$ for all $u \in \Dom{B}$ and a $\gamma>0$.

Since $B$ is densely defined, there exists the unique adjoint $B^*$.
The following theorem is due to von Neumann.
\begin{theorem}
\label{cauchy_thm_B*B}
Let $B: \Dom{B} \subset H \to H$ be a densely defined closed linear operator on
a Hilbert space $H$. Then, $B^*B$ is self--adjoint and $\Dom{B^*B}$ is a core of
$B$.
\end{theorem}

\begin{proof}
See \cite{Kato1980}, p. 275.
\end{proof}

We set $A:=B^*B$ and consider the abstract second order Cauchy problem
\begin{equation}
\label{cauchy_eq_2ndorder}
 \left \{ \begin{array}{l} \psi_{tt}(t)=-A \psi(t) \mbox{ for } t>0 \\
\psi(0)=\psi_0, \psi_t(0)=\psi_1 \end{array} \right. 
\end{equation}
for a function $\psi: [0,\infty) \to H$.

\paragraph{Function spaces}
Define $Y:=\Dom{B}$ and $\|u\|_Y:=\sqrt{(Bu|Bu)_H}$ for $u \in \Dom{B}$.

\begin{lemma} \label{cauchy_lem_YBanach}
The normed vector space $\NVS{Y}$ is a Banach space.
\end{lemma}

\begin{proof}
Let $(u_j)$ be a Cauchy sequence in $\NVS{Y}$, i.e. $(Bu_j)$ is a Cauchy
sequence in $H$.
$H$ is complete and therefore $(Bu_j)$ has a limit $f \in H$.
Since $(u|u)_H \leq \gamma^{-1} (Bu|Bu)_H = \gamma^{-1}\|u\|_Y^2$ for all 
$u \in Y$, $(u_j)$ is a Cauchy sequence in $H$ as well and hence it has a limit 
$u \in H$. 
Therefore, we have $u_j \to u$ and $Bu_j \to f$ in $H$.
Since $B$ is closed, it follows that $u \in \Dom{B}=Y$, $Bu=f$ and we have
$\|u-u_j\|_Y=\|B(u-u_j)\|_H=\|f-Bu_j\|_H \to 0$.
\end{proof}  

\begin{lemma}
\label{cauchy_lem_YcontH}
The Banach space $(Y, \|\cdot\|_Y)$ is continuously embedded in $(H,
\|\cdot\|_H)$, i.e. the inclusion map $i: (Y, \|\cdot\|_Y) 
\to (H, \|\cdot\|_H)$
defined by $i(u)=u$ for $u \in Y$ is bounded.
\end{lemma}

\begin{proof}
Let $u \in Y$ and calculate $\|i(u)\|_H^2=\|u\|_H^2 \leq \gamma^{-1} 
\|Bu\|_H^2 = \gamma^{-1}
\|u\|_Y^2$.
\end{proof}

\begin{remark}
We write $Y \hookrightarrow H$ whenever $(Y, \|\cdot\|_Y)$ embeds continuously
in $(H, \|\cdot\|_H)$.
\end{remark}
 
We define $\Dom{L}:=\Dom{A} \times \Dom{B}$ and $X:=Y \times H$.
Introducing the norm $\|(u,v)\|_X:=\sqrt{\|u\|_Y+\|v\|_H}$ for $(u,v) \in X$, 
$\NVS{X}$ becomes a Banach space thanks to Lemma \ref{cauchy_lem_YBanach}.
We define $L: \Dom{L} \subset X \to X$ by 
$L(u,v):=(v,-Au)$ for $(u,v) \in \Dom{L}$ and consider the Cauchy
problem
\begin{equation}
\label{cauchy_eq_1storder}
\left \{ \begin{array}{l} \frac{d}{dt}\mathbf{u}(t)=L \mathbf{u}(t) \mbox{ for }
t>0 \\
\mathbf{u}(0)=\mathbf{u_0} \end{array} \right.
\end{equation}
for $\mathbf{u}: [0, \infty) \to X$ and $\mathbf{u_0} \in X$ which is (formally)
equivalent to eq. (\ref{cauchy_eq_2ndorder}).
Then the following holds.

\begin{proposition}
\label{cauchy_prop_wp}
The evolution problem eq. (\ref{cauchy_eq_1storder}) is well--posed.
\end{proposition}

\subsection{Well--Posedness}
We prove Prop. \ref{cauchy_prop_wp} by showing that the operator $L$ generates a
strongly continuous one--parameter semigroup on $X$, thus we verify the
assumptions of the Hille--Yosida Theorem.

\paragraph{Analytic properties}

\begin{lemma}
\label{cauchy_lem_Ldense}
The operator $L$ is densely defined.
\end{lemma}

\begin{proof}
According to Theorem \ref{cauchy_thm_B*B}, $\Dom{A}$ is a core of $B$, i.e.
there exists a closeable operator $C: \Dom{C} \subset H \to H$ such that
$\Dom{C}=\Dom{A}$ and $\overline{C}=B$.
We define the $\emph{graph}$ $G(B)$ of $B$ by $G(B):=\{(u,Bu) \in H \times H: u
\in \Dom{B}\}$.
We equip $H \times H$ with the norm $\|(u,v)\|_{H \times
H}:=\sqrt{\|u\|_H^2+\|v\|_H^2}$ and hence, $\NVS{H \times H}$ is a Banach space.
The fact that $B$ is closed is equivalent to $G(B)$ being a closed subset of
$\NVS{H \times H}$, i.e. $G(B)=\overline{G(B)}$.
$G(C)$ is a subset of $G(B)$ and $\overline{G(C)}=G(B)$.

Now let $u \in Y=\Dom{B}$. Then, $(u, Bu) \in G(B)$ and, since $G(C)$ is dense 
in $G(B)$, there exists a sequence $((u_j, Cu_j)) \subset G(C)$ with 
$(u_j, Cu_j) \to (u,Bu)$ in $H \times H$.
Observe that $Cu_j=Bu_j$ for all $u_j \in \Dom{C}$ since $B$ is an extension of
$C$.
Therefore we have
$$ \|u-u_j\|_Y^2 \leq \|Bu-Bu_j\|_H^2+\|u-u_j\|_H^2=\|(u_j,Cu_j)-(u,Bu)\|_{H \times
H}^2 \to 0 $$
which shows that $\Dom{A}$ is dense in $\NVS{Y}$.

By assumption we have $\Dom{B}$ dense in $H$ and hence $\Dom{L}=\Dom{A} \times
\Dom{B}$ is dense in
$X=Y \times H$.
\end{proof}

\begin{lemma}
\label{cauchy_lem_Lclosed}
The operator $L$ is closed.
\end{lemma}

\begin{proof}
Let $((u_j,v_j)) \subset \Dom{L}$ be a sequence with $(u_j,v_j) \to (u,v)$ in
$X$ and
$L(u_j,v_j) \to (f,g)$ in $X$, i.e.  $u_j \to u$ in $Y$, $v_j \to v$ in $H$, 
$v_j \to f$ in $Y$ and $-Au_j \to g$ in $H$.
Convergence in $Y$ implies convergence in $H$ since $Y \hookrightarrow H$ by
Lemma \ref{cauchy_lem_YcontH} and hence we also have $v_j \to f$ in $H$ which
implies $f=v$ by uniqueness of limits. Therefore, $v \in Y=\Dom{B}$.

By the same argument we have $u_j \to u$ in $H$ and together with $-Au_j \to g$
in $H$ and the closedness of $A$ we conclude $u \in \Dom{A}$ and $-Au=g$.

Hence, we have shown that $(u,v) \in \Dom{L}=\Dom{A} \times \Dom{B}$ and
$L(u,v)=(f,g)$ which proves closedness of $L$.
\end{proof}

\paragraph{Spectral properties}
\begin{lemma}
\label{cauchy_lem_specA}
The spectrum $\sigma(A)$ of $A$ satisfies 
$\sigma(A) \subset [\gamma, \infty)$.
\end{lemma}

\begin{proof}
$A$ is self--adjoint and satisfies $(Au|u)_H=(Bu|Bu)_H \geq \gamma (u|u)_H$ for
all $u \in \Dom{A}$.
Invoking Prop. \ref{propsa_prop_semibound} finishes the proof.
\end{proof}

\begin{lemma}
\label{cauchy_lem_specL}
The spectrum $\sigma(L)$ of $L$ satifies $\sigma(L) \subset 
\{\lambda \in \mathbb{C}: -\lambda^2 \in \sigma(A)\}$, i.e. $\lambda \in
\sigma(L) \Rightarrow -\lambda^2 \in \sigma(A)$.
\end{lemma}

\begin{proof}
Consider the equation
\begin{equation}
\label{cauchy_eq_ev} (\lambda-L) \left ( \begin{array}{c}u\\v \end{array} \right ) = \left (
\begin{array}{c} f \\ g \end{array} \right ) 
\end{equation}
for $(u,v) \in \Dom{L}$ and $(f,g) \in X$ which is equivalent to the system
$$ \left \{ \begin{array}{l}v=\lambda u -f \\ (\lambda^2+A)u=\lambda f+g
\end{array} \right. .$$

Suppose $-\lambda^2 \in \rho(A)$. 
Then $\lambda^2+A: \Dom{A} \subset H \to H$ is invertible and we define
$u:=(\lambda^2+A)^{-1}(\lambda f + g)$ and $v:=\lambda u - f$ for given 
$f \in Y \subset H$ and $g \in H$. 
Then, $(u,v) \in \Dom{A} \times Y=\Dom{L}$ and $(\lambda-L)(u,v)=(f,g)$ which
shows that $\lambda-L$ is surjective.
Let $(u,v) \in \Dom{L}$ and $L(u,v)=0$. It follows that $(\lambda^2+A)u=0$ which
implies $u=0$ since $-\lambda^2 \in \rho(A)$. We conclude that $v=\lambda u=0$
and hence $(u,v)=(0,0)$ which proves injectivity of $\lambda-L$ and therefore,
$\lambda-L$ is bijective which implies $\lambda \in \rho(L)$. 

Thus, we have shown $-\lambda^2 \in \rho(A) \Rightarrow \lambda \in \rho(L)$ 
which is equivalent to 
$\lambda \in \sigma(L) \Rightarrow -\lambda^2 \in \sigma(A)$ and this is the claim.
\end{proof}

\begin{corollary}
\label{cauchy_cor_specL}
The interval $(0, \infty)$ is contained in the resolvent set of $L$, i.e. $(0,
\infty) \subset \rho(L)$.
\end{corollary}

\begin{proof}
Suppose $\lambda \in (0,\infty)$ and $\lambda \notin \rho(L)$, i.e. $\lambda \in
\sigma(L)$.
From Lemma \ref{cauchy_lem_specL} it follows that $-\lambda^2 \in \sigma(A)$
but this is a contradiction to Lemma \ref{cauchy_lem_specA} which states that
$\sigma(A) \subset [\gamma, \infty)$ and $\gamma>0$.
\end{proof}

\begin{lemma}
\label{cauchy_lem_resL}
The resolvent $R_L$ of $L$ satisfies $\|R_L(\lambda)\|_{\mathcal{B}(X)} \leq
\frac{1}{\lambda}$ for all $\lambda >0$.
\end{lemma}

\begin{proof}
Let $\lambda \in (0,\infty)$. From Corollary \ref{cauchy_cor_specL} we know that
$R_L(\lambda)=(\lambda-L)^{-1} \in \mathcal{B}(X)$ exists and we set
$(u,v):=R_L(\lambda)(f,g)$ for $(f,g) \in X$.
Then we have $Au+\lambda v=g$.
We take the inner product with $v$ and obtain $(Au|v)_H+\lambda
\|v\|_H^2=(g|v)_H$.
Substituting $v=\lambda u -f$ yields $\lambda (Au|u)_H+\lambda
\|v\|_H^2=(g|v)_H+(Au|f)_H$.
Using the Cauchy--Schwarz inequality we estimate
$$ \lambda (\|Bu\|_H^2+\|v\|_H^2)=(g|v)_H+(Au|f)_H=(g|v)_H+(Bu|Bf)_H $$ 
$$ \leq \|g\|_H \|v\|_H+\|Bu\|_H\|Bf\|_H \leq (\|Bu\|_H^2+\|v\|_H^2)^{1/2}
(\|Bf\|_H^2+\|g\|_H^2)^{1/2} $$
which yields $\lambda \|(u,v)\|_X \leq \|(f,g)\|_X$
and this is equivalent to $\|R_L(\lambda)(f,g)\|_X \leq \frac{1}{\lambda}
\|(f,g)\|_X$.
Since $(f,g) \in X$ was arbitrary this implies
$\|R_L(\lambda)\|_{\mathcal{B}(X)} \leq \frac{1}{\lambda}$.

\end{proof}

\paragraph{Generation of the semigroup}
Taking together Lemmas \ref{cauchy_lem_Ldense}, \ref{cauchy_lem_Lclosed},
\ref{cauchy_lem_resL} and Corollary \ref{cauchy_cor_specL} we have shown that
$L$ satisfies the requirements of the Hille--Yosida Theorem and hence, $L$
generates a strongly continuous one--parameter semigroup $S: [0,\infty) \to
\mathcal{B}(X)$ on $X$ satisfying $\|S(t)\|_{\mathcal{B}(X)} \leq 1$ for all
$t>0$.
Thus, the abstract Cauchy problem eq. (\ref{cauchy_eq_1storder}) is well--posed
as claimed in Prop. \ref{cauchy_prop_wp}.

\paragraph{Summary}
We summarize the results of this section in the following theorem.

\begin{theorem}
\label{cauchy_thm_gen}
Let $H$ be a Hilbert space and $B: \Dom{B} \subset H \to H$ a densely defined
closed linear operator which satisfies $(Bu|Bu)_H \geq \gamma (u|u)_H$ for all $u \in
\Dom{B}$ and a $\gamma>0$.
Define $A:=B^*B$, 
$\Dom{L}:=\Dom{A} \times \Dom{B}$, $\|u\|_Y:=\|Bu\|_H$ for $u \in
\Dom{B}$, $Y:=\Dom{B}$, $X:=Y \times H$
and $L: \Dom{L} \subset X \to X$ by
$$ L:=\left ( \begin{array}{cc} 0 & 1 \\ -A & 0 \end{array} \right ). $$
Then, $L$ generates a strongly continuous one--parameter semigroup $S: [0,\infty)
\to \mathcal{B}(X)$ on $X$ satisfying $\|S(t)\|_{\mathcal{B}(X)} \leq 1$ for all $t>0$.
In particular, the abstract evolution problem
$$ \left \{ \begin{array}{l} \frac{d}{dt}\mathbf{u}(t)=L \mathbf{u}(t) \mbox{ for }
t>0 \\
\mathbf{u}(0)=\mathbf{u_0} \end{array} \right.
$$
for $\mathbf{u}: [0, \infty) \to X$ and 
$\mathbf{u_0} \in X$ is well--posed.
\end{theorem} 

\subsection{A Simple Example}

\paragraph{The wave equation}
As a simple example we apply the generation result to the one--dimensional 
wave equation with
Dirichlet boundary conditions, i.e. we consider the Cauchy problem
$$ \left \{ \begin{array}{l} \psi_{tt}(t,x)=\psi_{xx}(t,x) \mbox{ for } (t,x)
\in (0,\infty) \times (0,1) \\
\psi(t,0)=0, \psi(t,1)=0 \mbox{ for } t>0 \\
\psi(0, x)=\psi_0(x), \psi_t(0,x)=\psi_1(x) \mbox{ for } x \in [0,1] 
\end{array} \right. $$

\paragraph{Operator formulation}
As a Hilbert space we take $H:=L^2(0,1)$ and set
$\Dom{B}:=\{u \in H^1(0,1): u(0)=u(1)=0\}$.
We define $B: \Dom{B} \subset H \to H$ by $Bu:=u'$ for 
$u \in \Dom{B}$.

\begin{lemma} 
The operator $B: \Dom{B} \subset H \to H$ is closed.
\end{lemma}

\begin{proof}
See e.g. \cite{Locker1986}, p. 29, Example 3.10.
\end{proof}
 
We claim that $-B^*Bu=u''$ for $u \in C^\infty_c(0,1)$ and thus, 
$$ \left \{ \begin{array}{l} \psi_{tt}(t)=-B^*B\psi(t) \mbox{ for } t>0 \\
\psi(0)=\psi_0, \psi_t(0)=\psi_1 \end{array} \right .
$$
is an operator version of the one--dimensional wave equation.

\begin{lemma}
For $u \in C^\infty_c(0,1)$ we have $-B^*Bu=u''$.
\end{lemma}

\begin{proof}
Integration by parts immediately yields $(Bu|v)_H=(u|-Bv)_H$ 
for all $u,v \in \Dom{B}$ and thus, 
$-B$ is adjoint to $B$, i.e. $-B \subset B^*$.
Let $u \in C^\infty_c(0,1)$. Then we have $-B^*Bu=-B^*u'=Bu'=u''$.
\end{proof}





Now we show well--posedness using Theorem \ref{cauchy_thm_gen}.

\begin{lemma}
The operator $B$ satisfies $(Bu|Bu)_H \geq (u|u)_H$ for all $u \in \Dom{B}$.
\end{lemma}

\begin{proof}
Let $u \in \Dom{B}$ and observe that 
$$ |u(x)|=\left | \int_0^x u'(s) ds \right | \leq \int_0^1 |u'(s)|ds \leq 
\left ( \int_0^1 |u'(s)|^2 ds \right )^{1/2}=\|u'\|_H $$
for all $x \in [0,1]$ by Cauchy--Schwarz.
Integration yields $(u|u)_H=\|u\|_H^2 \leq \|u'\|_H^2=(Bu|Bu)_H$.
\end{proof}  

Thus, our previous results in this section (Theorem \ref{cauchy_thm_gen}) 
imply that the first--order operator 
evolution
problem which is associated to the one--dimensional wave equation is well--posed.

\section{The Case $\gamma=0$}
In Theorem \ref{cauchy_thm_gen} the bound $\gamma$ is assumed to be strictly
positive.
The obvious question is whether this requirement can be weakened.
In this section we discuss the case $\gamma=0$, i.e. we assume that $B: \Dom{B}
\subset H \to H$ is a densely defined closed linear operator on a 
Hilbert space $H$ satisfying
$(Bu|Bu)_H \geq 0$.
The main difficulty one encounters in this case is the fact that the normed
vector space $\NVS{Y}$
defined by $Y:=\Dom{B}$ and $\|u\|_Y:=\|Bu\|_H$ is not complete.  
Hence, the semigroup cannot act on $Y \times H$ (which is the energy space).
To go around this problem one has to introduce a slightly different Banach space
in order to recover the existence of the semigroup.
This yields a well--posedness result but the growth estimate
becomes worse.
However, this can partly be compensated by energy conservation.

\subsection{Generation of the Semigroup}
Without loss of generality we can assume $B$ to be self--adjoint ($B^*B$ is
self--adjoint and nonnegative and hence, there exists the self--adjoint
nonnegative square root of $B^*B$, cf. sec. \ref{propsa_sec}). 
The idea is to consider the operator $B+i\varepsilon$ for $\varepsilon>0$
instead of $B$ and "repair" this defect by a bounded perturbation.

\paragraph{Splitting of the operator}
We define the operator $L$ by $L:=L_0+L'$ where $L_0$ and $L'$ are given by
$$ L_0:=\left ( \begin{array}{cc} 0 & 1 \\ -B^*B-\varepsilon^2 & 0
\end{array} \right ) \mbox{ and } 
L':=\left ( \begin{array}{cc} 0 & 0 \\ \varepsilon^2 & 0 \end{array} \right ).
$$
We show that $L_0$ generates a strongly continuous one--parameter
semigroup and apply the bounded perturbation theorem.

\begin{theorem}[Bounded Perturbation Theorem]
\label{wp_thm_bpt}
Let $L_0: \Dom{L_0} \subset X \to X$ be the generator of a strongly continuous 
one--parameter semigroup
$S_0: [0,\infty) \to \B{X}$ on a Banach space $X$ satisfying 
$$\|S_0(t)\|_{\B{X}} \leq e^{\omega t}$$
for an $\omega \in \mathbb{R}$ and all $t>0$.
If $L' \in \B{X}$ then $L:=L_0+L': \Dom{L_0} \subset X \to X$ is the 
generator of a
strongly continuous one--parameter semigroup $S: [0,\infty) \to \B{X}$
satisfying the estimate 
$$\|S(t)\|_{\B{X}} \leq e^{(\omega+\|L'\|_{\B{X}})t} $$
for all $t>0$.
\end{theorem}

\begin{proof}
See \cite{Engel2000}, p. 158.
\end{proof}

\paragraph{Generation of the semigroup}
Since $((B+i\varepsilon)u|(B+i\varepsilon)u)_H = (Bu|Bu)_H+\varepsilon ^2
(u|u)_H \geq \varepsilon^2 (u|u)_H$ for all $u \in \Dom{B}$,
Theorem \ref{cauchy_thm_gen} implies that the operator $L_0$
generates a strongly continuous one--parameter contraction 
semigroup on
$X^\varepsilon$.
Note that the space $X^\varepsilon$ on
which the semigroup acts depends on
$\varepsilon$ since the norm $\|\cdot\|_{X^\varepsilon}$ is given by
$\|(u,v)\|_{X^\varepsilon}^2=\|(B+i \varepsilon)u\|_H^2+\|v\|_H^2$ for
$(u,v) \in \Dom{B} \times H$! 
The perturbation $L'$ satisfies 
$$
\|L'(u,v)\|_{X^\varepsilon}=
\|(0,\varepsilon^2 u)\|_{X^\varepsilon}=\varepsilon^2 \|u\|_H  \leq
\varepsilon \|(B+i \varepsilon) u\|_H  \leq \varepsilon
\|(u,v)\|_{X^\varepsilon}
$$
for all $(u,v) \in X^\varepsilon$.
Thus, applying the bounded perturbation theorem we conclude that
$L=L_0+L'$ generates a strongly continuous one--parameter
semigroup $S^\varepsilon$ on $X^\varepsilon$ satisfying
$\|S^\varepsilon(t)\|_{\mathcal{B}(X^\varepsilon)} \leq
e^{\varepsilon t}$ for all $t>0$.

\subsection{Removing the $\varepsilon$--Dependence}
We show how to remove the bothersome $\varepsilon$--dependence
of the underlying Banach space.
We define $X:=X^1$, i.e.
$X=\Dom{B} \times H$ and
$\|(u,v)\|_X^2=\|(B+i)u\|_H^2+\|v\|_H^2$ for $(u,v) \in X$.
Since $\|(B+i\varepsilon)u\|_H^2=\|Bu\|_H^2+\varepsilon^2 \|u\|_H^2$, we have
$$ \varepsilon^2 \|(B+i)u\|_H^2 \leq \|(B+i\varepsilon)u\|_H^2 \leq
\|(B+i)u\|_H^2 $$
for all $u \in \Dom{B}$ and any $0<\varepsilon<1$.
Thus, we conclude that
$$ \varepsilon \|\mathbf{u}\|_X \leq \|\mathbf{u}\|_{X^\varepsilon} \leq
\|\mathbf{u}\|_X $$
for all $\mathbf{u} \in X$ and any $0<\varepsilon<1$.
Hence, we infer the estimate 
$$ \varepsilon  
\|S^\varepsilon(t)\mathbf{u}\|_X \leq
\|S^\varepsilon(t)\mathbf{u}\|_{X^\varepsilon} \leq 
e^{\varepsilon t}\|\mathbf{u}\|_{X^\varepsilon} \leq e^{\varepsilon t}
\|\mathbf{u}\|_X $$
for all $\mathbf{u} \in X$ which implies
$\|S^\varepsilon(t)\|_{\mathcal{B}(X)} \leq \frac{1}{\varepsilon}e^{\varepsilon
t}$ for any $0<\varepsilon<1$.
Furthermore, all $S^\varepsilon$ (for different $\varepsilon$) 
have the same generator $L$ which does not
depend on $\varepsilon$ and hence, they all coincide.
Thus, we can drop the superscript
$\varepsilon$ and, for any $0<\varepsilon<1$, we obtain
$\|S(t)\|_{\mathcal{B}(X)} \leq \frac{1}{\varepsilon}e^{\varepsilon
t}$  for all $t > 0$.
Finally, choosing $\varepsilon=\frac{1}{t}$ for $t>1$ we arrive at
\begin{equation}
\label{growth_eq_estS0}
 \|S(t)\|_{\mathcal{B}(X)} \leq t \mbox{ for all } t>1. 
\end{equation}

\subsection{Energy Conservation}
We define the \emph{energy "norm"} on $X$ by 
$\|(u,v)\|_E^2:=\|Bu\|_H^2+\|v\|_H^2$ for $(u,v) \in X$  
\footnote{In general this is only a seminorm 
since $0$
might be an eigenvalue of $B$ and thus, there might exist elements 
$(u,v) \in X$
with $u \not=0$ but $\|(u,v)\|_E=0$.}.
Thus, we have $\|(u,v)\|_E \leq \|(u,v)\|_X$ for all $(u,v) \in X$.
Consider the time evolution of initial data $(u_0,v_0) \in \Dom{L}$.
According to semigroup theory, the solution $(u(t),v(t)):=S(t)(u_0,v_0)$ stays
in $\Dom{L}$ for all $t>0$ and
satisfies the equation 
$$ \left \{ \begin{array}{l} \frac{d}{dt}u(t)=v(t) \\
\frac{d}{dt} v(t)=-B^*B u(t) \end{array} \right. $$
in the strong sense, i.e. it is a classical solution.
Recall that the norm $\|(u_0,v_0)\|_X$ is given by
$\|(u_0,v_0)\|_X^2=\|(B+i)u_0\|_H^2+\|v_0\|_H^2$ 
and note that $\|(B+i)u_0\|_H^2=\|Bu_0\|_H^2+\|u_0\|_H^2$.
Hence, we can write \footnote{Recall
the definition of the graph norm $\|\cdot\|_B$ given by
$\|u\|_B:=\|Bu\|_H+\|u\|_H$ for $u \in \Dom{B}$.} 
$\|(u_0,v_0)\|_X^2=\|u_0\|_B^2+\|v_0\|_H^2$  and this implies
that the derivative $\frac{d}{dt}u(t)$ exists with respect to the graph norm of $B$, i.e.
$\lim_{h \to 0}h^{-1}\|u(t+h)-u(t)\|_B$ exists for all $t>0$.
Therefore, since $B$ is a closed operator, 
we have $\frac{d}{dt}Bu(t)=B\frac{d}{dt}u(t)$. 
Having these observations in mind we readily calculate
\begin{multline*}
\frac{d}{dt}\left [ (Bu(t)|Bu(t))_H+(v(t)|v(t))_H \right ]=2 \Re
(B\dot{u}(t)|Bu(t))_H \\+2 \Re (v(t)|\dot{v}(t))_H=
2 \Re(Bv(t)|Bu(t))_H-2\Re (v(t)|B^*Bu(t))_H=0
\end{multline*}
for all $t>0$ where $\dot{}:=\frac{d}{dt}$.
Thus, the function $t \mapsto \|S(t)(u_0,v_0)\|_E$ is constant.
This shows that for all classical solutions the energy is conserved.

Now let $\mathbf{u_0} \in X$.
Then, there exists a sequence $\mathbf{u_0}_j \subset \Dom{L}$ such that
$\mathbf{u_0}_j \to \mathbf{u_0}$ in $X$.
Since $S(t)$ is bounded we conclude that $S(t)\mathbf{u_0}_j \to
S(t)\mathbf{u_0}$ in $X$ for any $t>0$.
Recall that $\|\cdot\|_E \leq \|\cdot\|_X$ which shows that $S(t)\mathbf{u_0}_j
\to S(t)\mathbf{u_0}$ with respect to $\|\cdot\|_E$ as well.
In particular we have
$$ \|S(t)\mathbf{u_0}\|_E=\lim_{j \to \infty} \|S(t)\mathbf{u_0}_j\|_E=
\lim_{j \to \infty}\|\mathbf{u_0}_j\|_E=\|\mathbf{u_0}\|_E $$
for all $t \geq 0$.
This shows that energy conservation holds for generalized solutions as well. 

\subsection{Summary}
We formulate the results of this section as a theorem.

\begin{theorem}
\label{gamma0_thm_gen}
Let $H$ be a Hilbert space and $A: \Dom{A} \subset H \to H$ a self--adjoint
operator satisfying $(Au|u)_H \geq 0$ for all $u \in \Dom{A}$. 
Define $X:=\Dom{A^{1/2}} \times H$ and
$\|(u,v)\|_X^2:=\|u\|_{A^{1/2}}^2+\|v\|_H^2$ for $(u,v) \in X$.
Then, the operator $L: \Dom{L}\subset X \to X$, 
defined by
$\Dom{L}:=\Dom{A} \times \Dom{A^{1/2}}$ and
$$ L:=\left ( \begin{array}{cc} 0 & 1 \\ -A & 0 \end{array} \right ), $$
generates a strongly continuous one--parameter semigroup $S: [0, \infty) \to
\mathcal{B}(X)$ on $X$ satisfying 
$$ \|S(t)\|_{\mathcal{B}(X)} \leq t $$
for all $t > 1$.

Furthermore, if $(u_0,v_0) \in X$ and $(u(t),v(t)):=S(t)(u_0,v_0)$, 
the function $t \mapsto \|A^{1/2}u(t)\|_H+\|v(t)\|_H$ is constant for all $t \geq 0$. 
\end{theorem}

\section{The Inhomogeneous Problem}
\label{inhomog_sec}

As a next step we discuss semigroup theory for inhomogeneous evolution
problems.
This approach relies on the notion of an integral of a semigroup whose 
definition requires a
little background in measure theory which will be outlined first.

\subsection{Basic Aspects of Measure Theory}
We briefly discuss the construction of the Lebesgue integral for Banach space valued
functions on intervals.
An introduction to measure theory can be found in e.g. \cite{Bauer2001}.
For the definition of the Bochner integral we also refer to \cite{Yosida1980}.

\paragraph{$\sigma$--algebra, measure}
Let $\Omega$ be a set and $\mathcal{A} \subset \mathcal{P}(\Omega)$ where
$\mathcal{P}(\Omega)$ denotes the power set of $\Omega$, i.e. the set of all
subsets of $\Omega$.
$\mathcal{A}$ is said to be a \emph{$\sigma$--algebra} if
\begin{itemize}
\item $\emptyset \in \mathcal{A}$
\item $A \in \mathcal{A} \Rightarrow \Omega \backslash A \in \mathcal{A}$
\item $A_n \in \mathcal{A}$ for $n \in \mathbb{N}$ $\Rightarrow \bigcup_{n \in
\mathbb{N}} A_n \in \mathcal{A}$
\end{itemize}
Let $\mathcal{C} \subset \mathcal{P}(\Omega)$.
The smallest $\sigma$--algebra which contains $\mathcal{C}$ is denoted by
$\sigma(\mathcal{C})$.
We remark that $\sigma(\mathcal{C})$ exists for any $\mathcal{C} \subset
\mathcal{P}(\Omega)$ since $\mathcal{P}(\Omega)$ is a $\sigma$--algebra itself.

A mapping $\mu: \mathcal{A} \to [0,\infty]$ is called a \emph{measure} on
$\mathcal{A}$ if $\mu(\emptyset)=0$ and $\mu(\bigcup_{n \in
\mathbb{N}}A_n)=\sum_{n=1}^\infty \mu(A_n)$ for $A_n \in \mathcal{A}$, $n \in
\mathbb{N}$ and $A_i
\cap A_j=\emptyset$ for $i \not=j$.

\paragraph{Borel $\sigma$--algebra, Lebesgue measure}
Let $\Omega=\mathbb{R}$ and set $\mathcal{H}:=\{\emptyset\} \cup \{(a,b]: a,b
\in \mathbb{R}, a<b\}$.
Then, $\mathcal{B}:=\sigma(\mathcal{H})$ is called the \emph{Borel
$\sigma$--algebra}.
One can show that there exists a unique measure $\lambda$ on $\mathcal{B}$ 
such that $\lambda((a,b])=b-a$ for all half--open intervals $(a,b] \in
\mathcal{H}$.
The measure $\lambda$ is called the \emph{Lebesgue measure}.
A similar construction can be applied to $\Omega=\mathbb{R}^n$.

\paragraph{Measurable functions, simple functions}
Let $\mathcal{B}$ be the Borel $\sigma$--algebra on $\mathbb{R}$, 
$\Omega$ a set and $\mathcal{A} \subset \mathcal{P}(\Omega)$ a $\sigma$--algebra.
A function $f: \mathbb{R} \to \Omega$ is said to be \emph{measurable} if
$f^{-1}(A) \in \mathcal{B}$ for all $A \in \mathcal{A}$.

Now we restrict ourselves to real--valued functions.
We denote the characteristic function of a set $A$ by $\chi_A$, i.e.
$$ \chi_A(x)=\left \{ \begin{array}{l} 1 \mbox{ for } x \in A \\
0 \mbox{ for } x \notin A \end{array} \right.$$
A function $u: \mathbb{R} \to \mathbb{R}$ is called \emph{simple} if there exist
$A_j \in \mathcal{B}$ for $j \in \mathbb{N}$ which satisfy 
$\bigcup_{j \in \mathbb{N}} A_j=\mathbb{R}$ and $A_i \cap A_j=\emptyset$ for $i
\not= j$ and $u=\sum_{j=1}^n c_j \chi_{A_j}$ where 
$c_j \in [0,\infty)$, $n \in \mathbb{N}$.

\paragraph{Lebesgue integral}
One defines the Lebesgue integral $\int u d\lambda$ over a simple function 
$u: \mathbb{R} \to \mathbb{R}$ with $u=\sum_{j=1}^n c_j \chi_{A_j}$ by 
$$ \int u d\lambda:=\sum_{j=1}^n c_j \lambda(A_j) $$ 
where $\lambda$ denotes the
Lebesgue measure.

Let $f: \mathbb{R} \to \mathbb{R}$ be a nonnegative measurable function.
One can show that there exists a sequence of simple functions $(u_n)$ such that
$u_n \leq u_{n+1}$ for all $n \in \mathbb{N}$ and $f(x)=\lim_{n \to \infty}
u_n(x)=\sup_{n \in \mathbb{N}}u_n(x)$.
Then one defines 
$$ \int f d\lambda:=\sup_{n \in \mathbb{N}} \int u_n d\lambda. $$
For general measurable functions $f: \mathbb{R} \to \mathbb{R}$ one sets
$f^-(x):=\max\{-f(x),0\}$ and $f^+(x):=\max\{f(x),0\}$.
One easily shows that the nonnegative functions $f^-$ and $f^+$ are 
measurable.
If $\int f^+ d\lambda < \infty$ and $\int f^- d\lambda < \infty$ one says
that $f$ is \emph{integrable} and defines
$$ \int f d\lambda:=\int f^+ d\lambda - \int f^- d\lambda. $$

The generalization to complex--valued functions is obtained by considering the
real and imaginary parts separately.

\paragraph{Lebesgue's theorem on dominated convergence}
We have the following important convergence theorem.

\begin{theorem}[Dominated convergence theorem]
Let $(\Omega, \mathcal{A}, \mu)$ be a measure space (i.e. $\mathcal{A} \subset
\mathcal{P}(\Omega)$ is a $\sigma$--algebra and $\mu$ a measure on
$\mathcal{A}$) and $f,f_n: \Omega \to \mathbb{R}$ measurable functions.
Furthermore, assume that $f_n \to f$ pointwise almost everywhere and there
exists a nonnegative measurable function $g$ with $\int g d\mu < \infty$ and
$|f_n| \leq g$ almost everywhere for all $n \in \mathbb{N}$.
Then, $f_n$ is integrable for all $n \in \mathbb{N}$ and we have
$$ \lim_{n \to \infty} \int f_n d\mu=\int f d\mu. $$
\end{theorem}

\paragraph{Banach space valued functions}
Let $X$ be a Banach space.
Again, $\mathcal{B}$ denotes the Borel $\sigma$--algebra on $\mathbb{R}$ and
$\lambda$ the Lebesgue measure.

A function $u: \mathbb{R} \to X$ is said to be \emph{simple} if there exist
$A_j \in \mathcal{B}$, $j \in \mathbb{N}$ with $\bigcup_{j \in
\mathbb{N}}A_j=\mathbb{R}$, $A_i \cap A_j=\emptyset$ for $i \not=j$ and 
$x_j \in X$ such that $u=\sum_{j=1}^n  x_j \chi_{A_j}$, $n \in \mathbb{N}$. 

\paragraph{Measurable functions}
A function $u: \mathbb{R} \to X$ is said to be \emph{weakly measurable} if,
for any $f \in X^*$, the
complex--valued function $t \mapsto f(u(t)): \mathbb{R} \to \mathbb{C}$ is
measurable.
$u$ is said to be \emph{strongly measurable} if there exists a sequence $(u_n)$
of simple functions and a $B_0 \in \mathcal{B}$ with $\lambda(B_0)=0$ such that 
$\|u(t)-u_n(t)\|_X \to 0$ for all $t \in \mathbb{R} \backslash B_0$, i.e.
$u_n(t) \to u(t)$ in $X$ for almost all $t \in \mathbb{R}$. 
It turns out that, if $X$ is separable, the notions "weakly measurable" and
"strongly measurable" are equivalent. 

\paragraph{The Bochner integral}
Let $u: \mathbb{R} \to X$ be a simple function with $u=\sum_{j=1}^n x_j
\chi_{A_j}$.
Then, the Bochner integral $\int u d\lambda \in X$ of $u$ is defined by
$$ \int u d\lambda:=\sum_{j=1}^n \lambda(A_j)x_j. $$ 
Let $u: \mathbb{R} \to X$ be a strongly measurable function.
One can show that the real--valued function $t \mapsto \|u(t)\|_X: \mathbb{R} \to
\mathbb{R}$ is measurable. 
The function $u$ is said to be 
\emph{Bochner integrable} if there exists a sequence $(u_n)$ of simple functions
such that 
$$ \lim_{n \to \infty} \int \|u-u_n\|_X d\lambda=0. $$
It turns out that this condition implies the existence of the limit of the
sequence $\left ( \int u_n d\lambda \right )$ in $X$.
Then, one defines the \emph{Bochner integral} of $u$ by
$$ \int u d\lambda:=\lim_{n \to \infty} \int u_n d\lambda. $$

\begin{theorem}[Bochner] Let $\NVS{X}$ be a Banach space.
A strongly measurable function $u: \mathbb{R} \to X$ is Bochner integrable if
and only if the function $t \mapsto \|u(t)\|_X: \mathbb{R} \to \mathbb{R}$ is 
integrable.
In this case the estimate
$$ \left \| \int u d\lambda \right \|_X \leq \int \|u\|_X d\lambda $$
holds.
\end{theorem} 

\paragraph{Notation}
To improve readability we will adapt the usual "Riemann--like" notation for
integrals, i.e. consider a function $u: I \to X$ where $I \subset \mathbb{R}$ is
some interval and $X$ a Banach space. Then we define a function 
$\tilde{u}: \mathbb{R} \to X$ by
$$ \tilde{u}(t):=\left \{ \begin{array}{l} u(t) \mbox{ for } t \in I \\
0 \in X \mbox{ for } t \notin I \end{array} \right. $$
and set
$$ \int_I u(t)dt:=\int \tilde{u} d\lambda. $$

\subsection{The Abstract Problem}
We consider the inhomogeneous abstract Cauchy problem, see \cite{Engel2000}.
We also refer to \cite{Sell2002} for an extensive treatment of this subject.

\paragraph{Statement of the problem}
Let $\NVS{X}$ be a separable \footnote{The assumption of separability 
is introduced here for convenience only, it is by no means necessary.} 
Banach space, 
$L: \Dom{L} \subset X \to X$ 
a linear operator
and $f: [0,\infty) \to X$ a function.
Consider the abstract inhomogeneous evolution problem
\begin{equation}
\label{inhomog_eq_absin}
\left \{ \begin{array}{l} \frac{d}{dt}u(t)=Lu(t)+f(t) \mbox{ for } t>0 \\
u(0)=u_0 \end{array} \right.
\end{equation}
for a function $u: [0,\infty) \to X$ and initial data $u_0 \in X$.
Now we extend the notion of well--posedness to the inhomogeneous problem.

\begin{definition}
A function $u: [0,\infty) \to \Dom{L} \subset X$ is called a 
\emph{classical solution} of eq.
(\ref{inhomog_eq_absin}) if it is continuously differentiable and satisfies eq.
(\ref{inhomog_eq_absin}).

The abstract evolution problem eq. (\ref{inhomog_eq_absin}) is said to be
\emph{well--posed} if for every $u_0 \in \Dom{L}$ there exists a unique classical
solution $u: [0, \infty) \to \Dom{L}$ of eq. (\ref{inhomog_eq_absin}) such that 
the
mapping $u_0 \mapsto u(t): \Dom{L} \subset X \to X$ is uniformly continuous in
$t$ on compact intervals $[0,T]$ for any $T>0$.
\end{definition}

\paragraph{Construction of solutions}
Suppose that $L$ generates a strongly continuous one--parameter semigroup $S: [0,
\infty) \to \mathcal{B}(X)$ on $X$ satisfying $\|S(t)\|_{\mathcal{B}(X)} \leq
C(t)$
for a continuous positive function $C$ and all $t>0$, i.e. the associated 
homogeneous problem ($f \equiv 0$) is
well--posed.
Assume further that $f: [0,\infty) \to X$ is continuous.
Then, the function $s \mapsto S(t-s)f(s): [0,t] \to X$ is continuous and hence,
for any $g \in X^*$, the function 
$s \mapsto g(S(t-s)f(s)): [0,t] \to \mathbb{C}$ is continuous and thus
measurable.
This shows that $s \mapsto S(t-s)f(s): [0,t] \to X$ is weakly measurable.
Since $X$ is separable we conclude that $s \mapsto S(t-s)f(s)$ is strongly 
measurable.
Moreover, for $s \in [0,t]$ we have $\|S(t-s)f(s)\|_X \leq C(t-s)\|f(s)\|_X \leq
\sup_{s \in [0,t]} C(s)\|f(s)\|_X < \infty$ which shows that
$$ \int_0^t \|S(t-s)f(s)\|_X ds $$ 
exists.
Thus, Bochner's Theorem implies that $s \mapsto S(t-s)f(s)$ is 
Bochner integrable.

Define
\begin{equation}
\label{inhomog_eq_mild} 
u(t):=S(t)u_0 + \int_0^t S(t-s)f(s)ds.
\end{equation}
Then, $u$ is called the \emph{mild solution} of eq. (\ref{inhomog_eq_absin}).

\paragraph{Classical solutions}
It is easy to show that every classical solution is a mild solution.
In particular, this implies that a classical solution is unique.
It turns out that, if $f$ satisfies certain additional regularity conditions, a
classical solution can be constructed by formula eq. (\ref{inhomog_eq_mild}).
If, for instance, $f$ has a weak $t$--derivative which is integrable and 
$u_0 \in \Dom{L}$ then $u$ defined
by eq. (\ref{inhomog_eq_mild}) is the classical solution of eq.
(\ref{inhomog_eq_absin}).
More precise, we have the following theorem.

\begin{theorem}
\label{inhomog_thm_classic}
Let $L$ be the generator of a strongly continuous one--parameter semigroup $S:
[0,\infty) \to \mathcal{B}(X)$ on $X$. If $u_0 \in \Dom{L}$ and $f \in
W^{1,1}([0,\infty), X)$ then $u: [0,\infty) \to X$ defined by eq.
(\ref{inhomog_eq_mild}) is the unique classical solution of eq.
(\ref{inhomog_eq_absin}).
\end{theorem}

\begin{proof}
See \cite{Engel2000}, p. 439.
\end{proof}

\begin{remark}
The space $W^{1,1}([0,\infty),X)$ is a Sobolev space of Banach space valued
functions $u: [0,\infty) \to X$ analogous to $W^{1,1}(0,\infty)$ for
complex--valued functions.
One can define $W^{1,1}([0,\infty),X)$ as follows.
Consider the vector space of functions $u:
[0,\infty) \to X$ such that $u$ is integrable (in the sense of Bochner) and
there exists an integrable function $v: [0,\infty) \to X$ such that $u$ can be
written as
$$ u(t)=u(t_0)+\int_{t_0}^t v(s) ds $$
for a $t_0 \in [0,\infty)$.
Two such functions are identified if they coincide for almost all $t \in
[0,\infty)$.
Then, the quotient space defined by this equivalence relation is denoted by
$W^{1,1}([0,\infty),X)$.
\end{remark}

\paragraph{Well--posedness of the inhomogeneous problem}
We summarize the results of this section in a theorem.

\begin{theorem}
\label{inhomog_thm_wp}
Let $X$ be a separable Banach space, $L: \Dom{L} \subset X \to X$ a linear
operator and $f \in W^{1,1}([0,\infty), X)$.
If $L$ generates a strongly continuous one--parameter semigroup $S: [0,\infty)
\to \mathcal{B}(X)$ then the abstract inhomogeneous Cauchy problem eq.
(\ref{inhomog_eq_absin}) is well--posed.
Its unique classical solution $u$ is given by eq. (\ref{inhomog_eq_mild}) for
$u_0 \in \Dom{L}$.
\end{theorem}

\section{A Nonlinear Problem}
In this section we consider existence and uniqueness of an abstract nonlinear
evolution equation for a function $u: [0,\infty) \to X$ on a separable
Banach space $X$ 
given by 
\begin{equation}
\label{nonlinear_eq_nonlinear}
\left \{ \begin{array}{l}
\frac{d}{dt}u(t)=Lu(t)+g(u(t))
 \\
u(0)=u_0 \end{array} \right. 
\end{equation}
where 
$u_0 \in X$ and $g: X \to X$ is 
Lipschitz--continuous, i.e. there exists a $C>0$ 
such that$\|g(x)-g(y)\|_X \leq C\|x-y\|_X$ for all $x,y \in X$. 
We assume that $L$ generates a strongly continuous semigroup $S: [0,\infty) \to
\B{X}$ satisfying $\|S(t)\|_X \leq e^{\omega t}$ for all $t \geq 0$ and an 
$\omega \in
\mathbb{R}$.



\subsection{Existence and Uniqueness}
Let $T>0$ and denote the vector space of continuous functions 
$u: [0,T] \to X$ by $C([0,T],X)$.
For $u \in C([0,T],X)$ the real--valued function $t \mapsto
\|u(t)\|_X$ is continuous and hence attains its maximum on the compact
interval $[0,T]$.
We define a norm $\|\cdot\|_{C([0,T],X)}$ on $C([0,T],X)$ by
$$ \|u\|_{C([0,T],X)}:=\sup_{t \in [0,T]} \|u(t)\|_X. $$ 
It is easy to show that $\NVS{C([0,T],X)}$ is a Banach space.

Now let $u_0 \in X$ be fixed and define a mapping 
$K: C([0,T],X) \to C([0,T],X)$ by 
\begin{equation}
\label{nonlinear_eq_K}
 (K(v))(t):=S(t)u_0+\int_0^t S(t-s)g(v(s))ds 
 \end{equation}
for $t \in [0,T]$.
The function $s \mapsto S(t-s)g(v(s))$ is continuous since $g: X \to X$
is continuous.
Hence, it is Bochner integrable and the above definition of $K$ makes sense. 

Note that the function $u:=Kv$ is the unique mild solution of the
linear inhomogeneous problem
$$ 
\left \{ \begin{array}{l} \frac{d}{dt}u(t)=L u(t)
+g(v(t)) 
\mbox{ for }
t \in [0,T] \\
u(0)=u_0 \end{array} \right.
$$
as discussed in sec. \ref{inhomog_sec}.
Thus, a fixed point $u$ of $K$ ($u=K(u)$) is a
mild solution of the nonlinear problem eq. (\ref{nonlinear_eq_nonlinear}).
Hence, to show existence of solutions of eq. (\ref{nonlinear_eq_nonlinear}) it
suffices to show existence of a fixed point of the mapping $K$.
To do so we will invoke the Banach fixed point theorem.

\begin{theorem}[Banach fixed point theorem]
Let $\NVS{Z}$ be a Banach space and $K: Z \to Z$ a mapping that satisfies
$$ \|K(x)-K(y)\|_Z \leq k \|x-y\|_Z $$
for all $x,y \in Z$ and a $k<1$.
Then, there exists a unique fixed point of $K$.
\end{theorem}

With the help of this theorem we can prove existence and uniqueness of 
mild solutions
of eq. \ref{nonlinear_eq_nonlinear} for small times.

\begin{lemma}
\label{nonlinear_lem_fp}
There exists a $T>0$ such that the mapping $K: C([0,T],X) \to C([0,T],X)$ 
defined by eq.
(\ref{nonlinear_eq_K}) has a unique fixed point.
\end{lemma}

\begin{proof}
Let $T>0$, $u, v \in C([0,T],X)$ and observe that
$$ \|K(u)-K(v)\|_{C([0,T],X)} \leq \sup_{t \in [0,T]} \int_0^t
\|S(t-s)[g(u(s))-g(v(s))]\|_X ds $$
$$ \leq Ce^{\omega T} \int_0^T \|u(s)-v(s)\|_X ds
\leq CTe^{\omega T} \sup_{s \in [0,T]} \|u(s)-v(s)\|_X $$
where we have used $\|S(t)\|_{\mathcal{B}(X)} \leq e^{\omega t}$ for all 
$t \geq 0$ and the
Lipschitz--continuity of $g$.
Thus, we have
$$ \|K(u)-K(v)\|_{C([0,T],X)} \leq
CTe^{\omega T} \|u-v\|_{C([0,T],X)}. $$
Hence, choosing $T$ such that $Te^{\omega T}<1/C$ shows that $K$ satisfies the contraction property
required by the Banach fixed point theorem.
Therefore, for such a $T$, $K$ has a unique fixed point.
\end{proof}

Thus, we have shown that eq. (\ref{nonlinear_eq_nonlinear}) has a unique solution
$u: [0,T] \to X$ for any $T$ satisfying $Te^{\omega T}<1/C$.

From the local result we immediately obtain a global result.

\begin{lemma}
There exists a unique global (mild) solution $u \in C([0, \infty),X)$ of eq.
(\ref{nonlinear_eq_nonlinear}).
\end{lemma}

\begin{proof}
Applying Lemma \ref{nonlinear_lem_fp} we obtain a constant $T>0$ and a
local--in--time solution $u_1 \in C([0,T],X)$ of eq.
(\ref{nonlinear_eq_nonlinear}) for given initial data $u_0 \in X$.
Now we set $u_0:=u_1(T)$ and apply Lemma
\ref{nonlinear_lem_fp} again to obtain a solution $u_2 \in C([0,T],X)$
satisfying $u_2(0)=u_1(T)$.
Repeating this process yields a sequence $(u_j)$ of solutions of eq.
(\ref{nonlinear_eq_nonlinear}) in $C([0,T],X)$ satisfying
$u_j(0)=u_{j-1}(T)$ for $j \in \mathbb{N}$.
Now we define $u(t):=u_j(t-(j-1)T)$ for $t \in [(j-1)T,jT)$, 
$j \in \mathbb{N}$.
By construction, $u \in C([0,\infty),X)$, $u$ satisfies
$\frac{d}{dt}u(t)=Lu(t)$ for each $t > 0$ and
$u(0)=u_0$.
\end{proof}
 
\subsection{Dependence on Data and Growth Estimates}
By construction, the global solution $u \in C([0,\infty), X)$ of eq.
(\ref{nonlinear_eq_nonlinear})
satisfies the equation
\begin{equation}
\label{nonlinear_eq_nonlinint}
u(t)=S(t)u_0+\int_0^t S(t-s)g(u(s))ds 
\end{equation}
for all $t \geq 0$.
In order to obtain an estimate for $u(t)$ in terms of the initial data
$u_0$ we apply Gronwall's inequality.

\begin{lemma}[Gronwall's inequality]
Let $\xi: [0,T] \to \mathbb{R}$ be a nonnegative integrable function which
satisfies
$$ \xi(t) \leq C_2 + C_1 \int_0^t \xi(s)ds $$
for constants $C_1, C_2 \geq 0$ and almost all $t \in [0,T]$.
Then, 
$$ \xi(t) \leq C_2 \left ( 1+C_1te^{C_1t} \right ) $$
for almost all $t \in [0,T]$.
\end{lemma}

\begin{proof}
See e.g. \cite{Evans1998}, p. 625.
\end{proof}
 
\begin{lemma}
\label{nonlinear_lem_globest}
The unique global solution $u \in C([0,\infty),X)$ satisfies the
estimate 
$$ \|u(t)\|_X \leq e^{\omega t}\|u_0\|_X 
\left (1+C t e^{\omega t} e^{Cte^{\omega t}} \right ) $$
for all $t \geq 0$ and a constant $C>0$.
\end{lemma}

\begin{proof}
Using eq. (\ref{nonlinear_eq_nonlinint}), $\|S(t)\|_{\mathcal{B}(X)} \leq
e^{\omega t}$ for
all $t \geq 0$ and the Lipschitz--continuity of $g$ we readily estimate
$$ \|u(t')\|_X \leq e^{\omega t} \|u_0\|_X + 
C e^{\omega t} \int_0^{t'} \|u(s)\|_X ds $$
for all $0 \leq t' \leq t$.
Application of Gronwall's inequality yields
$$ \|u(t')\|_X \leq e^{\omega t}\|u_0\|_X (1+C t' e^{\omega t} e^{C t' e^{\omega
t}}) $$
for all $0 \leq t' \leq t$.
Setting $t'=t$ we infer the desired result since $t \geq 0$ is arbitrary.  
\end{proof}

Although the estimate stated in Lemma \ref{nonlinear_lem_globest} 
is very weak, it suffices to ensure continuous dependence on the initial data. 

\subsection{Summary}
We reformulate the results on the nonlinear problem as a theorem.

\begin{theorem}
Let $L: \Dom{L} \subset X \to X$ be a linear operator on a Banach space $X$
which generates a strongly continuous one--parameter
semigroup $S: [0,\infty) \to \B{X}$ satisfying $S(t) \leq e^{\omega t}$ for all
$t \geq 0$ and an $\omega \in \mathbb{R}$.
Moreover, let $g: X \to X$ be a Lipschitz--continuous function.
Then, the nonlinear abstract evolution problem
$$ \left \{ \begin{array}{l}
\frac{d}{dt}u(t)=Lu(t)+g(u(t)) \mbox{ for } t \geq 0 \\
u(0)=u_0 \end{array} \right . $$
for $u_0 \in X$ has a unique mild solution $u: [0,\infty) \to X$
which depends continuously on $u_0$.
\end{theorem}

\chapter{The Cauchy Problem for Wave Maps}
\thispagestyle{empty}
\label{wpwm_ch}

We formulate the Cauchy problem for the wave maps system eq.
(\ref{wmeq_eq_wmsys})
$$ \Box \Phi^A + \eta^{\mu \nu} \Gamma^A{}_{BC}(\Phi)(\partial_\mu
\Phi^B)(\partial_\nu \Phi^C)=0 $$
and state some basic results.
As a technical requirement we need fractional Sobolev
spaces which are introduced first.
Then, we state a general result concerning local well--posedness of nonlinear
wave equations which can be applied to the wave maps system.
We also mention some recent developments in connection with global existence
for solutions with small data.
\section{Fractional Sobolev Spaces}
We introduce noninteger Sobolev spaces by using the Fourier transform (cf.
e.g. \cite{Evans1998}, \cite{Yosida1980}).

\paragraph{Lebesgue spaces}
We have already defined Lebesgue spaces for functions defined on intervals.
The generalization to complex--valued mappings on open subsets of $\mathbb{R}^n$ is
similar and goes as follows. 
Let $U \subset \mathbb{R}^n$ be open and consider the set 
$C^\infty_c(U)$ of smooth 
functions from $U$ to $\mathbb{C}$ having compact support.
For $p \geq 1$ we define a norm $\|\cdot\|_{L^p(U)}$ on $C^\infty_c(U)$ by
$$ \|u\|_{L^p(U)}:=\left ( \int_{U} |u(x)|^p d^nx \right
)^{1/p} $$
where integration is understood with respect to the ordinary Lebesgue measure on
$\mathbb{R}^n$.
The Lebesgue space $L^p(U)$ is defined as the completion of $C^\infty_c(U)$ with
respect to $\|\cdot\|_{L^p(U)}$.
We also define local versions $L^p_\mathrm{loc}(U)$ by 
$$L^p_\mathrm{loc}(U):=\{u
\in L^p(V): V \subset U, V \mbox{ compact}\}.$$

\paragraph{The Fourier transform}
Now suppose $u \in L^2(\mathbb{R}^n) \cap L^1(\mathbb{R}^n)$ and define 
$$ \hat{u}(\xi):=\frac{1}{(2 \pi)^{n/2}} \int_{\mathbb{R}^n} u(x)e^{-i\xi \cdot
x}d^nx $$
where $\xi \cdot x:=\sum_{j=1}^n \xi^j x^j$ for $x=(x^1,\dots,x^n),\xi=(\xi^1,
\dots, \xi^n) \in \mathbb{R}^n$.
Since $u \in L^1(\mathbb{R}^n)$ and $|e^{-i\xi \cdot x}|=1$, $\hat{u}$ is
well--defined and it is called the \emph{Fourier transform} of $u$.
It turns out that the mapping $\mathcal{F}: u \mapsto \hat{u}$ can be 
extended to $L^2(\mathbb{R}^n)$.
We list some important properties.
\begin{itemize}
\item $\mathcal{F}: L^2(\mathbb{R}^n) \to L^2(\mathbb{R}^n)$ is an isometric
isomorphism, i.e. it is linear, invertible and norm--preserving (Plancherel's
theorem). 
\item For $\hat{u} \in L^1(\mathbb{R}^n) \cap L^2(\mathbb{R}^n)$ the inverse
$\mathcal{F}^{-1}$ is given by 
$$ \mathcal{F}^{-1}\hat{u}(x)=\frac{1}{(2 \pi)^{n/2}} \int_{\mathbb{R}^n} 
\hat{u}(\xi)e^{i\xi \cdot x}d^n\xi $$
\item "$\mathcal{F}$ maps derivatives into multiplication", i.e. 
$\mathcal{F}D^\alpha u(\xi)=(i\xi)^\alpha \mathcal{F}u(\xi)$ if $D^\alpha u \in
L^2(\mathbb{R}^n)$ where $\alpha=(\alpha_1, \dots, \alpha_n) 
\in \mathbb{N}_0^n$, $D^\alpha
u:=\partial^{\alpha_1}_1 \cdots
\partial^{\alpha_n}_n u$ and $\xi^\alpha:=\Pi_{j=1}^n (\xi^j)^{\alpha_j}$ for
$\xi=(\xi^1,\dots ,\xi^n) \in \mathbb{R}^n$.
\end{itemize}

\paragraph{Fractional Sobolev spaces}
For $s \in \mathbb{R}$ we define the Sobolev space $H^s(\mathbb{R}^n)$ by 
$$ H^s(\mathbb{R}^n):=\{u \in L^2(\mathbb{R}^n): \xi \mapsto (1+|\xi|^s)
\hat{u}(\xi) \in L^2(\mathbb{R}^n)\} $$
where $|\xi|$ is the Euclidean norm of the vector $\xi \in \mathbb{R}^n$, i.e.
$|\xi|^2:=\sum_{j=1}^n |\xi^j|^2$.
Furthermore, we set
$$ \|u\|_{H^s(\mathbb{R}^n)}:=\left (
\int_{\mathbb{R}^n}|(1+|\xi|^s)\hat{u}(\xi)|^2 d^n \xi \right )^{1/2} $$
for $u \in H^s(\mathbb{R}^n)$.
Note that $H^0(\mathbb{R}^n)=L^2(\mathbb{R}^n)$.
Equipped with this norm, $H^s(\mathbb{R}^n)$ becomes a Banach space.
Another commonly used notion is the \emph{homogeneous Sobolev space 
$\dot{H}^s(\mathbb{R}^n)$}
which is defined as
$$ \dot{H}^s(\mathbb{R}^n):=\{u \in L^2(\mathbb{R}^n): \xi \mapsto |\xi|^s
\hat{u}(\xi) \in L^2(\mathbb{R}^n)\} $$
and 
$$ \|u\|_{\dot{H}^s(\mathbb{R}^n)}:=\left ( \int_{\mathbb{R}^n} |\xi|^{2s}
|\hat{u}(\xi)|^2 d^n\xi \right )^{1/2}. $$
For brevity we will write $H^s$ instead of $H^s(\mathbb{R}^n)$.

\section{Local Well--Posedness}
We state some known results concerning local well--posedness of nonlinear wave
equations and wave maps.

\subsection{Semilinear Wave Equations}
We consider the Cauchy problem for a nonlinear wave equation of the form
\begin{equation}
\label{locwp_eq_nweq}
\Box \psi=f(\psi, \partial \psi), \:\: \psi|_{t=0}=\psi_0, \partial_t 
\psi|_{t=0}=\psi_1 
\end{equation}
for a function $\psi: \mathbb{R} \times \mathbb{R}^n \to \mathbb{R}$
where $\psi_0,\psi_1: \mathbb{R}^n \to \mathbb{R}$, $f$ is smooth and 
satisfies $f(0)=0$.
$f(\psi, \partial \psi)$ is a shorthand notation to indicate that $f$ might
depend on the function $\psi$ and its first partial derivatives.
The necessity to prescribe $\psi|_{t=0}$ \emph{and} $\partial_t \psi|_{t=0}$
as initial data follows from the fact that the equation is second order in time:
One has
to know
the function and its first time derivative on the initial surface $t=0$
to be able to \emph{formally} calculate higher derivates.

By a (local) solution of the Cauchy problem eq. (\ref{locwp_eq_nweq}) we mean a
function $\psi \in C([0,T],H^1)$ with $\partial_t \psi \in
(C[0,T], L^2)$ that satisfies
\begin{multline}
\int_0^T \int_{\mathbb{R}^n} \psi(t,x) \Box \varphi(t,x) d^n x
dt-\int_{\mathbb{R}^n} \partial_t \psi(t,x)\varphi(0,x)d^n x \\+\int_{\mathbb{R}^n}
\psi(t,x)\partial_t\varphi(0,x)d^nx=
\int_0^T \int_{\mathbb{R}^n} f(\psi,\partial \psi)(t,x)\varphi(t,x) d^n x dt
\end{multline}
for all test functions $\varphi \in C^\infty_c((-T,T) \times \mathbb{R}^n)$ and a
constant $T>0$ (which might depend on $\psi_0$ and $\psi_1$).
For simplicity we write $\psi(t,x)$ instead of $\psi(t)(x)$.
We state the classical local well--posedness theorem.

\begin{theorem}
\label{locwp_thm_class}
Let $f: \mathbb{R} \times \mathbb{R}^n \to \mathbb{R}$ be smooth and $f(0)=0$.
Then, for any $s>\frac{n}{2}+1$ and $(\psi_0,\psi_1) \in H^s \times
H^{s-1}$ there exists a $T>0$ such that the Cauchy problem 
\begin{equation*}
\Box \psi=f(\psi, \partial \psi), \:\: \psi|_{t=0}=\psi_0, \partial_t 
\psi|_{t=0}=\psi_1 
\end{equation*}
has a unique solution $(\psi, \partial_t \psi) \in C([0,T],H^s) \times
C([0,T],H^{s-1})$.
Moreover, $T$ is bounded below by a strictly positive continuous function of
$\|\psi_0\|_{H^s}+\|\psi_1\|_{H^{s-1}}$ and the
mapping $(\psi_0,\psi_1) \mapsto (\psi, \partial_t \psi):
H^s\times H^{s-1} \to 
C([0,T],H^s) \times C([0,T],H^{s-1})$ is continuous.
\end{theorem}
We say that the Cauchy problem eq. (\ref{locwp_eq_nweq}) is \emph{locally
well--posed in $H^s$ for any $s>\frac{n}{2}+1$}.
The proof (see e.g. \cite{Klainerman2002}) of Theorem \ref{locwp_thm_class} 
relies on a fixed point iteration
together with energy estimates for the
linear wave equation, a Sobolev embedding, the so--called \emph{Moser
inequality} and the Gronwall inequality.
We remark that Theorem \ref{locwp_thm_class} is equally valid for systems of
equations, i.e. vector--valued $\psi$ and $f$.

\subsection{Wave Maps}
Observe that the wave maps equation is in fact a system of semilinear wave 
equations and
the involved nonlinearity satisfies the
requirements of Theorem \ref{locwp_thm_class}.
Thus, we immediately obtain local well--posedness 
of the Cauchy problem for wave maps in $H^s$ for any
$s>\frac{n}{2}+1$.
At this point we should remark that wave maps have orginally been required to 
be $C^\infty$ (cf. ch. \ref{wm_ch}).
However, from the point of view of partial differential equations this is an
unnecessary restrictive assumption and therefore we relax it.  

It turns out that this local well--posedness result can
be improved.
The important observation in this respect is the fact that the nonlinearity in
the wave maps system is not generic.
It satisfies the so--called \emph{null condition} (cf. \cite{Klainerman1986}).
Exploiting this special algebraic structure it is possible to show local
well--posedness of the Cauchy problem for wave maps in $H^s$ for $s >
\frac{n}{2}$ and $n \geq 2$ (see \cite{Klainerman2002} and references therein). 
This result is sharp in the sense that there exist wave map systems which are
not locally well--posed (i.e. \emph{ill--posed}) 
in $H^s$ for $s \leq \frac{n}{2}$, see \cite{Georgiev2004}.

\section{Global Results}

In general, global existence of solutions for semilinear wave equations is  
expected to hold
only if the data satisfy certain "smallness" conditions.
We cite some results in this direction for wave maps.
On the other hand, if the solution does not exist for all times, 
the question arises how the
breakdown occurs.
For our particular wave map model this issue will be studied in the next
chapter.

There are many recent results concerning the global well--posedness for the 
Cauchy
problem of wave maps and we are unable to mention them all (see e.g.
\cite{Tataru2005}, \cite{Krieger2005}, \cite{Klainerman2002a} and references
therein).
However, of most interest for our purposes is Tao's work \cite{Tao2001}
which deals with wave maps from ($n+1$)--dimensional Minkowski space to the
($m-1$)--sphere for $m,n \geq 2$.
In this work it is shown that, roughly speaking, a wave map with smooth initial datum 
$(\psi_0,\psi_1)$ 
which is small in
the homogeneous Sobolev space $\dot{H}^{n/2} \times \dot{H}^{n/2-1}$ extends
globally in time and stays smooth.
Thus, for time evolutions starting from smooth data $(\psi_0,\psi_1)$, blow up 
can only occur if $\|\psi_0\|_{\dot{H}^{n/2}}^2+\|\psi_1\|_{\dot{H}^{n/2-1}}^2$
is sufficiently large.

Finally, we mention two earlier global existence results for wave maps from 
($3+1$) Minkowski space to the three--sphere which have been obtained by 
Kovalyov \cite{Kovalyov1987} and Sideris \cite{Sideris1989}.
We also refer to the work of Shatah and Tahvildar--Zadeh \cite{Shatah1994} for 
the special case of equivariant wave maps.

\chapter{Self--Similar Solutions}
\thispagestyle{empty}
\label{css_ch}

\section{Blow Up}
\label{blowup_sec}
We discuss solutions of the wave map problem eq. (\ref{wm_eq_wm}) with smooth
initial data which become singular after a finite time.
Such a behaviour is called \emph{blow up}.
We remark that this phenomenon can already be observed for ordinary differential 
equations.
Consider for example the equation $u'=u^2$ with initial data $u(0)=1/T$.
The solution is given by $u(t)=(T-t)^{-1}$ and thus it ceases to exist at $t=T$.
We have already mentioned regularity results which state that the solution is
smooth for all times provided the
initial data are smooth and small in some Sobolev space.
Hence, blow up can only occur if the data are large enough.

Technically we note that the derivations in this section have an informal 
character, i.e. we relax the mathematical rigor and restrict ourselves to
a heuristic discussion.

\subsection{Scaling and Criticality Class}
We consider the \emph{conserved energy}
\begin{equation}
\label{blowup_eq_E} 
E_\psi(t):=\int_0^\infty \left ( \psi_t^2(t,r)+\psi_r^2(t,r)+
\frac{2 \sin^2(\psi(t,r))}{r^2} \right ) r^2 dr. 
\end{equation}
of the wave map equation 
\begin{equation}
\label{blowup_eq_wm}
\psi_{tt} - \psi_{rr} -\frac{2}{r} \psi_r + \frac{\sin(2 \psi)}{r^2}=0.
\end{equation}
A mapping $(t,r) \mapsto (t/\lambda, r/\lambda)$ for a constant $\lambda>0$ is
called a \emph{dilation}.
Note that eq. (\ref{blowup_eq_wm}) is invariant under dilations:
Suppose $\psi$ solves eq. (\ref{blowup_eq_wm}).
Then,
$\psi_\lambda$ defined by $\psi_\lambda(t,r):=\psi(t/\lambda, r/\lambda)$ is
also a solution of eq. (\ref{blowup_eq_wm})
provided that $r$ in eq. (\ref{blowup_eq_wm}) is substituted 
by $r/\lambda$.
This scale invariance can be used to classify conserved quantities of the
equation.
Let $\psi$ be a solution of eq. (\ref{blowup_eq_wm}).
The energy eq. (\ref{blowup_eq_E}) scales as 
$E_{\psi_\lambda}(t)=\lambda^\alpha E_\psi(t/\lambda)$ for $\alpha=1$.
One says that the scaling of energy is \emph{subcritical}, \emph{critical} or 
\emph{supercritical} if
$\alpha<0$, $\alpha=0$ or $\alpha>0$, respectively.
Thus, the energy for the wave map equation eq. (\ref{blowup_eq_wm}) is
supercritical.
An informal principle states that solutions of energy 
supercritical equations develop singularities for large initial data while they
stay regular for small ones.
Based on the criticality classification one gains a heuristic understanding of
certain aspects of the dynamics.
In the supercritical case it is favourable for solutions to shrink since this
process is
connected with a decrease of the local energy. 
Such a shrinking may eventually lead to singularity formation.
Conversely, in the subcritical case shrinking is forbidden since it requires a larger
and larger amount of energy. 
Hence, we expect eq. (\ref{blowup_eq_wm}) to possess blow up solutions.

\subsection{Characteristics and Finite Speed of Propagation}
We discuss a fundamental feature of wave equations: Finite speed of
propagation of information.

\paragraph{Characteristics}
It turns out that information propagates along certain curves in spacetime 
which are called
\emph{characteristics}.
In what follows we will explain what is meant by this statement.
Consider a first--order differential equation 
$$ u_t+Au_r+f(u)=0 $$
for a vector--valued function $u=(u_1, \dots, u_n)$ and a 
$n \times n$--matrix $A$
depending on $t$ and $r$.
We assume $A$ to be diagonalizable with eigenvalues 
$\{\lambda_i: i=1, \dots, n\}$ (the $\lambda_i$'s are functions of $t$ and $r$
as well).
Hence, there exists an invertible matrix $V$ (depending on $t$ and $r$) 
such that $V^{-1}AV$ is diagonal.
Defining $v:=V^{-1}u$ we can write the system in component form 
\begin{equation}
\label{blowup_eq_diag}
 \partial_t v_i+\lambda_i \partial_r v_i+g(v)_i=0 
 \end{equation}
where $g(v):=V^{-1}(f(Vv)+AV_rv+V_tv)$. 
Now let $r_i$ be a function of $t$ such that $\dot{r}_i(t)=\lambda_i(t, r_i(t))$ for all
$t$ where $\dot{ }:=\frac{d}{dt}$.
Then, the spacetime curve $t \mapsto (t,r_i(t))$ is called a
\emph{characteristic}.

\paragraph{Finite speed of propagation}
Let $v$ be a solution of eq. (\ref{blowup_eq_diag}).
We calculate the directional derivative of $v$ along the characteristic 
$t \mapsto (t, r_i(t))$.
$$ \frac{d}{dt}v_i(t, r_i(t))=\partial_t v_i(t,r_i(t))+\partial_r v_i(t,r_i(t))
\dot{r}_i(t)$$
$$=-\lambda_i(t,r_i(t))\partial_r v_i(t,r_i(t))+\dot{r}_i(t)
\partial_r v_i(t,r_i(t))-g(v(t, r_i(t)))_i $$
$$ =-g(v(t,r_i(t)))_i. $$ 
Integrating this equation yields
$$ v_i(t,r_i(t))=v_i(0,r_i(0))-\int_0^t g(v(s,r_i(s)))_i ds. $$
Thus, the solution at a spacetime point $(t,r)$ depends solely on the value of the
solution along the characteristics through that point.
If $\dot{r}_i(t) < \infty$ for all $i$ and $t$ it follows that information 
encoded in the initial data propagates with finite speed.
Hence, initial data given on a compact subset of the initial surface 
can only influence a compact spacetime domain in the future, the \emph{domain of
influence}.
On the other hand, the value of the solution at a fixed spacetime point depends
solely on the value of the field in a compact region of spacetime in the past, 
the \emph{domain of dependence}. 

Note that the notion of domain of dependence is of fundamental importance 
for the
numerical treatment of wave equations.
Discretizing the equation in a way which is compatible with the characteristic 
structure is absolutely necessary in order to obtain a stable scheme.

\paragraph{Characteristics of the Wave Map Equation}
We set $u_1:=\psi$, $u_2:=\psi_t$, $u_3:=\psi_r$ and write eq.
(\ref{blowup_eq_wm}) in the form
$$ u_t+Au_r+f(u)=0 $$
where 
$$ A=\left ( \begin{array}{ccc} 0 & 0 & 0 \\
0 & 0 & -1 \\
0 & -1 & 0 \end{array} \right ) $$
and 
$$ f(u)(t,r)=\left ( \begin{array}{c} -u_2(t,r) \\
-\frac{2}{r}u_3(t,r)+\frac{\sin(2 u_1(t,r))}{r^2} \\
0 \end{array} \right ). $$
The eigenvalues of $A$ are $-1,1,0$ and hence, there are three characteristics,
$r_1(t)=r_1(0)-t$, $r_2(t)=r_2(0)+t$ and $r_3(t)=r_3(0)$. 
In a spacetime diagram the characteristics $r_1$ and $r_2$ are lines with 
slope $1$ or
$-1$ and hence, the speed of propagation of information is limited by $1$.

Finally, we remark that the notion of characteristics can be defined in the more
general context of a fully nonlinear first--order partial differential
equation, see e.g. \cite{Evans1998}.
 
\subsection{Blow Up Solutions}
We intend to construct an explicit example of a solution of eq.
(\ref{blowup_eq_wm}) with smooth initial data which develops a singularity in
finite time.
The most promising strategy is to look for self--similar solutions.
We have already discussed the dilation invariance of the wave map equation
(\ref{blowup_eq_wm}).
Hence, it is natural to look for a solution which shares this invariance, i.e.
we seek solutions $\psi$ with the property $\psi(\lambda t, \lambda
r)=\psi(t,r)$ for any $\lambda>0$.
Such solutions are called \emph{self--similar}.
To this end we plug the ansatz $\psi(t,r)=f(\frac{r}{T-t})$ in eq.
(\ref{blowup_eq_wm}) and obtain
\begin{equation}
\label{blowup_eq_css}
f''+\frac{2}{\rho}f'-\frac{\sin(2f)}{\rho^2(1-\rho^2)}=0
\end{equation}
where $\rho:=\frac{r}{T-t}$, $':=\frac{d}{d\rho}$ and $T>0$ is an arbitrary
constant.
Shatah \cite{Shatah1988} showed that eq. (\ref{blowup_eq_css}) has a smooth
solution.
This solution has been found in closed form by Turok and Spergel 
\cite{Turok1990}
and is given by 
$$ f_0(\rho):=2 \arctan \rho. $$
Set $\psi_0(t,r):=f_0(\frac{r}{T-t})$.
Then, $\psi_0$ is perfectly smooth for $t<T$ but 
$\partial_r \psi_0(t,0)=(T-t)^{-1}$ and hence the spatial derivative at
the center $r=0$ blows up for $t \to T-$. 
Thus, $\psi_0$ is an explicit example of a solution of eq.
(\ref{blowup_eq_wm}) with smooth initial data $\psi_0(0,\cdot)$ and 
$\partial_t \psi_0(0, \cdot)$ which develops a singularity in finite time.

One might argue that this example is of no physical relevance since the solution
$\psi_0$ is not a finite energy solution, i.e. $E_{\psi_0}(0)=\infty$.
However, as we have seen in the previous section, the speed of propagation of
information is limited by $1$.
This fact can be used to construct a blow up solution with finite energy.
Consider smooth initial data which equal $\psi_0(0,\cdot)$ and $\partial_t
\psi_0(0,\cdot)$ for $0 \leq r \leq T$ and are identically zero for $r>2T$.
These data have finite energy and due to finite speed of  
propagation the singularity at $r=0$ will form before any information of the
region $r>T$ reaches the center. 
Hence, the existence of self--similar solutions together with finite propagation
speed implies the existence of a physically relevant blow up solution.

\subsection{Some Numerics}
A natural question is whether blow up occurs for generic initial data or the
example given above is an "exceptional case" in a certain sense.
In order to answer this question we employ some very simple numerics.

\paragraph{Coordinate transformation}
Since interesting things are expected to happen around $r=0$ it is useful to
make a coordinate change $r \mapsto x:=\log(\alpha +r)$, $\alpha>0$, in order to 
gain a better resolution near the center.
The value of the constant $\alpha$ determines the quality of this primitive mesh
refinement.
The wave map equation (\ref{blowup_eq_wm}) in these new coordinates reads
\begin{equation}
\label{blowup_eq_wmx}
\psi_{tt}-e^{-2x}\psi_{xx}-e^{-2x}\frac{e^x+\alpha}{e^x-\alpha}\psi_x+
\frac{\sin(2\psi)}{(e^x-\alpha)^2}=0.
\end{equation}

\paragraph{Characteristics}
We calculate the characteristics of this equation.
Setting $u_1:=\psi$, $u_2:=\psi_t$ and $u_3:=\psi_x$ we write eq.
(\ref{blowup_eq_wmx}) in first--order form
$$ u_t+Au_r+f(u)=0 $$
where 
$$ A=\left ( \begin{array}{ccc} 0 & 0 & 0 \\
0 & 0 & -e^{-2x} \\
0 & -1 & 0 \end{array} \right ).
$$
The eigenvalues of $A$ are $-e^{-x},e^{-x},0$ and hence the
characteristic curves $x_1$ and $x_2$ satisfy the differential equations
$\dot{x}_1(t)=-e^{-x_1(t)}$ and $\dot{x}_2(t)=e^{-x_2(t)}$.
Thus, the \emph{characteristic speeds} $\dot{x}_1$ and $\dot{x}_2$ at the 
center $x=\log \alpha$ are given by
$\pm \frac{1}{\alpha}$ and therefore they diverge as $\alpha$ approaches zero.

\paragraph{Discretization}
We choose one of the simplest second--order accurate schemes to discretize eq.
(\ref{blowup_eq_wmx}), the centered space and centered time method.
We label discrete spacetime points by $(n,k)$ where $n,k$ are natural numbers
including zero and
use the following approximations for derivatives of $\psi$.
$$ \psi_{tt}(n \Delta t, k \Delta x) \approx
\frac{\psi^{n+1}_k-2\psi^n_k+\psi^{n-1}_k}{\Delta t^2} $$
$$ \psi_{xx}(n \Delta t, k \Delta x) \approx
\frac{\psi^n_{k+1}-2\psi^n_k+\psi^n_{k-1}}{\Delta x^2} $$
$$ \psi_{x}(n \Delta t, k \Delta x) \approx
\frac{\psi^n_{k+1}-\psi^n_{k-1}}{2 \Delta x} $$
where we have applied the usual abbreviation 
$\psi^n_k:=\psi(n \Delta t, k \Delta x)$.
Plugging these approximations in eq. (\ref{blowup_eq_wmx}) and solving for
$\psi^{n+1}_k$ we obtain
\begin{equation}
\label{blowup_eq_intscheme}
\begin{split}
 \psi^{n+1}_k=a(k \Delta x)\Delta t^2 
\frac{\psi^n_{k+1}-2\psi^n_k+\psi^n_{k-1}}{\Delta x^2}+b(k \Delta x)\Delta t^2 
\frac{\psi^n_{k+1}-\psi^n_{k-1}}{2 \Delta x} \\
+c(k \Delta x)\Delta t^2 \sin(2 \psi^n_k)+2 \psi^n_k -\psi^{n-1}_k
\end{split}
\end{equation} 
where $a(x):=-e^{-2x}$, $b(x):=-e^{-2x}\frac{e^x+\alpha}{e^x-\alpha}$, 
$c(x):=\frac{1}{(e^x-\alpha)^2}$.

We use a finite spatial grid which covers the interval $[0,X]$ for some large 
$X>0$ and impose Dirichlet conditions $\psi(t,0)=0$ and $\psi(t,X)=0$ for all
$t$ at the endpoints. 
Since the discretization scheme (\ref{blowup_eq_intscheme}) requires two time
steps in the past we have to calculate the first time step by Taylor expansion.
This yields the initialization formula
\begin{equation*}
\begin{split}
\psi^1_k=f_k+g_k \Delta t+\frac{\Delta t^2}{2} \left (
a(k \Delta x)\frac{f_{k+1}-2f_k+f_{k-1}}{\Delta x^2} \right. \\ + \left.
b(k \Delta x)\frac{f_{k+1}-f_{k-1}}{2 \Delta x}-c(k \Delta x)\sin
(2 f_k) \right ) 
\end{split}
\end{equation*}
where $f_k:=\psi(0,k \Delta x)$ and $g_k:=\psi_t(0,k \Delta x)$ are the initial
data.

\paragraph{Domain of dependence}
As we have already mentioned, the characteristic speeds at the center are given
by $\pm \frac{1}{\alpha}$.
In order to obtain a stable discretization scheme it is necessary to make sure
that the physical domain of dependence is included in
the numerical domain of dependence.
This requirement is known as the \emph{Courant--Friedrichs--Lewy condition} (cf.
\cite{Gustafsson1995}).
Hence, we have to choose $\Delta x$, $\Delta t$ and $\alpha$ in such a way that 
the condition $\frac{\Delta t}{\Delta x} \leq \alpha$ is satisfied.
We observe that in order to improve the quality of the mesh refinement (i.e. 
decrease $\alpha$) we have to decrease $\Delta t$ as well if $\Delta x$ is kept
fixed.
Thus, one cannot increase the resolution around the center without
increasing the computational effort.  

\begin{figure}[h]
\centering
\includegraphics[totalheight=7cm,angle=-90]{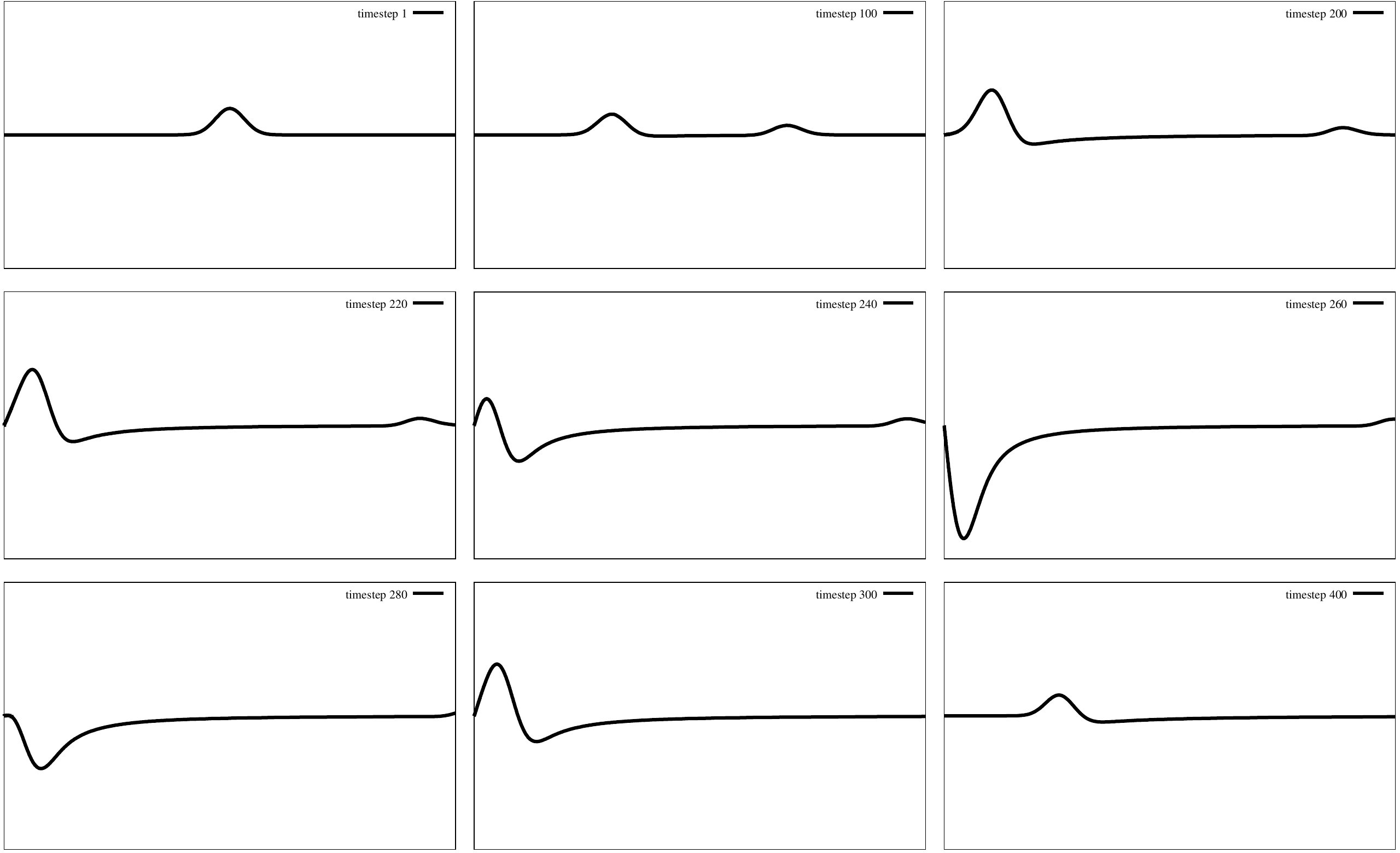}
\caption{Time evolution of a Gau\ss{} pulse}
\label{blowup_fig_te}
\end{figure}

\begin{figure}[h]
\centering
\psfrag{psi_x}{$\max_t \psi_x(t,\log \alpha)$}
\psfrag{A}{$A$}
\includegraphics[totalheight=5cm,angle=-90]{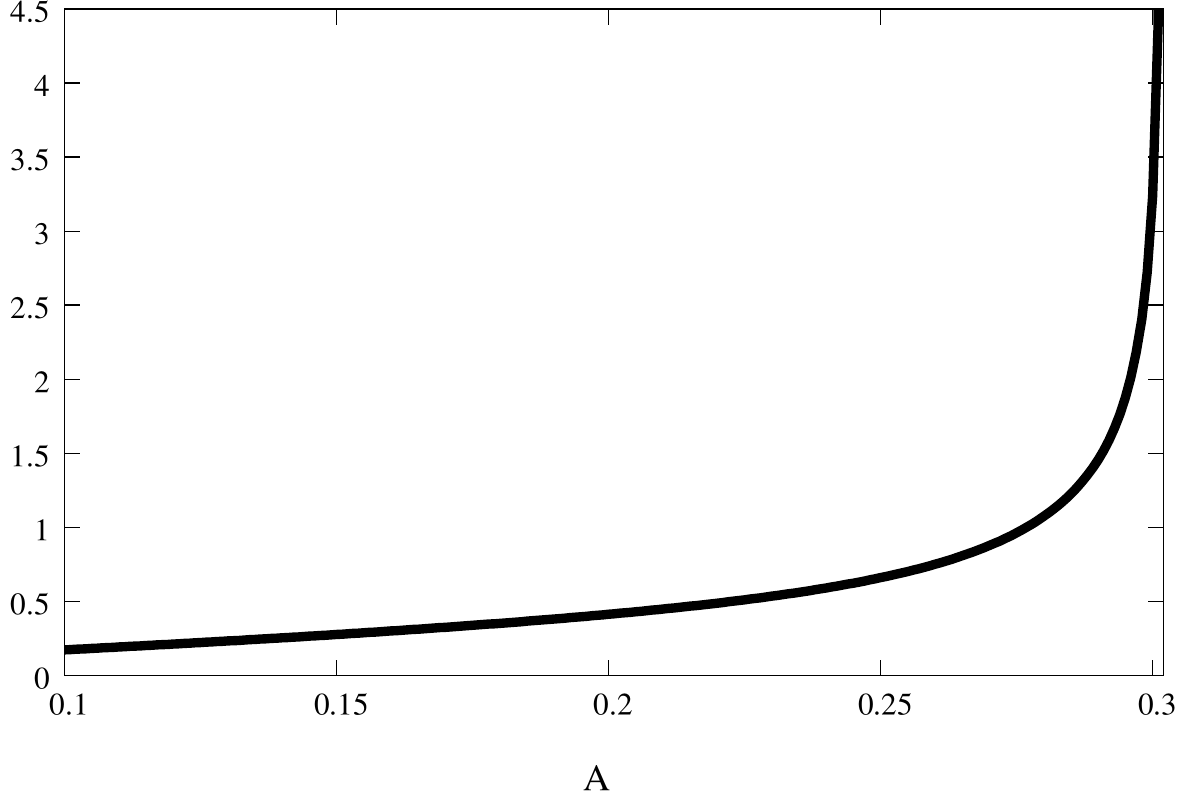}
\caption{Blow up}
\label{blowup_fig_blowup}
\end{figure}

\paragraph{Numerical results}
We choose a Gaussian with amplitude $A$ as initial data.
This pulse splits into an ingoing and an outgoing wave packet.
We make the grid large enough and focus on the ingoing pulse since
nothing interesting happens to the outgoing one.
Fig. \ref{blowup_fig_te} shows the time evolution of a Gau\ss{} pulse with small
amplitude in the original $(t,r)$--coordinates.
Then we successively increase the amplitude $A$ and monitor the $r$--derivative
of the solution $\psi$ at the center, given by 
$\alpha^{-1} \psi_x(t,\log \alpha)$.
Fig. \ref{blowup_fig_blowup} shows $\max_t \psi_x(t,\log \alpha)$ 
of the solution $\psi$ plotted against $A$.

We observe that the spatial derivative of the solution at the center diverges 
when $A$ approaches a critical value from below.
A similar result can be produced when using other types of initial data.
From this observation we conclude that the blow up is a generic phenomenon which
occurs whenever the initial data are large enough.

\subsection{Universality of Blow Up}
\begin{figure}[h]
\centering
\psfrag{psi_x}{$\max_t \psi_x(t,\log \alpha)$}
\psfrag{A}{$A$}
\includegraphics[totalheight=7cm,angle=-90]{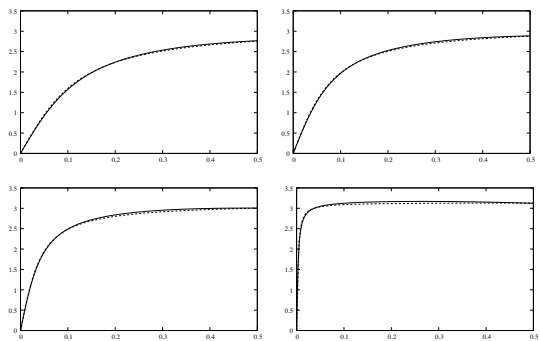}
\caption{Self--similar blow up via $\psi_0$}
\label{blowup_fig_cssblowup}
\end{figure}

Bizo\'n et. al. \cite{Bizon2000}
have studied eq. (\ref{blowup_eq_wm}) numerically and
based on their observations they have formulated some conjectures
concerning the blow up.
They claim that there exists a large open set of initial data which lead to blow
up and, in addition, the asymptotic shape of the blow up solution 
approaches the Turok Spergel solution $\psi_0$ as $t \to T-$ locally near the
center $r=0$.
Hence, the self--similar blow up behaviour defined by the solution 
$\psi_0$ is conjectured to be universal in this sense.
Furthermore, families of initial data depending on a parameter
$p$ which 
interpolate between dispersion and blow up have been studied.
There exists a critical value $p^*$ of the parameter $p$ such that initial data
with $p<p^*$ lead to dispersion while data with $p>p^*$ blow up. 
Considering initial data which lie exactly at the boundary, i.e. $p=p^*$, 
another self--similar solution $\psi_1$ which plays the role of an
"intermediate attractor" 
has been identified.
This means that the solution approaches $\psi_1$ locally around the center for a
certain time and eventually disperses or blows up via $\psi_0$ since $p=p^*$
exactly is numerically impossible. 
With the help of the code developed in the previous section it is possible to
reproduce these results. 
Fig. \ref{blowup_fig_cssblowup} for example 
shows the last stages of the self--similar blow up of a solution
$\psi$
with initial data of the form
$$ \psi(0,r)=\left \{ \begin{array}{c} A\sin r \mbox{ for } 0 \leq r <
\frac{\pi}{2} \\
A\exp(-\frac{1}{2} (r-\frac{\pi}{2})^2) \mbox{ for } r \geq \frac{\pi}{2}
\end{array} \right.
$$
and $\psi_t(0,\cdot) \equiv 0$.
The dashed line is a plot of the Turok Spergel solution $\psi_0$ with
appropriately chosen $T$.
One sees that for small $r$ the two solutions $\psi$ and $\psi_0$ 
coincide.

Hence, the simple equation (\ref{blowup_eq_wm}) shows very interesting
behaviour and a better (mathematically rigorous) understanding of these 
phenomena is desireable.

\section{Properties of Self--Similar Solutions}
\label{selfsimilar_sec}

Numerical studies of eq. (\ref{blowup_eq_wm}) suggest that self--similar
solutions play an important role in the dynamics of time evolution.
Hence, it is necessary to take a closer look at the equation
\begin{equation}
\label{selfsimilar_eq_css}
f''+\frac{2}{\rho}f'-\frac{\sin(2f)}{\rho^2(1-\rho^2)}=0.
\end{equation} 

\subsection{Existence of Self--Similar Solutions}
Eq. (\ref{selfsimilar_eq_css}) can be solved numerically using a shooting and
matching technique.
This has been done first by \AA{}minneborg and Bergstr\"om \cite{Aminneborg1995}
and later in \cite{Bizon2000} and \cite{Liebling2000}.
The singular behaviour of eq. (\ref{selfsimilar_eq_css}) at $\rho=0$ and 
$\rho=1$ yields the regularity requirements $f(0)=0$ and $f(1)=\frac{\pi}{2}$
for smooth solutions $f$.
One imposes these boundary conditions and integrates the equation away from the
singularities towards $\rho=\frac{1}{2}$ with a standard ODE integrator.
By this, one obtains two solutions $f_l$ and $f_r$ on $[0,\frac{1}{2}]$ and
$[\frac{1}{2},1]$, respectively.
Varying the free parameters $f_l'(0)$ and $f_r'(1)$ one tries to smoothly 
match the two solutions at $\rho=\frac{1}{2}$.
It turns out that there exists a countable family $\{f_n: n=0,1,2,\dots \}$
of different self--similar solutions whose existence has been proved by Bizo\'n
\cite{Bizon2000a}.

\begin{theorem}
\label{selfsimilar_thm_bizon}
There exists a countable family of smooth solutions $f_n$ of eq.
(\ref{selfsimilar_eq_css}) satisfying the boundary conditions $f_n(0)=0$ and
$f_n(1)=\frac{\pi}{2}$.
The index $n=0,1,2,\dots$ denotes the number of intersections of $f_n$ with the
line $f=\frac{\pi}{2}$ (the equator of $S^3$) on $\rho \in [0,1)$.
\end{theorem}

Additionally it is shown in \cite{Bizon2000a} that, for $n \to \infty$, 
the solutions $f_n$ converge 
to the limiting solution $f_\infty \equiv \frac{\pi}{2}$ pointwise for all 
$\rho \in (0,1]$.
Furthermore, we note that the self--similar solutions $f_n$ can be interpreted 
as harmonic
maps from the hyperbolic space $H^3$ to $S^3$ (cf. \cite{Cazenave1998}). 

\subsection{Stability Properties}
Stability of self--similar solutions is an important issue.
Highly unstable solutions are not expected to play a role in the time evolution
of generic initial data since they cannot be approached while stable solutions
may act as attractors.
In order to make these ideas more precise it is useful to transform 
the wave map equation (\ref{blowup_eq_wm}) to the new space coordinate
$\rho=\frac{r}{T-t}$ which yields
\begin{equation}
\label{selfsimilar_eq_wmtrho}
(T-t)^2 \psi_{tt}+2\rho (T-t)\psi_{t\rho}- (1-\rho^2) \psi_{\rho \rho}
-\frac{2(1-\rho^2)}{\rho}\psi_\rho+\frac{\sin(2
\psi)}{\rho^2}=0. 
\end{equation}
Hence, all self--similar solutions with blow up time $T$ are static solutions of
eq. (\ref{selfsimilar_eq_wmtrho}).
Based on the numerical observations of sec. \ref{blowup_sec} we expect the Turok
Spergel solution $\psi_0$ to act as a static attractor for solutions of eq.
(\ref{selfsimilar_eq_wmtrho}).
However, the situation is more delicate.
Since the Turok Spergel solution is not a single solution but a 
family of solutions (depending on the parameter $T$), we expect a solution of
eq. (\ref{selfsimilar_eq_wmtrho}) with blow up initial data 
to converge to a certain $\psi_0$ for an appropriate $T$.
Hence, if one fixes $T$, not all blow up solutions will converge to $\psi_0$ but
only the ones with the "right" blow up time.
Numerically this can be tested by considering a family of
initial data depending on a parameter $p$.
Then, it should be possible to adjust the parameter $p$ such that the
fine--tuned solution converges to $\psi_0$ with a prescribed $T$.

\subsection{Hyperbolic Coordinates}
Eq. (\ref{selfsimilar_eq_wmtrho}) is not well--suited for a rigorous mathematical
analysis since the coefficients depend on $t$.
Furthermore, the mixed derivative is bothersome.
Hence, we intend to introduce a new time coordinate $\sigma$ in order to
simplify the structure of the equation.
To this end we interpret the new coordinates $\sigma$, $\rho$ as 
functions of $t$, $r$ and
calculate the
new coordinate vector fields $\partial_\sigma$, $\partial_\rho$ with the help
of the
equations $\partial_t=\sigma_t \partial_\sigma+\rho_t \partial_\rho$ and
$\partial_r=\sigma_r \partial_\sigma + \rho_r \partial_\rho$.
In order to avoid bothersome off--diagonal terms we require the new coordinates
to be orthogonal, i.e. $\eta(\partial_\sigma, \partial_\rho)=0$ where $\eta$ is
the (coordinate representation of the) Minkowski metric.
Setting $\rho(t,r)=\frac{r}{T-t}$ this yields the partial differential equation
$$ \sigma_t-\frac{T-t}{r}\sigma_r=0 $$
which is a transport equation for the function $\sigma$.
The general solution is given by
$$ \sigma(t,r)=g(t^2-2Tt-r^2+c) $$
where $g$ is a free function and $c$ an arbitrary constant.
We transform eq. (\ref{selfsimilar_eq_wmtrho}) to the new time coordinate
$\sigma$.
By construction, the mixed derivative $\psi_{\sigma \rho}$ drops out.
We use the remaining freedom to make the coefficients
of the equation $\sigma$--independent.
This yields $c=T^2$ and $g=\alpha \log$ where $\alpha$ is a constant.
Therefore, we obtain $\sigma(t,r)=\alpha \log ((T-t)^2-r^2)$.
Since we want $\sigma$ to increase with $t$ we choose $\alpha$ negative and it
turns out that $\alpha=-\frac{1}{2}$ is convenient.
Hence, we arrive at $\sigma=-\log \sqrt{(T-t)^2-r^2}$.
The wave map equation (\ref{selfsimilar_eq_wmtrho}) transforms to
\begin{equation}
\label{selfsimilar_eq_wmsigrho}
\psi_{\sigma \sigma}-2 \psi_\sigma -(1-\rho^2)^2 \psi_{\rho \rho}-\frac{2
(1-\rho^2)^2}{\rho}\psi_\rho +\frac{(1-\rho^2)\sin(2 \psi)}{\rho^2}=0.
\end{equation}
We will refer to $(\sigma, \rho)$ as \emph{hyperbolic self--similar 
coordinates} or simply \emph{hyperbolic coordinates} since the lines
$\sigma=const$ are hyperbolae in a spacetime diagram. 
\begin{figure}[h]
\centering
\includegraphics[totalheight=7cm,angle=-90]{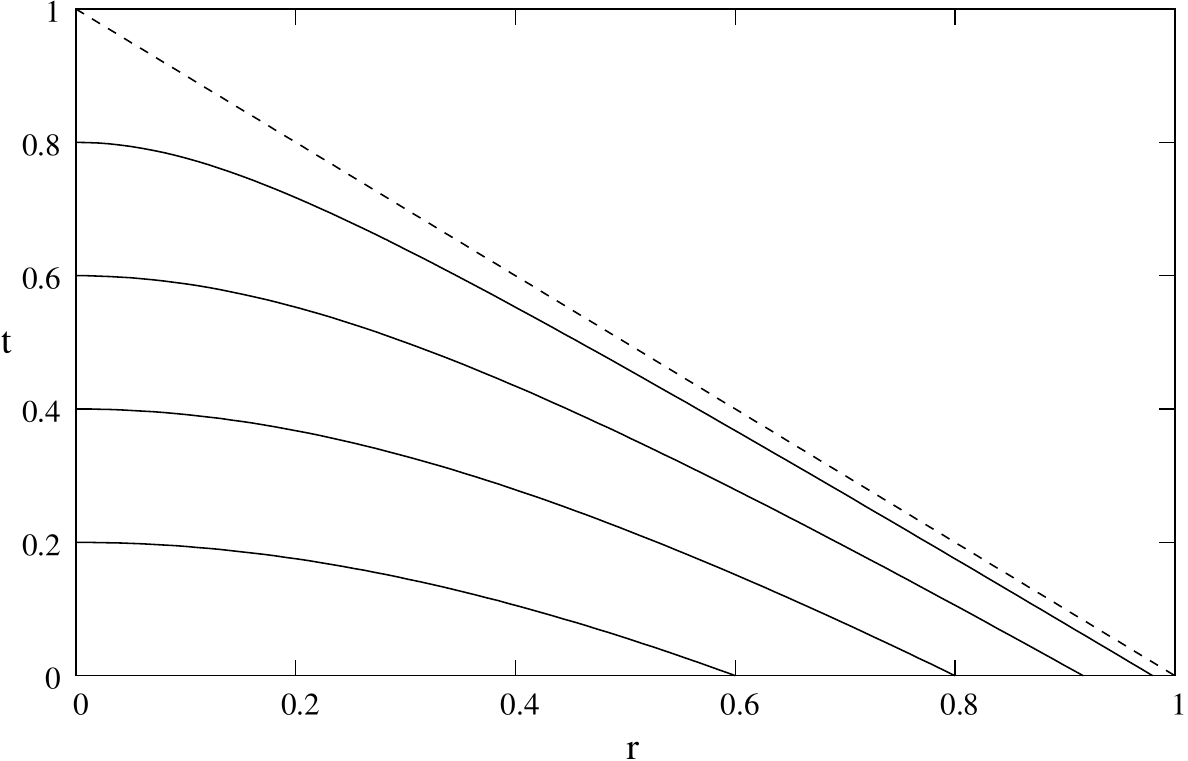}
\caption{Hyperbolic coordinates}
\label{selfsimilar_fig_hyperb}
\end{figure}
Note that $\sigma$ is only defined for $r<T-t$ and hence, the hyperbolic
coordinates cover the interior of the backward lightcone $r=T-t$ of the
blow up point $(t,r)=(T,0)$. 
The inverse transformation is given by
$$ t=T-\frac{1}{e^\sigma \sqrt{1-\rho^2}} \mbox{ and } r= \frac{\rho}
{e^\sigma \sqrt{1-\rho^2}}. $$
Fig. \ref{selfsimilar_fig_hyperb} shows lines of constant $\sigma$ in a
spacetime diagram where $T=1$. 
The dashed line is the backward lightcone of the blow up point.

\section{Well--Posedness}
\label{wp_sec}
We prove well--posedness of the Cauchy problem for the linearization of 
eq. (\ref{selfsimilar_eq_wmsigrho}) which governs the linear flow around 
a self--similar solution $f_n$.

\subsection{Linearization}

\paragraph{Linearized flow around $f_n$}
We substitute the ansatz $\psi(\sigma,\rho)=f_n(\rho)+\phi(\sigma,\rho)$ in
eq. (\ref{selfsimilar_eq_wmsigrho}) where $f_n$ is a self--similar solution of
eq. (\ref{blowup_eq_wm}).
Expanding the nonlinear term in powers of
$\phi$ and neglecting all
terms of order higher than $1$ yields
the linear evolution equation 
\begin{equation}
\label{wp_eq_wmlinear}
\phi_{\sigma \sigma}-2 \phi_\sigma -(1-\rho^2)^2 \phi_{\rho \rho}-\frac{2
(1-\rho^2)^2}{\rho}\phi_\rho +\frac{2(1-\rho^2)\cos(2 f_n)}{\rho^2}\phi=0
\end{equation}
for the perturbation $\phi$.
The wave map $f_n$ is said to be \emph{linearly stable} if solutions of eq.
(\ref{wp_eq_wmlinear}) do not grow (with respect to some suitable norm) as
$\sigma$ increases.
However, we emphasize that our discussion is still on a heuristic level since
we
have not specified what we mean by a solution so far. 
Furthermore, 
we note that it is by no means clear 
whether the nonlinear flow around $f_n$ can be approximated by this linear equation.
There are explicit examples of nonlinear evolution equations where certain
phenomena cannot be treated by linear perturbation theory, although the
perturbations are small in a certain sense (e.g. \cite{Bizon2007}).
However, it is generally believed that instabilities in the linearized problem
lead to instabilities in the nonlinear case and therefore
it is useful to study the linearized equation.

\paragraph{Symmetries}
We informally discuss the role of the time translation symmetry of eq.
(\ref{blowup_eq_wm}).
Let $\psi$ be a solution of the original wave map equation (\ref{blowup_eq_wm})
in $(t,r)$--coordinates.
Define a function $\psi^\varepsilon$ by
$\psi^\varepsilon(t,r):=\psi(t-\varepsilon, r)$.
Then, $\psi^\varepsilon$ is also a solution of eq. (\ref{blowup_eq_wm}).
Hence, the mapping $\Phi_\varepsilon: \psi \mapsto \psi^\varepsilon$ maps solutions
to solutions and satisfies $\Phi_0=\mathrm{id}$,
$\Phi_{\varepsilon+\delta}=\Phi_\varepsilon \circ \Phi_\delta$ and
$\Phi_\varepsilon^{-1}=\Phi_{-\varepsilon}$.
Let $\psi_n$ be a self--similar solution with blow up time $T$ given by
$\psi_n(t,r)=f_n \left ( \frac{r}{T-t} \right )$.
Then, $\Phi_\varepsilon (\psi_n)=\psi^\varepsilon_n$ and
$\psi^\varepsilon_n(t,r)=f_n \left (\frac{r}{T+\varepsilon - t} \right)$.
Thus, $\Phi_\varepsilon$ maps self--similar solutions with blow up time $T$ to
self--similar solutions with blow up time $T':=T+\varepsilon$.
The \emph{generator} of the \emph{orbit} 
$\{\Phi_\varepsilon (\psi): \varepsilon \in \mathbb{R}\}$ is
given by $\frac{d}{d\varepsilon}|_{\varepsilon=0} \Phi_\varepsilon(\psi)$ and
we readily calculate 
$\frac{d}{d \varepsilon}|_{\varepsilon=0}\Phi_\varepsilon(\psi)=-\psi_t$.
For a self--similar solution $\psi_n$ in $(\sigma,\rho)$--coordinates we 
obtain 
$$ \left. \phi_n^G(\sigma,\rho):=-\frac{d}{d\varepsilon} \right |_{\varepsilon=0}
\Phi^\varepsilon(\psi_n)(\sigma,\rho)=e^\sigma \rho \sqrt{1-\rho^2} f_n'(\rho).
$$ 
By direct calculation one easily verifies that $\phi^G_n$ solves eq.
(\ref{wp_eq_wmlinear}) and hence, the time translation symmetry of eq.
(\ref{blowup_eq_wm}) is reflected by an exponentially growing solution of the
linearized equation (\ref{wp_eq_wmlinear}).
This exponential instability is referred to as the \emph{gauge instability}.
 
Although $\phi^G_n$ is not differentiable at $\rho=1$ (the backward lightcone of
the singularity) we expect
this solution to play a role in dynamical time evolution since $\rho=1$ is not
in the domain covered by the hyperbolic coordinates (only $0<\rho<1$ is valid,
cf. fig. \ref{selfsimilar_fig_hyperb}). 
This fact spoils the linear stability analysis to a certain degree since
the best we can expect is a growth estimate like 
$\|\phi(\sigma,\cdot)\| \leq Ce^\sigma \|\phi(0, \cdot)\|$ for solutions of
eq. (\ref{wp_eq_wmlinear}) which would only rule out the existence of solutions
that grow faster than the gauge instability.

\subsection{The Operator $A$}
Our aim is to give a rigorous operator formulation of the evolution problem eq.
(\ref{wp_eq_wmlinear}) and apply the semigroup theory developed in sec.
\ref{cauchy_sec} to show well--posedness.

\paragraph{Simplifications} 
First of all we make the simple transformation $\phi \mapsto \tilde{\phi}$ where
$\tilde{\phi}(\sigma,\rho):=e^{-\sigma} \phi(\sigma,\rho)$ to get rid of the first
order term $-2\phi_\sigma$ in eq. (\ref{wp_eq_wmlinear}).
The transformed equation reads
\begin{equation}
\label{wp_eq_linwmtilde}
\tilde{\phi}_{\sigma \sigma}-(1-\rho^2)^2
\tilde{\phi}_{\rho
\rho}-\frac{2(1-\rho^2)^2}{\rho}\tilde{\phi}_\rho+\frac{2(1-\rho^2)\cos(2
f_n)-\rho^2}{\rho^2}\tilde{\phi}=0.
\end{equation}
We split the "potential" and write eq. (\ref{wp_eq_linwmtilde}) as
\begin{equation}
\label{wp_eq_linwmtildes}
\tilde{\phi}_{\sigma \sigma}-(1-\rho^2)^2
\tilde{\phi}_{\rho
\rho}-\frac{2(1-\rho^2)^2}{\rho}\tilde{\phi}_\rho+\frac{2(1-\rho^2)^2}{\rho^2}
\tilde{\phi}+g_n \tilde{\phi}=0 
\end{equation}
where 
$$ g_n(\rho):=\frac{2(1-\rho^2)\cos(2f_n(\rho))-\rho^2-2(1-\rho^2)^2}{\rho^2}. $$
Note that $g_n$ is regular at $\rho=0$, i.e. $g_n \in C[0,1]$ which can easily  
be checked using de l'Hospital's rule.
The idea now is to give a well--posed operator formulation 
of eq. (\ref{wp_eq_linwmtildes}) 
without $g_n \tilde{\phi}$ and to apply  
a perturbation argument.
Thus, we first consider the formal differential expression $a$ given by
$$ a:=\frac{(1-\rho^2)^2}{\rho^2} \left 
( -\frac{d}{d\rho}\rho^2 \frac{d}{d\rho}+2 \right ). $$
We set $w(\rho):=\frac{\rho^2}{(1-\rho^2)^2}$ and define the Hilbert space $H$
by $H:=L^2_w(0,1)$.

\paragraph{The method of Frobenius}
We give a brief description of a well--known method which 
provides us 
with asymptotic estimates for solutions of the equation
$au=0$.  
The method of Frobenius is a standard approach for obtaining series expansions
for
solutions of
linear second order ordinary differential equations with meromorphic coefficients 
around singular points.
Consider a differential equation $u''+pu'+qu=0$ in $\mathbb{C}$ 
where $p$ and $q$ are
holomorphic functions except for a set of isolated points in the complex plane.
Suppose that $z=z_0$ is a \emph{regular singular point} which means that 
$p_0:=\lim_{z \to z_0}(z-z_0)p(z)$ and $q_0:=\lim_{z \to z_0}(z-z_0)^2q(z)$ exist.
Then, there exist two linearly independent solutions $u_0$, $u_1$ whose
asymptotic behaviour for $z \to z_0$ can be stated explicitly:
We denote the solutions of the so--called \emph{indicial equation} 
$s(s-1)+p_0s+q_0=0$ by $s_-$
and $s_+$ (the \emph{indices}) where $|s_-| \leq |s_+|$.
\begin{itemize}
\item
If the difference $s_+-s_-$ is not an integer then
$u_0(z)=(z-z_0)^{s_+}\widetilde{u_0}(z)$ and
$u_1(z)=(z-z_0)^{s_-}\widetilde{u_1}(z)$ where 
$\widetilde{u_0}$ and $\widetilde{u_1}$
are holomorphic around $z=z_0$ with $\widetilde{u_0}(z_0)\not=0$ and 
$\widetilde{u_1}(z_0)\not=0$.
\item
If the difference is an integer then we have $u_0(z)=(z-z_0)^{s_+}\widetilde{u_0}(z)$
and $u_1(z)=c u_0(z) \log(z-z_0)+(z-z_0)^{s_-}\tilde{u_1}(z)$ where again 
$\widetilde{u_0}$ and $\widetilde{u_1}$ are holomorphic around $z=z_0$ with 
$\widetilde{u_0}(z_0) \not=0$ and $\widetilde{u_1}(z_0) \not=0$.
The constant $c$ may also be zero and thus it is possible that the logarithmic 
term does not appear.
\end{itemize}
In either case these representations are valid in the largest open circle 
around $z_0$ 
which contains no other singularity of $p$ or $q$. 

\paragraph{Endpoint classification}
Consider the Sturm--Liouville problem $au=0$ on $(0,1)$.
The endpoints $\rho=0$ and $\rho=1$ are regular singular points and therefore,
we can apply the method of Frobenius to obtain asymptotic estimates for
solutions.
Around $\rho=0$ the indices are $-2$ and $1$.
Thus, there is only one solution which belongs to $H$ near $\rho=0$ and 
Sturm--Liouville
theory (sec. \ref{sl_sec}) tells us that $\rho=0$
is in the limit--point case.
Around $\rho=1$ the indices are both equal to $\frac{1}{2}$ which shows that
there does not exist a solution which belongs to $H$ near $\rho=1$. 
Thus, $\rho=1$ is in the limit--point case as well.

\paragraph{Definition of the operator $A$}
We set $p(\rho):=\rho^2$ and $q(\rho) \equiv 2$.
Then, $au=0$ reads
$$ \frac{1}{w}\left ( -(pu')'+qu \right )=0. $$
We define $\Dom{A}:=\{u \in H: u,pu' \in AC_\mathrm{loc}(0,1), au \in H\}$ and
$Au:=au$ for $u \in \Dom{A}$.
According to Lemma \ref{sl_lem_lplp}, the operator $A: \Dom{A} \subset H \to H$
is self--adjoint.

\subsection{Properties of $A$}
We claim that $A: \Dom{A} \subset H \to H$ satisfies the estimate 
$(Au|u)_H \geq \gamma (u|u)_H$
for some $\gamma>0$ and all $u \in \Dom{A}$.
In order to show this we apply Hardy's inequality.

\begin{lemma}[Hardy's inequality]
Let $u \in C^1[a,b]$ and $u(a)=0$. Then,
$$ \int_a^b \frac{|u(x)|^2}{(x-a)^2}dx \leq 4 \int_a^b |u'(x)|^2dx. $$
Similarly, if $u(b)=0$ then
$$ \int_a^b \frac{|u(x)|^2}{(x-b)^2}dx \leq 4 \int_a^b |u'(x)|^2dx. $$
\end{lemma}

\begin{proof}
Let $u \in C^1[a,b]$ with $u(a)=0$.
Integration by parts yields
$$ \int_a^b \frac{|u(x)|^2}{(x-a)^2}dx=
\left. -\frac{|u(x)|^2}{(x-a)} \right |_a^b + \int_a^b
\frac{u'(x)\overline{u(x)}+u(x) \overline{u'(x)}}{x-a}dx $$
Using de l'Hospital's rule and $u(a)=0$ we conclude that
$$ \lim_{x \to a+} \frac{|u(x)|^2}{(x-a)}=0. $$
Observe that $-\frac{|u(b)|^2}{(b-a)} \leq 0$ and hence we have
\begin{equation*}
\begin{split} \int_a^b \frac{|u(x)|^2}{(x-a)^2}dx \leq 2 \int_a^b \frac{|u(x)|}{x-a}|u'(x)|dx
\\
\leq 2 \left ( \int_a^b \frac{|u(x)|^2}{(x-a)^2}dx \right )^{1/2} \left (\int_a^b
|u'(x)|^2 dx \right )^{1/2}
\end{split}
\end{equation*}
by Cauchy--Schwarz.
Dividing by $\left ( \int_a^b \frac{|u(x)|^2}{(x-a)^2}dx \right )^{1/2}$ and
squaring the resulting inequality yields
the claim.
The same calculation can be applied for the case $u(b)=0$.
\end{proof}

\begin{lemma}
\label{wp_lem_estB0}
There exists a $\gamma>0$ such that the operator 
$A: \Dom{A}\subset H \to H$ satisfies the estimate
$$ (Au|u)_H \geq \gamma (u|u)_H $$
for all $u \in \Dom{A}$.
\end{lemma}

\begin{proof}
Suppose $u \in \Dom{A}$ has compact support.
Then, $u \in C^1[0,1]$, $u(0)=u(1)=0$ and 
integration by parts shows
$$ (Au|u)_H=\int_0^1 \rho^2 |u'(\rho)|^2  d\rho+2 \int_0^1 
|u(\rho)|^2 d\rho $$
Observe that
\begin{equation*}
\begin{split}
\int_0^{1/2}  \rho^2 |u'(\rho)|^2 d\rho+2 \int_0^{1/2}
|u(\rho)|^2 d\rho \geq 2 \int_0^{1/2}
|u(\rho)|^2 d\rho \geq \\
\geq C \int_0^{1/2} \frac{\rho^2}{(1-\rho^2)^2}|u(\rho)|^2 d\rho.
\end{split}
\end{equation*}
We use the letter $C$ for a generic real constant greater than zero which is
\emph{not} assumed to have the same value every time it appears.
Furthermore, using Hardy's inequality we estimate
\begin{equation*}
\begin{split}
\int_{1/2}^1 \rho^2|u'(\rho)|^2  d\rho+2 \int_{1/2}^1
|u(\rho)|^2 d\rho \geq 
C \int_{1/2}^1 |u'(\rho)|^2 d\rho \\
\geq C \int_{1/2}^1 \frac{|u(\rho)|^2}{(1-\rho)^2}d\rho \geq 
C \int_{1/2}^1  \frac{\rho^2}{(1-\rho^2)^2}|u(\rho)|^2d\rho
\end{split}
\end{equation*}
Adding up these two estimates we arrive at
$$ \int_0^1 \rho^2 |u'(\rho)|^2  d\rho+2 \int_0^1
|u(\rho)|^2 d\rho 
\geq C \int_0^1 \frac{\rho^2}{(1-\rho^2)^2}|u(\rho)|^2 d\rho=C(u|u)_H. $$
Since $\{u \in \Dom{A}: u \mbox{ has compact support}\}$ is a core for $A$
(Corollary \ref{sl_cor_closAtilde}) this inequality is valid for all $u \in
\Dom{A}$ and we conclude that $(Au|u)_H \geq \gamma (u|u)_H$ for a $\gamma >0$
and all $u \in \Dom{A}$.
\end{proof}

\paragraph{Well--posedness}
We define $Y:=\Dom{A^{1/2}}$, $X:=Y
\times H$, $\Dom{\tilde{L}}:=\Dom{A} \times \Dom{A^{1/2}}$, 
$\tilde{L}(u,v):=(v,-Au)$ for $(u,v)
\in \Dom{\tilde{L}}$. Applying Theorem \ref{cauchy_thm_gen} we conclude that 
$\tilde{L}$
generates a strongly continuous one--parameter semigroup $\tilde{S}: [0,\infty) \to
\B{X}$ on $X$ satisfying $\|\tilde{S}(\sigma)\|_{\B{X}} \leq 1$. 
In particular, the Cauchy problem
$$ \left \{ \begin{array}{l}
\frac{d}{d\sigma}\mathbf{u}(\sigma)=\tilde{L}\mathbf{u}(\sigma) 
\mbox{ for } \sigma>0 \\
\mathbf{u}(0)=\mathbf{u_0} 
\end{array} \right. $$
for $\mathbf{u}: [0,\infty) \to X$ with initial data $\mathbf{u_0} \in X$ which
is an operator formulation of the unperturbed equation 
$$ 
 \tilde{\phi}_{\sigma \sigma}-(1-\rho^2)^2
\tilde{\phi}_{\rho
\rho}-\frac{2(1-\rho^2)^2}{\rho}\tilde{\phi}_\rho+\frac{2(1-\rho^2)^2}{\rho^2}
\tilde{\phi}=0 
$$
is
well--posed. 

\subsection{Bounded Perturbations}
Now we turn to the full linear equation
$$ 
\tilde{\phi}_{\sigma \sigma}-(1-\rho^2)^2
\tilde{\phi}_{\rho
\rho}-\frac{2(1-\rho^2)^2}{\rho}\tilde{\phi}_\rho+\frac{2(1-\rho^2)^2}{\rho^2}
\tilde{\phi}+g_n \tilde{\phi}=0.
$$
An operator formulation can be given as follows.
We adopt the notation of the last section, i.e. 
$$ \left \{ \begin{array}{l}
\frac{d}{d\sigma}\mathbf{u}(\sigma)=\tilde{L}\mathbf{u}(\sigma) 
\mbox{ for } \sigma>0 \\
\mathbf{u}(0)=\mathbf{u_0} 
\end{array} \right. $$
for $\mathbf{u}: [0,\infty) \to X$ with initial data $\mathbf{u_0} \in X$
is an operator formulation of the unperturbed equation ($g_n
\equiv 0$).
Now we introduce a perturbation operator $L_n': X \to X$ defined by $L_n'
(u,v):=(0,-g_n u)$ for $(u,v) \in X$.
This definition makes sense since $g_n \in C[0,1]$ and hence, $-g_n u \in H$ if
$u \in Y \subset H$.
We readily estimate $\|L_n' \mathbf{u}\|_X\leq \|g_n\|_{C[0,1]} \|\mathbf{u}\|_X$
for all $\mathbf{u} \in X$.
Thus, $L_n'$ is a bounded operator on $X$ and $\|L_n'\|_{\B{X}} \leq
\|g_n\|_{C[0,1]}$ for any $n=0,1,2,\dots$.
Therefore, 
\begin{equation}
\label{wp_eq_linwmop}
\left \{ \begin{array}{l}
\frac{d}{d\sigma}\mathbf{u}(\sigma)=\tilde{L}\mathbf{u}(\sigma)+L_n'\mathbf{u}(\sigma) 
\mbox{ for } \sigma>0 \\
\mathbf{u}(0)=\mathbf{u_0} 
\end{array} \right. 
\end{equation}
for $\mathbf{u}: [0,\infty) \to X$ with initial data $\mathbf{u_0} \in X$
is an operator formulation of the full linearized problem eq.
(\ref{wp_eq_linwmtildes}).

We apply the bounded perturbation theorem \ref{wp_thm_bpt} 
to the operator $\tilde{L}+L_n'$ defined above
which shows that $\tilde{L}+L_n'$ generates a strongly continuous 
one--parameter semigroup
$S_n: [0,\infty) \to \B{X}$ on $X$ satisfying 
\begin{equation}
\label{wp_eq_estS}
\|S_n(\sigma)\|_{\B{X}} \leq e^{\|g_n\|_{C[0,1]} \sigma}
\end{equation}
for all $\sigma>0$.
It follows that the Cauchy problem eq. (\ref{wp_eq_linwmop}) describing the 
linearized
flow around a self--similar solution $f_n$ in hyperbolic coordinates is
well--posed.

We note that the above estimate eq. (\ref{wp_eq_estS}) is not very
satisfactory because for the Turok Spergel solution $f_0$
we have $\|g_0\|_{C[0,1]}=15$ which yields 
$$ \|S_0(\sigma)\|_{\B{X}} \leq e^{15 \sigma} $$  
while the original intention was to rule out solutions
that grow faster than the gauge instability.
Translated to the semigroup approach this would require the estimate
$\|S_0(\sigma)\|_{\B{X}} \leq 1$ (remember the
transformation $\phi \mapsto \tilde{\phi}$) for the semigroup $S_0$ 
generated by $\tilde{L}+L_0'$.
In order to achieve this we have to study the spectrum of the operator 
$L_n:=\tilde{L}+L_n'$ in more detail which is the topic of the next chapter.

\chapter{The Spectrum of $L_0$}
\thispagestyle{empty}
\label{specl0_ch}

We study in detail the spectrum of the operator $L_0$ which is the generator of
the semigroup describing the linearized flow around the Turok Spergel solution
$f_0(\rho)=2 \arctan(\rho)$.
This will lead to a significant refinement of the growth estimate
\ref{wp_eq_estS} and eventually we will be able to derive a result
which rules out the existence of solutions of the linearized equation 
(\ref{wp_eq_wmlinear}) that
grow faster than the gauge instability.

\section{The Operators $A_n$ and $L_n$}
\label{opa_sec}

We adopt the notation of sec. \ref{wp_sec}, i.e. $H:=L^2_w(0,1)$ with
$w(\rho):=\frac{\rho^2}{(1-\rho^2)^2}$, $p(\rho):=\rho^2$, $q(\rho) \equiv 2$, 
$au:=\frac{1}{w}(-(pu')'+qu)$ and $\Dom{A}:=\{u \in H: u,pu' \in 
AC_\mathrm{loc}(0,1), au \in H\}$, $Au:=au$. 
First we discuss properties of the operator $A_n$ defined by
$\Dom{A_n}:=\Dom{A}$ and
$$ A_nu:=Au+g_nu $$
for $u \in \Dom{A_n}$ and 
$$ g_n(\rho):=\frac{2(1-\rho^2)\cos(2f_n(\rho))-
\rho^2-2(1-\rho^2)^2}{\rho^2} $$
where $f_n$ is the $n$--th self--similar wave map (cf. Theorem 
\ref{selfsimilar_thm_bizon}).
Hence, $L_n: \Dom{L} \subset X \to X$ is given by 
$L_n(u,v)=(v,-A_nu)$ for $(u,v) \in
\Dom{L_n}=\Dom{A} \times \Dom{A^{1/2}}$ where $X:=\Dom{A^{1/2}} \times H$ with 
$\|(u,v)\|_X^2:=\|A^{1/2}u\|_H^2+\|v\|_H^2$ for $(u,v) \in X$.
Thus,
 $A_n$ is the essential nontrivial part of $L_n$.

First of all we note that the operator $A_n$ is self--adjoint.
This can either be shown directly via Sturm--Liouville theory or it
follows from the fact that $A$ is self--adjoint, 
the boundedness of the symmetric multiplication operator 
$u \mapsto g_nu$ on $H$ and the following theorem.

\begin{theorem}
Let $A: \Dom{A} \subset H \to H$ be a self--adjoint operator on a Hilbert space
$H$. If $B: H \to H$ is bounded and symmetric then $A+B$ is self--adjoint.
\end{theorem}

\begin{proof}
See \cite{Kato1980}, p. 287.
\end{proof} 

Hence, we see that the spectrum of $A_n$ is real.
The following lemma shows how the spectra of $A_n$ and $L_n$ are related.

\begin{lemma}
\label{opa_lem_specAL}
Let $\lambda \in \mathbb{C}$. Then, $\lambda \in \sigma(L_n)$ if and only if 
$-\lambda^2 \in \sigma(A_n)$.
\end{lemma}

\begin{proof}
Invoking Lemma \ref{cauchy_lem_specL} we conclude that $\sigma(L_n) \subset 
\{\lambda \in \mathbb{C}: -\lambda^2 \in \sigma(A_n)\}$.
Hence, it remains to show that $-\lambda^2 \in \sigma(A_n)$ implies $\lambda \in 
\sigma(L_n)$ which is equivalent to $\lambda \in \rho(L_n) \Rightarrow 
-\lambda^2 \in \rho(A_n)$.

Suppose $\lambda \in \rho(L_n)$, i.e. $(\lambda-L_n)^{-1}: X \to X$ exists 
as a bounded operator on $X$. 
Let $f \in H$ and define $(u,v):=(\lambda-L_n)^{-1}(0,-f)$.
Then, $(0,-f)=(\lambda-L_n)(u,v)=(\lambda u-v, A_nu+\lambda v)$ and hence, 
$(-\lambda^2-A_n)u=f$ which shows that 
$(-\lambda^2-A_n): \Dom{A} \subset H \to H$ is surjective.
Now let $u \in \Dom{A}$ with $(-\lambda^2-A_n)u=0$.
We have $(\lambda-L_n)(u,\lambda u)=(\lambda u -\lambda u, \lambda^2 u+A_n u)
=(0,0)$.
However, since $\lambda-L_n$ is injective, we conclude that $u=0$ which 
shows that 
$-\lambda^2-A_n$ is injective as well.
Thus,  we infer $-\lambda^2 \in \rho(A_n)$.
\end{proof} 

The above result shows that the spectrum of $L_n$ can be calculated from the 
spectrum of $A_n$ and therefore, it suffices to study the operator $A_n$.
Furthermore, since $-\lambda^2 \in \mathbb{R}$ if $\lambda \in 
\sigma(L_n)$, we observe that $\sigma(L_n)$ is a subset of the union of 
the real and imaginary axis.

\section{The Spectrum of $A_0$}
\label{speca_sec}
We explicitly calculate the spectrum of $A_0$.

\subsection{Prerequisites}

\paragraph{A technical lemma}
In what follows we will frequently encounter two types of 
singular behaviour.
Terms of the form $f(x) \int_0^x g(s)ds$ where $f$ becomes unbounded for 
$x \to 0$ and
expressions like $f(x) \int_x^1 g(s)ds$ where the integral is divergent for $x
\to 0$ while $f$ goes to zero.
In either case the singular behaviour of one factor might be compensated by the
other one.
Note that we cannot treat these problems with de l'Hospital's rule since
normally the function $g$ belongs to some Lebesgue space only 
and hence it cannot be
evaluated at single points.
The following technical lemma shows how to deal with such problems.
\begin{lemma}
\label{speca_lem_singint}
Let $f \in L^2(0,1)$. 
\begin{enumerate}
\item If $\alpha > -\frac{1}{2}$ then $u$, defined by
$u(x):=x^{-\alpha-1} \int_0^x s^\alpha f(s)ds$, belongs to $L^2(0,1)$.
\item If $\alpha \geq \frac{1}{2}$ then $u$, defined by 
$u(x):=x^{\alpha-1} \int_x^1 s^{-\alpha} f(s)ds$, belongs to $L^2(0,1)$.
\end{enumerate}
\end{lemma}

\begin{proof}
Let $f \in L^2(0,1)$.
\begin{enumerate}
\item We define $v(x):=\int_0^x s^\alpha f(s)ds$ where $\alpha>-\frac{1}{2}$.
Using integration by parts and Cauchy's inequality with $\varepsilon$ we readily
calculate
\begin{multline*}
\int_0^1 |u(x)|^2dx=\int_0^1 x^{-2\alpha-2}|v(x)|^2dx= \\
\frac{1}{-2\alpha-1}\left (
\left. x^{-2\alpha-1} |v(x)|^2 \right |_0^1 
- \int_0^1 x^{-2\alpha-1}\Re{[v'(x)v(x)]}dx \right ) \\
\leq C \left. x^{-2\alpha-1} |v(x)|^2 \right |_0^1
 + C\int_0^1 \left |x^{-\alpha-1}v(x) 
\right| \left |x^{-\alpha}v'(x) \right |dx \\
\leq C \left. x^{-2\alpha-1} |v(x)|^2 \right |_0^1 + \frac{1}{2} 
\int_0^1 x^{-2 \alpha-2}|v(x)|^2dx +
C\int_0^1 x^{2\alpha} |v'(x)|^2dx \\
=C \left. x^{-2\alpha-1} |v(x)|^2 \right |_0^1
 + \frac{1}{2} \int_0^1 |u(x)|^2dx
+C \int_0^1 |f(x)|^2dx
\end{multline*}
and hence, we arrive at the inequality
$$ \|u\|_{L^2(0,1)}^2 \leq C \|f\|_{L^2(0,1)}^2
+C \left. x^{-2\alpha-1} |v(x)|^2 \right |_0^1. $$
However, using the Cauchy--Schwarz inequality we estimate
$$ |v(x)|^2 \leq \int_0^x s^{2\alpha} ds \int_0^x |f(s)|^2 ds \leq Cx^{2\alpha+1}
\|f\|_{L^2(0,1)}^2 $$ 
which implies 
$$ \lim_{x \to 0}x^{-2\alpha-1}|v(x)|^2 < \infty $$
and the claim follows.
\item Define $v(x):=\int_x^1 s^{-\alpha} f(s)ds$. For $\alpha>\frac{1}{2}$ the
very same calculation as above can be applied.
For $\alpha=\frac{1}{2}$ one encounters logarithmic terms but nevertheless the
same reasoning goes through.
\end{enumerate}
\end{proof}

\begin{remark}
Of course, the choice of the interval $(0,1)$ in Lemma \ref{speca_lem_singint} 
is completely arbitrary and
therefore analogous results are true for a general finite interval $(a,b)$ 
at either endpoint $a$ or $b$.
\end{remark}

\paragraph{The $\sim$ notation}
In connection with asymptotic estimates it is useful to introduce a common
notation.
Let $u: (a,b) \to \mathbb{C}$ and $c \in [a,b]$. 
We write $u(x) \sim v(x)$ for $x \to c$ if
$\limsup _{x \to c}\left | \frac{u(x)}{v(x)} \right | < \infty$ \footnote{$u(x)
\sim v(x)$ is sometimes denoted as $u=\mathcal{O}(v)$ where $\mathcal{O}$ is a
so--called Landau symbol.}.
Note that due to the usage of $\limsup$ in the definition, $\left |
\frac{u(x)}{v(x)} \right |$ is not required to converge for $x \to c$.
For instance we have $\sin \frac{1}{x} \sim 1$ for $x \to 0$.
Simpler examples are $e^x \sim 1$ and $\sin x \sim x$ for $x \to 0$.
But note carefully that according to our definition we also have 
$\sin x \sim 1$ for $x \to 0$.

\subsection{The Point Spectrum of $A_0$}
\paragraph{Singular points}
We consider
perturbations around the Turok Spergel solution $f_0(\rho)=2 \arctan(\rho)$.
The function $g_0$ is given by
$$ g_0(\rho)=\frac{2(1-\rho^2)\cos(2f_0(\rho))-
\rho^2-2(1-\rho^2)^2}{\rho^2}. $$
However, the term $\cos(4 \arctan(\rho))$ can be written as a rational function
$$ \cos(4 \arctan(\rho))=\frac{1-4\rho^2+\rho^4}{(1+\rho^2)^2} $$
which reveals two singularities $\rho=\pm i$ in the complex plane.
Thus, $g_0$ is meromorphic and the equation $(\lambda-A_0)u=0$ has six regular
singular points $\rho=0,\pm 1, \pm i, \infty$.
With the help of the transformation $\rho \mapsto z:=\rho^2$ this number can be
reduced by two. 
Hence, the general (formal) \footnote{We remark that the term "formal" in this
context refers to
the fact that an actual solution must belong to $\Dom{A_0}$. We do not consider
so--called formal power series solutions, i.e. series solutions whose
convergence radius is zero.} solution of $(\lambda-A_0)u=0$ can be given in terms of
Heun's functions which are the solutions of the general Fuchsian equation with
four regular singular points (cf. \cite{Ronveaux1995}).
However, since the study of Heun's functions relies heavily on numerical
techniques, this observation is not very useful for us at the present stage.

\paragraph{Asymptotic estimates}
Nevertheless we can apply Frobenius' method to obtain asymptotic estimates of
(formal) solutions of $(\lambda-A_0)u=0$ around $\rho=0$ and $\rho=1$. 
The equation $(\lambda-A_0)u=0$ reads
$$ u''+\frac{2}{\rho}u'+\left (\frac{1+\lambda}{(1-\rho^2)^2}-\frac{2
\cos(2f_0)}{\rho^2(1-\rho^2)} \right ) u=0 $$
and thus, the indices at $\rho=0$ and $\rho=1$ are $\{-2,1\}$ and 
$\{\frac{1-\sqrt{-\lambda}}{2}, \frac{1+\sqrt{-\lambda}}{2}\}$, respectively.
It is clear that the solution which behaves as $\rho^{-2}$ for
$\rho \to 0$ does not belong to $H$ and hence, eigenfunctions $u$ are
holomorphic around $\rho=0$ and satisfy $u(0)=0$.
Furthermore, the requirement $u \in H$ yields $\lim_{\rho \to 1}u(\rho)=0$.
To conclude, we have the result that every eigenfunction $u$ of $A_0$ is
in $C^1(0,1)$ and satisfies $u(0)=u(1)=0$. 

From the asymptotic estimates around $\rho=1$ we immediately infer that
there are no eigenvalues $\lambda \geq 0$ since neither of the two solutions is 
in $H$ for $\lambda \geq 0$ because 
$\Re{\frac{1-\sqrt{-\lambda}}{2}}=\Re{\frac{1+\sqrt{-\lambda}}{2}}=\frac{1}{2}$.
Hence, we have shown that $\sigma_p(A_0) \subset (-\infty,0)$.

\paragraph{Nonexistence of eigenvalues smaller than 0}
We apply a standard oscillation argument to show that there are no negative 
eigenvalues
of $A_0$.
The proof is based on the observation that the \emph{gauge mode} $\theta$,
defined by 
$$ \theta(\rho):=\rho \sqrt{1-\rho^2}f_0'(\rho)=\frac{2 \rho
\sqrt{1-\rho^2}}{1+\rho^2}$$ (formally) 
satisfies $A_0\theta=0$ as already discussed in connection with the gauge
instability (cf. sec. \ref{wp_sec}).
However, note that $\theta \notin H$ and hence, it is only a formal solution.
We remark that $\theta$ has no zeros in $(0,1)$ which will be crucial now.

\begin{lemma}
\label{speca_lem_Sturm}
The equation $(\lambda-A_0)u=0$ has no nontrivial solutions for $\lambda < 0$.
\end{lemma}

\begin{proof}
Suppose $u \in \Dom{A_0}$ is a nontrivial solution of $(\lambda-A_0)u=0$ for $\lambda<0$.
It follows that 
$u(\rho) \sim \rho$ for $\rho \to
0$ and $u(\rho) \sim (1-\rho)^\alpha$ for $\rho \to 1$ where
$\alpha:=\frac{1+\sqrt{-\lambda}}{2}>\frac{1}{2}$ by Frobenius' method.
The function $u$ satisfies the equation
$$ \lambda
u=-(1-\rho^2)^2u''-\frac{2(1-\rho^2)^2}{\rho}u'
+\frac{2(1-\rho^2)\cos(2f_0)-\rho^2}{\rho^2}u $$
for $\rho \in (0,1)$.
Let $[a,b] \subset (0,1)$.
We multiply by $\frac{\rho^2}{(1-\rho^2)^2}\theta(\rho)$ and integrate by 
parts to obtain
$$ \lambda \int_a^b \frac{\rho^2}{(1-\rho^2)^2}u(\rho)\theta(\rho)d\rho=
b^2 W(u,\theta)(b)-a^2 W(u,\theta)(a) $$
where $W(u,\theta):=u \theta'-u' \theta$ denotes the \emph{Wronskian} of the
functions $u$ and $\theta$.
For $a \to 0$ we obtain
$$ \lambda \int_0^b \frac{\rho^2}{(1-\rho^2)^2}u(\rho)\theta(\rho)d\rho=
b^2 W(u,\theta)(b). $$
Note that the limit $b \to 1$ of the left--hand side exists thanks to the
asymptotic behaviour $u(\rho)\theta(\rho) \sim (1-\rho)^{1/2+\alpha}$ and
$\frac{1}{2}+\alpha>1$.
Without loss of generality we assume $u'(0)>0$.
Suppose $u$ has its first zero at $b \in (0,1]$.
It follows that $u'(b) \leq 0$
and
$$ \lambda \int_0^b \frac{\rho^2}{(1-\rho^2)^2}u(\rho)\theta(\rho)d\rho=-b^2
u'(b)\theta(b) $$
which is a contradiction since the left--hand side is strictly negative
($\lambda<0$!) while the right--hand side is either zero or positive.
Hence, $u$ does not have a zero in the interval $(0,1]$ which is a contradiction
to $u(\rho) \sim (1-\rho)^\alpha$ for $\rho \to 1$ and $\alpha>\frac{1}{2}$. 
\end{proof}

\begin{remark} We note that this observation has first been made by Bizo\'n
\cite{Bizon2000}.
\end{remark} 

Combining this result with $\sigma_p(A_0) \subset (-\infty,0)$ we conclude that 
there are no eigenfunctions at all and hence,
the point spectrum of $A_0$ is empty! 
We formulate this result as a proposition.

\begin{proposition}
\label{speca_prop_spA0}
The point spectrum $\sigma_p(A_0)$ of $A_0$ is empty.
\end{proposition}

\subsection{Invertibility of $\lambda-A_0$ for $\lambda<0$}
Next, we study the continuous spectrum of $A_0$, i.e.
we consider the inhomogeneous equation $(\lambda-A_0)u=f$ for a given $f
\in H$.
We are interested in solutions defined on the open interval $(0,1)$ and 
fix $\lambda < 0$.
Let $u_0$ and $u_1$ denote nontrivial (formal) solutions of the homogeneous equation
$(\lambda-A_0)u=0$ with the asymptotic behaviour $u_0(\rho) \sim \rho$ for $\rho
\to 0$ and $u_1(\rho) \sim (1-\rho)^\alpha$ for $\rho \to 1$ where
$\alpha=\frac{1+\sqrt{-\lambda}}{2}>\frac{1}{2}$.
The solutions $u_0$, $u_1$ exist by Frobenius' method and they are both defined
on $(0,1)$ since the minimal distance from $\rho=0$ (resp. $\rho=1$) 
to the next singularity of the equation in
the complex plane is $1$.  
The two solutions $u_0$ and $u_1$ are linearly independent as the following
lemma shows.

\begin{lemma}
The functions $u_0$ and $u_1$ are linearly independent.
\end{lemma}

\begin{proof}
Suppose
$u_0$ and $u_1$ are
linearly dependent.
Then, the asymptotic behaviour of $u_1$ is $u_1(\rho) \sim \rho$ for
$\rho \to 0$ which means that $u_1 \in \Dom{A_0}$.
Thus, $u_1$ is an eigenfunction of $A_0$ which is a contradiction to Prop.
\ref{speca_prop_spA0} which states that $\sigma_p(A_0)=\emptyset$.
\end{proof}

By normalization we can always have $[u_0,u_1]_p(x) \equiv 1$ which will be 
assumed from now on.

We apply the variation of constants formula to obtain a formal solution of
$(\lambda-A_0)u=f$ for $f \in H$.
The general solution can be written as
\begin{multline}
\label{speca_eq_vocf}
u(\rho)=c_0u_0(\rho)+c_1u_1(\rho)-u_0(\rho) \int_{\rho_0}^\rho
u_1(\xi)f(\xi)w(\xi)d\xi \\+u_1(\rho) \int_{\rho_1}^\rho
u_0(\xi)f(\xi)w(\xi)d\xi 
\end{multline}
for $\rho \in (0,1)$ where $c_0,c_1 \in \mathbb{C}$ and $\rho_0,\rho_1 \in 
(0,1)$ are free constants.
The so--defined $u$ satisfies $u,pu' \in AC_\mathrm{loc}(0,1)$ and by
differentiation one easily verifies that it (formally) satisfies 
$(\lambda-A_0)u=f$.
Thus, the question is whether the constants $c_0,c_1,\rho_0,\rho_1$ can be
adjusted in such a way that the resulting $u$ belongs to $H$.
The answer is yes as the following lemma shows.

\begin{lemma}
For a given $f \in H$ define
$$ u(\rho):=u_0(\rho)\int_\rho^1 u_1(\xi)f(\xi)w(\xi)d\xi
+u_1(\rho)\int_0^\rho u_0(\xi)f(\xi)w(\xi)d\xi $$
for $\rho \in (0,1)$.
Then, $u \in \Dom{A_0}$ and $(\lambda-A_0)u=f$.
\end{lemma}

\begin{proof}
With the help of the asymptotic estimates $u_0(\rho) \sim \rho$, 
$u_1(\rho) \sim \rho^{-2}$ for $\rho \to 0$ and
$u_0(\rho) \sim (1-\rho)^\beta$, 
$u_1(\rho) \sim (1-\rho)^\alpha$ 
for $\rho \to 1$ where $\alpha:=\frac{1+\sqrt{-\lambda}}{2} > \frac{1}{2}$, 
$\beta:=\frac{1-\sqrt{-\lambda}}{2}<\frac{1}{2}$ together with
Lemma \ref{speca_lem_singint} one easily observes that
$\sqrt{w}u$ belongs to $L^2(0,1)$.
\end{proof}

Hence, for any $\lambda<0$ and $f \in H$ we can explicitly construct a function
$u \in \Dom{A_0}$ such that $(\lambda-A_0)u=f$ which shows that the operator
$\lambda-A_0$ is surjective.
Combining this result with $\sigma_p(A_0)=\emptyset$ we arrive at the following
proposition.

\begin{proposition}
\label{speca_prop_rhoA0}
The set $(-\infty,0)$ is contained in the resolvent set $\rho(A_0)$ of the
operator $A_0$, i.e. $(\lambda-A_0)^{-1} \in \B{H}$ exists for $\lambda < 0$.
\end{proposition}

\subsection{The Operator $\lambda-A_0$ for $\lambda \geq 0$}
Finally, we study invertibility of the operator $\lambda-A_0$ for $\lambda \geq
0$.
We fix $\lambda \geq 0$ and consider the homogeneous equation
$(\lambda-A_0)u=0$.
The method of Frobenius implies the existence of two linearly independent
solutions around $\rho=1$ which both behave as 
$(1-\rho)^{1/2}$ for $\rho \to 1$.
Furthermore, around $\rho=0$ we have a solution which behaves as $\rho$ for
$\rho \to 0$.
Thus, we can find two linearly independent solutions $u_0$ and $u_1$ on $(0,1)$
which satisfy $u_0(\rho) \sim \rho$ for $\rho \to 0$ and $u_1(\rho) \sim
(1-\rho)^{1/2}$ for $\rho \to 1$.
Again, by normalization we can assume that $[u_0,u_1]_p(x) \equiv 1$.
According to the variation of constants formula eq. (\ref{speca_eq_vocf}),
solutions $u$ of 
$(\lambda-A_0)u=f$ for given $f \in H$ have the form
\begin{multline*}
u(\rho)=c_0u_0(\rho)+c_1u_1(\rho)-u_0(\rho) \int_{\rho_0}^\rho
 u_1(\xi)f(\xi)w(\xi)d\xi 
\\+u_1(\rho) \int_{\rho_1}^\rho
\ u_0(\xi)f(\xi)w(\xi)d\xi 
\end{multline*} 
for $\rho \in (0,1)$. 
Now we try to adjust the constants $c_0,c_1,\rho_0,\rho_1$ in such a way that
the resulting $u$ belongs to $\Dom{A_0}$.
Since $u_1(\rho) \sim \rho^{-2}$ for $\rho \to 0$ we are forced to
choose $c_1=-\int_{\rho_1}^0 u_0(\xi)f(\xi)w(\xi)d\xi$ to compensate this 
bad behaviour.
Hence, we have
\begin{equation}
\label{speca_eq_proofcontspec}
u(\rho)=c_0u_0(\rho) -u_0(\rho) \int_{\rho_0}^\rho  u_1(\xi)f(\xi)w(\xi) d\xi+
u_1(\rho) \int_0^\rho  u_0(\xi)f(\xi)w(\xi)d\xi. 
\end{equation} 
For any choice of $\rho_0 \in (0,1)$ we can certainly find an $f \in H$ satisfying 
$$\int_{\rho_0}^1 
u_1(\xi)f(\xi)w(\xi)d\xi=c_0 \mbox{ and } \int_0^1 u_0(\xi)f(\xi)w(\xi)d\xi \not=0. $$
For such an $f$ we have $u(\rho) \sim u_1(\rho)$ for $\rho \to 1$ 
since the first two terms in eq. (\ref{speca_eq_proofcontspec})
cancel in the limit $\rho \to 1$.
Obviously, $u \notin H$ and hence 
there exists an $f \in H$ which is not in the range of $\lambda-A_0$. 
Thus, the operator $\lambda-A_0$ is not surjective for $\lambda \geq 0$.
This shows that every $\lambda \geq 0$ belongs to the continuous 
spectrum of $A_0$ since self--adjoint operators do not have residual spectra
and hence, we have
$\sigma_c(A_0)=[0,\infty)$.

This result concludes our discussion of the spectrum of $A_0$ and eventually we
arrive at the following theorem.

\begin{theorem}
\label{speca_thm_speca}
The spectrum of the operator $A_0$ is given by 
$\sigma(A_0)=\sigma_c(A_0)=[0, \infty)$.
\end{theorem}

\section{Improved Growth Estimate}
\label{growth_sec}
With the results of the previous sections at hand we are able to improve the
insufficient growth estimate 
$$ \|S_0(\sigma)\|_{\mathcal{B}(X)} \leq e^{15 \sigma} $$
derived in sec. \ref{wp_sec} for the semigroup $S_0$ governing the 
linearized flow around the Turok Spergel solution in hyperbolic coordinates.

Since $\inf \sigma(A_0)=0$ we conclude that $(A_0u|u)_H \geq 0$ 
for all $u \in \Dom{A_0}$
(Lemma 
\ref{propsa_prop_semibound}).
Thus, Theorem \ref{gamma0_thm_gen} tells us that $L_0$ generates a strongly 
continuous one--parameter semigroup $S_0: [0,\infty) \to \B{X}$ satisfying 
$S_0(\sigma) \leq \sigma$ for all $\sigma>1$.

Consider the conserved energy given by $\|(u(\sigma),v(\sigma)\|_H^2:=
\|A_0^{1/2}u(\sigma)\|_H^2
+\|v(\sigma)\|_H^2$ for 
a classical solution $(u(\sigma),v(\sigma)):=S_0(\sigma)(u_0,v_0)$ 
with initial data 
$(u_0,v_0) \in \Dom{L_0}$.
Since $0 \notin \sigma_p(A_0)$ we see that there does not exist a 
$u \in \Dom{A_0}$, $u \not=0$ with $A_0^{1/2}u=0$.
Hence, the energy is a norm on $\Dom{L}$ and we 
have $\|(u(\sigma),v(\sigma))\|_E \equiv \|(u_0,v_0)\|_E$ for a classical
solution with initial data $(u_0,v_0) \in \Dom{L_0}$.
Actually we are interested in the function $e^{\sigma}(u(\sigma),v(\sigma))$
(recall the rescaling $\phi \mapsto \tilde{\phi}$) and hence,
we conclude that there
does not exist a classical solution which grows faster than the gauge 
instability.
According to the discussion in sec. \ref{wp_sec} this is the best 
result we could have expected.
It shows that the Turok Spergel solution is as stable as it can possibly be 
in these coordinates.
Hence, this result strongly supports the conjecture of linear stability of 
$f_0$.

\section{Discussion}
We give a physical interpretation of the spectral behaviour of the operator
$L_0$.

\subsection{The Spectrum of $L_0$}
According to Lemma \ref{opa_lem_specAL}, the
spectrum $\sigma(L_0)$ of $L_0$ is given by $\sigma(L_0)=\{\lambda \in
\mathbb{C}: -\lambda^2 \in \sigma(A_0)\}$.
Thus, since $\sigma(A_0)=[0,\infty)$, it follows that
$\sigma(L_0)=i\mathbb{R}$, i.e. the spectrum of $L_0$ fills the imaginary
axis.
However, we emphasize that the point spectrum $\sigma_p(L_0)$ is empty and
hence, there are no eigenvalues.
Nevertheless, this spectrum is expected to lead to an exponential growth of
solutions of the equation 
\begin{equation}
\label{disc_eq_perturb}
\phi_{\sigma \sigma}-2 \phi_\sigma -(1-\rho^2)^2 \phi_{\rho \rho}-\frac{2
(1-\rho^2)^2}{\rho}\phi_\rho +\frac{2(1-\rho^2)\cos(2 f_0)}{\rho^2}\phi=0
\end{equation} 
which describes the time evolution of linear perturbations of the Turok Spergel
solution (remember the rescaling $\tilde{\phi}(\sigma,
\rho)=e^{-\sigma}\phi(\sigma, \rho)$ where $L_0$ drives the evolution of
$\tilde{\phi}$).
We have already discussed that the gauge instability is supposed to lead to
solutions that grow like $e^\sigma$ for $\sigma \to \infty$.
However, this instability cannot be responsible for the whole spectrum.
Hence, the spectral behaviour of $L_0$ calls for a physical (or more intuitive)
explanation.

\subsection{Physical Explanation}
It turns out that the structure of the spectrum of $L_0$ is closely related to
the nature of the coordinate system $(\sigma, \rho)$.
A heuristic explanation of the instabilities caused by the spectrum of $L_0$
can be given by the following argument.
Consider a perturbation which has the form of a Gaussian.
The time evolution will lead to outgoing wave packets which travel 
from smaller $\rho$ to larger $\rho$.
However, they cannot leave the backward lightcone of the singularity
since the hyperbolic coordinates break down 
at $\rho=1$.
Hence, the wave will eventually cumulate near $\rho=1$ and one 
observes exponential growth of the solution.
Thus, this seemingly unstable behaviour is due to a defect of the coordinate
system $(\sigma, \rho)$. 

\begin{figure}[h]
\centering
\includegraphics[totalheight=7cm,angle=-90]{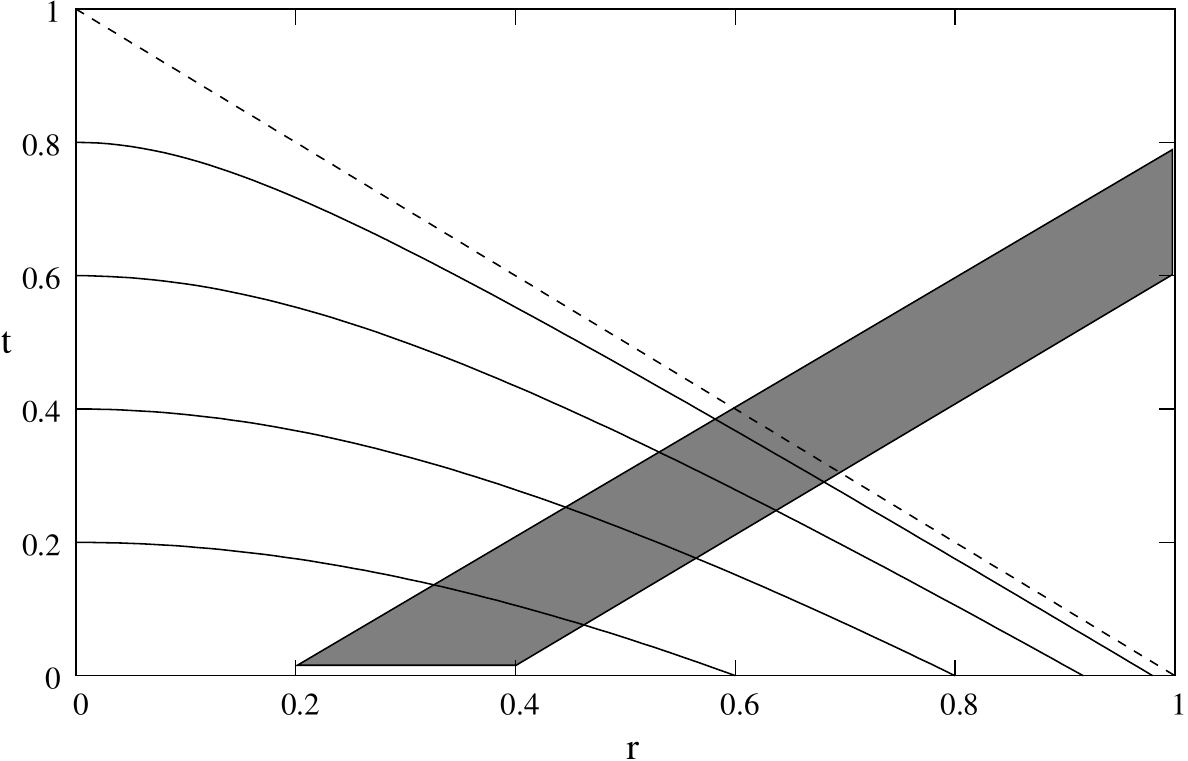}
\caption{Outgoing wave packet and hyperbolic coordinates}
\label{disc_fig_hyperb}
\end{figure}

Fig. \ref{disc_fig_hyperb} illustrates this phenomenon.
The hyperbolic lines are sections $\sigma=const$.
The shaded region represents an outgoing wave which eventually leaves the
backward lightcone (the dashed line). In $(\sigma,\rho)$ coordinates
the wave comes closer and closer to $\rho=1$ but never reaches it.

\subsection{Numerical Verification}
We employ some simple numerics to integrate the equation
$$ \tilde{\phi}_{\sigma \sigma}-(1-\rho^2)^2
\tilde{\phi}_{\rho
\rho}-\frac{2(1-\rho^2)^2}{\rho}\tilde{\phi}_\rho+\frac{2(1-\rho^2)\cos(2
f_0)-\rho^2}{\rho^2}\tilde{\phi}=0
$$ 
in order to illustrate the behaviour of solutions of the
perturbation equation (\ref{disc_eq_perturb}).
We discretize the equation using the same scheme as in sec. \ref{blowup_sec}.
The characteristic speeds are given by $\pm (1-\rho^2)$ and hence, they attain
their maximal absolute value $1$ at $\rho=0$ which implies that 
$\Delta \sigma=0.9 \Delta \rho$ is
sufficient to satisfy the CFL--condition.

\begin{figure}[h]
\centering
\includegraphics[totalheight=7cm,angle=-90]{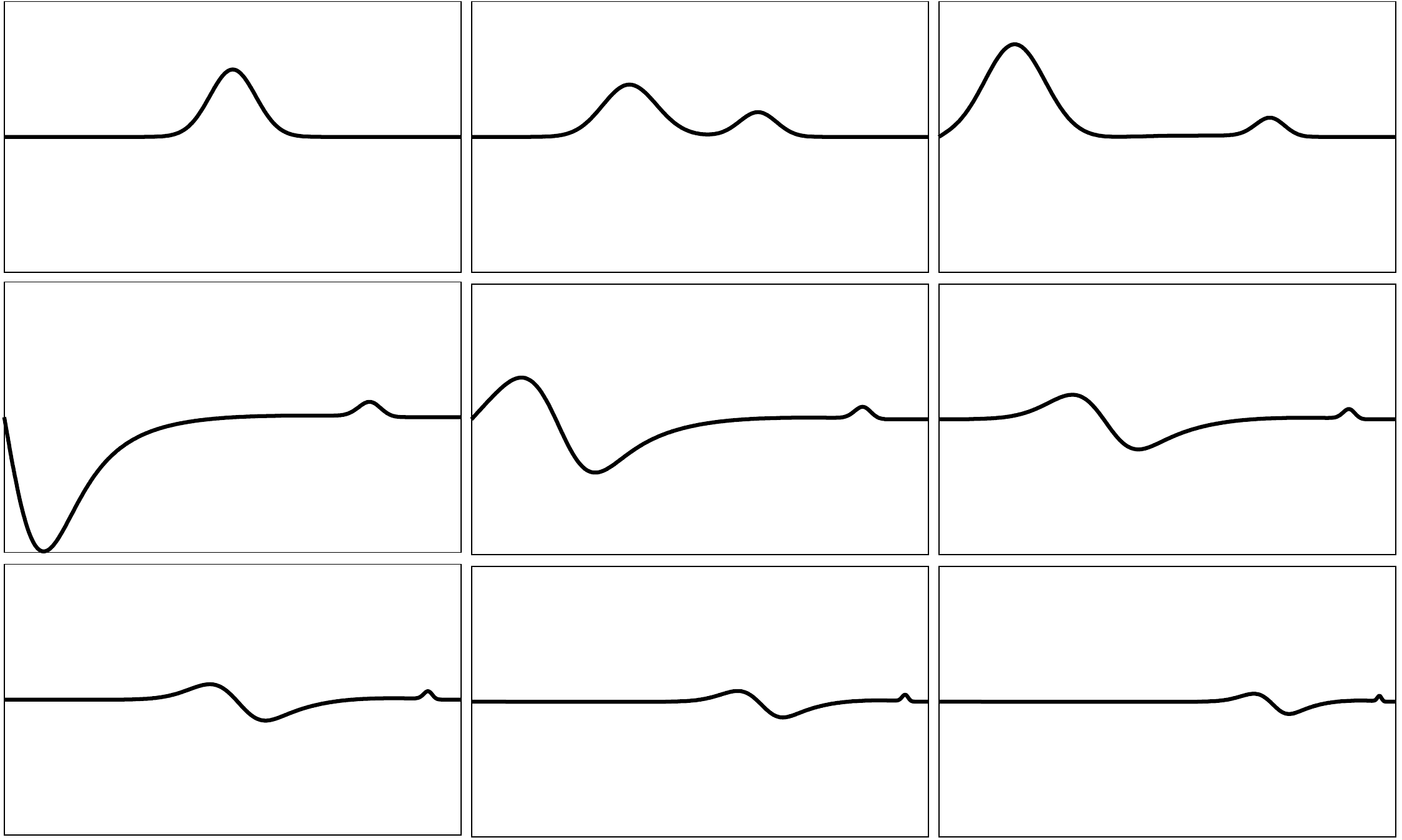}
\caption{Time evolution of a Gau\ss{} pulse}
\label{disc_fig_te}
\end{figure}

\begin{figure}[h]
\centering
\includegraphics[totalheight=7cm,angle=-90]{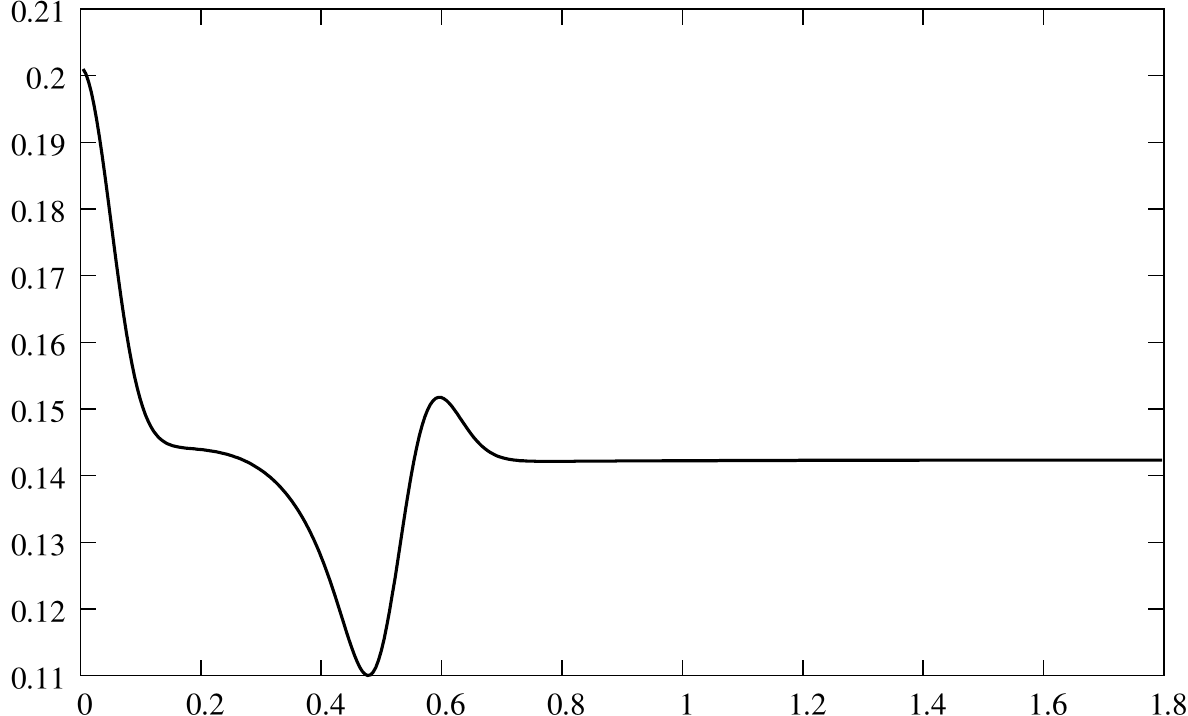}
\caption{Plot of $\sigma \mapsto \|\tilde{\phi}(\sigma, \cdot)\|_H$ against
$\sigma$}
\label{disc_fig_norm}
\end{figure}

Fig. \ref{disc_fig_te} shows the time evolution of a Gau\ss{} pulse.
As expected, the wave packets slow down and eventually freeze as they 
approach the
backward lightcone at $\rho=1$.
Fig. \ref{disc_fig_norm} shows a plot of the function
$\sigma \mapsto \|\tilde{\phi}(\sigma,\cdot)\|_H$ of the same time evolution.
For late times the norm behaves like a constant and hence, the 
corresponding $\phi(\sigma,\rho)=e^\sigma \tilde{\phi}(\sigma, \rho)$ grows
exponentially.
This instability is caused by the continuous spectrum of the operator $L_0$ as
discussed in the previous section.

\chapter{The Functional Calculus}
\thispagestyle{empty}
\label{funcalc_ch}

For the sake of completeness we discuss another, completely different method
for the treatment of ordinary differential equations on Banach spaces.
This method is the standard operator theoretic approach in quantum mechanics 
for studying dynamics of the Schr\"odinger equation.
It relies on the concept of functions of self--adjoint operators, i.e. the
functional calculus provides a method to define an operator $f(A)$ for a
complex--valued function $f$ and a self--adjoint operator $A$.
Thus, solutions of ordinary differential equations on Banach spaces
can be given in "explicit"
form as functions of self--adjoint operators.
However, this approach is not an equivalent substitute to semigroup theory since
it is by construction restricted to self--adjoint operators while
the notion of
self--adjointness does not play a role in the formulation of semigroup theory.
However, the examples we have studied so far have been formulated in a
self--adjoint manner
and hence, the functional calculus provides an alternative approach to these
problems.

\section{The Spectral Theorem}
The material outlined in this section can be found in much more detail in 
standard textbooks, e.g.
\cite{Yosida1980}.

\subsection{Spectral Families, Measures}
\label{specfam_sec}
\paragraph{Spectral Families}
Let $H$ be a Hilbert space and consider a family $\{E(\lambda) \in
\mathcal{B}(H): \lambda \in
\mathbb{R}\}$ of \emph{orthogonal projections} $E(\lambda)$ depending on a real
parameter $\lambda$.
Hence, for each $\lambda \in \mathbb{R}$, $E(\lambda): H \to H$ is a linear
bounded self--adjoint operator on $H$ which satisfies $E(\lambda)^2=E(\lambda)$.
The family $\{E(\lambda) \in \mathcal{B}(H): \lambda \in \mathbb{R}\}$ is called
a \emph{spectral family} or \emph{resolution of the identity} if it satisfies
\begin{itemize}
\item $E(\lambda_1)E(\lambda_2)=E(\lambda_2)E(\lambda_1)
=E(\min\{\lambda_1,\lambda_2\})$ for all $\lambda_1, \lambda_2 \in \mathbb{R}$
\item $\lim_{\lambda \to -\infty}E(\lambda)u=0$ and
$\lim_{\lambda \to \infty}E(\lambda)u=u$ for all $u \in H$
\item $\lim_{\varepsilon \to 0+}E(\lambda+\varepsilon)u=E(\lambda)u$ for all $u
\in H$
\end{itemize}

\paragraph{Measures}
Once there is given a spectral family it is possible to construct certain
measures.
Fix $u \in H$ and define $F_u(\lambda):=(E(\lambda)u|u)_H$.
Then, $F_u: \mathbb{R} \to \mathbb{R}$ is monotonically increasing and right
continuous, i.e. $F_u(\lambda_1)\leq F_u(\lambda_2)$ if 
$\lambda_1 \leq \lambda_2$
and $\lim_{\varepsilon \to 0+}F_u(\lambda+\varepsilon)=F_u(\lambda)$.
Hence, $F_u$ is a \emph{distribution function} in the sense of measure theory.
Given a distribution function $F_u$ one can show that there exists a 
unique measure
$\mu_u: \mathcal{B} \to
[0,\infty]$ which satisfies
$\mu_u((\lambda_1,\lambda_2])=F_u(\lambda_2)-F_u(\lambda_1)$ for $\lambda_1 \leq
\lambda_2$ where $\mathcal{B}$
denotes the Borel $\sigma$--algebra on $\mathbb{R}$.
Note that $\mu_u(\mathbb{R})=\|u\|_H^2 < \infty$ and hence, the measure
is finite.
Moreover, observe that due to the special properties of $E(\lambda)$
we have
$\mu_u((\lambda_1,\lambda_2])=\|[E(\lambda_2)-E(\lambda_1)]u\|_H^2$ for
$\lambda_1 \leq \lambda_2$.

For fixed $u,v \in H$ we define another, complex--valued finite 
measure $\mu_{u,v}: \mathcal{B} \to \mathbb{C}$ by
$$ \mu_{u,v}(B):=\frac{1}{4}\left (
\mu_{u+v}(B)-\mu_{u-v}(B)+i\mu_{u-iv}(B)-i\mu_{u+iv}(B) \right ) $$
for all $B \in \mathcal{B}$.
Invoking the polarization identity it follows that
$\mu_{u,v}((\lambda_1,\lambda_2])=(E(\lambda_2)u|v)_H-(E(\lambda_1)u|v)_H$ for
$\lambda_1 \leq \lambda_2$. 

\subsection{Operators Defined via Measures}
\label{opmeasure_sec}

\paragraph{Projections}
Fix $u \in H$, $B \in \mathcal{B}$ and consider the mapping $F_{u,B}: H \to
\mathbb{C}$ defined by $F_{u,B}(v):=\mu_{u,v}(B)$.
First, we claim that $F_{u,B}(\alpha v)=\overline{\alpha}F_{u,B}(v)$ 
for any $\alpha \in
\mathbb{C}$.
Since the measure $\mu_{u,v}$ is uniquely determined by its values on 
half--open
intervals it suffices to show properties for $B:=(\lambda_1,\lambda_2]$ where
$\lambda_1 \leq \lambda_2$ and they automatically remain 
true for arbitrary $B \in
\mathcal{B}$.
Hence, the claim follows immediately from the formula
$\mu_{u,v}((\lambda_1,\lambda_2])=(E(\lambda_2)u|v)_H-(E(\lambda_1)u|v)_H$.
Furthermore, we have $F_{u,B}(v+w)=F_{u,B}(v)+F_{u,B}(w)$ for $v,w \in H$ and 
$|F_{u,B}(v)| \leq \|u\|_H \|v\|_H$.
Therefore, $\overline{F_{u,B}}: H \to \mathbb{C}$ is a bounded linear functional 
on $H$.
Invoking the Riesz representation theorem \ref{hilbert_thm_Riesz} we conclude
that there exists a unique $w \in H$ such that 
$\overline{F}_{u,B}(v)=(v|w)_H$ for all $v \in H$.
Hence, for any $B \in \mathcal{B}$ there exists a well--defined unique mapping 
$P(B): H \to H$ such that
$\overline{F_{u,B}}(v)=(v|P(B)u)_H$ for all $u,v \in H$.
It is easy to see that $P(B)$ is linear and bounded.
Note that by construction we have
$P((\lambda_1,\lambda_2])=E(\lambda_2)-E(\lambda_1)$ and hence, we have extended
these projections to arbitrary Borel sets.
Another way to write $P(B)$ is $(P(B)u|v)_H=\mu_{u,v}(B)$ for all $u,v \in H$.

\paragraph{Properties of $P$}
Note that $(u|P(B)v)_H=\overline{(P(B)v|u)_H}=\overline{\mu_{v,u}(B)}
=\mu_{u,v}(B)=(P(B)u|v)_H$ by definition of $\mu_{u,v}$ and hence, $P(B)$ is 
self--adjoint.
Furthermore, $(P(B)P(C)u|v)_H=\mu_{P(C)u,v}(B)=\mu_{u,v}(B \cap C)
=(P(B \cap C)u|v)_H$ for all $v \in H$ and $B,C$ half--open intervals.
By uniqueness of the measure we have $P(B)P(C)=P(B \cap C)$ for arbitrary 
$B \in \mathcal{B}$ and half--open intervals $C$.
However, using the self--adjointness of $P(B)$ we can interchange the role of 
$B$ 
and $C$ and hence, $P(B)P(C)=P(B \cap C)$ holds for all $B,C \in \mathcal{B}$.
In particular it follows that $P(B)^2=P(B)$ and thus, $P(B)$ is an orthogonal 
projection.

\paragraph{The domain of $A_f$}
With the help of the measures $\mu_u$ and $\mu_{u,v}$ we can define 
an operator $A_f: \Dom{A_f} \subset H \to H$ for a given measurable function $f:
\mathbb{R} \to \mathbb{C}$ in the following way.
Set $$ \Dom{A_f}:=\left \{u \in H: \int |f|^2 d\mu_u < \infty 
\right \}. $$
Then, $u \in \Dom{A_f}$ implies $\alpha u \in \Dom{A_f}$ for any $\alpha \in
\mathbb{C}$ since by definition we have $\mu_{\alpha
u}((\lambda_1,\lambda_2])=|\alpha|^2 \mu_u((\lambda_1,\lambda_2])$ for $\lambda_1 \leq
\lambda_2$ and, by uniqueness of the measure $\mu_u$, it follows that
$\mu_{\alpha u}=|\alpha|^2 \mu_u$.
Similarly, for $u,v \in H$ we have $\mu_{u+v} \leq 2 ( \mu_u+\mu_v)$ by the
triangle inequality and hence, $u,v \in \Dom{A_f}$ implies $u+v \in \Dom{A_f}$
which shows that $\Dom{A_f}$ is a subspace of $H$.
 
Furthermore, $\Dom{A_f}$ is dense in $H$ which can be seen as follows.
Define $B_n:=|f|^{-1}((-\infty,n])$ for $n \in \mathbb{N}$.
Since $f$ is measurable, $B_n \in \mathcal{B}$ for all $n \in \mathbb{N}$.
Let $u \in H$ and set $u_n:=P(B_n)u$.
Then we have 
$$ \int |f|^2 d\mu_{u_n}=\int |f|^2 \chi_{B_n} d\mu_u \leq n^2 \|u\|^2 < \infty$$
since $\mu_{u_n}(C)=\mu_{P(B_n)u}(C)=\mu_u(B_n \cap C)$ for 
all $C \in \mathcal{B}$.
Therefore, $u_n \in \Dom{A_f}$ for all $n \in \mathbb{N}$ and one easily sees
that $u_n \to u$ in $H$.

\paragraph{The operator $A_f$}
Fix $u \in \Dom{A_f}$ and consider the mapping $F: H \to \mathbb{C}$ defined by
$$ F(v):=\int f d\mu_{u,v}. $$
One readily observes that $\overline{F}$ is a bounded linear functional on $H$ and
hence, by the Riesz representation theorem there exists a well--defined
operator $A_f: \Dom{A_f} \subset H \to H$ such that $F(v)=(A_fu|v)_H$
for all $v \in H$.
It is easily seen that $A_f$ is linear and
symbolically one usually writes
$$ A_f=\int_{\mathbb{R}} f(\lambda)dE(\lambda). $$  

\subsection{Spectral Families and Self--Adjoint Operators}

\paragraph{The spectral theorem}
It turns out that to every self--adjoint operator there is associated a unique 
spectral family. 
This result is known as the \emph{spectral theorem}.
More precise we have the following statement.

\begin{theorem}
Let $A: \Dom{A} \subset H \to H$ be a self--adjoint operator on a
Hilbert space $H$. Then, there exists a unique spectral family $\{E(\lambda):
\lambda \in \mathbb{R}\}$ such that 
$$ A=\int_{\mathbb{R}} \lambda dE(\lambda). $$
\end{theorem}

\paragraph{Representations}
All the information concerning the spectrum of a self--adjoint operator 
$A: \Dom{A} \subset H \to H$ is encoded in its spectral family $\{E(\lambda):
\lambda \in \mathbb{R}\}$.

To see this we note that for any spectral family $\{E(\lambda): \lambda \in
\mathbb{R}\}$ we have the representation
$$ I=\int_\mathbb{R} dE(\lambda) $$
for the identity operator $I: H \to H$ on $H$.
This can be immediately seen by inserting the definitions.
Consider the operator $A_1$ (cf. sec. \ref{opmeasure_sec}). 
Then we have $\Dom{A_1}=H$ and 
$(A_1u|v)_H=\int d\mu_{u,v}=\mu_{u,v}(\mathbb{R})=(u|v)_H$ for all
$u,v \in H$ and hence, $A_1=I$.
Thus, the operator $\lambda_0-A$ is given by
$$ \lambda_0-A=\int_\mathbb{R} (\lambda_0-\lambda)dE(\lambda) $$
where $\{E(\lambda): \lambda \in \mathbb{R}\}$ is the unique spectral family
associated to $A$.
Furthermore, one can show that 
$$ \|(\lambda_0-A)u\|_H^2=\int_\mathbb{R} (\lambda_0-\lambda)^2 d\mu_u(\lambda) $$
where $\mu_u$ is the measure associated to $\{E(\lambda)\}$ defined in sec.
\ref{specfam_sec}.

\paragraph{The spectrum of $A$}
We have $\lambda_0 \in \sigma_p(A)$ if and only if there exists a $u \in \Dom{A}$,
$u \not=0$ such that
$$ 0=\|(\lambda_0-A)u\|_H^2=\int_\mathbb{R} (\lambda_0-\lambda)^2
d\mu_u(\lambda). $$
We conclude that the distribution function $F_u$ given by
$F_u(\lambda)=(E(\lambda)u|u)_H$ has to be constant except for a possible
discontinuity at $\lambda=\lambda_0$.
The requirements $\lim_{\lambda \to -\infty} F_u(\lambda)=0$ and $\lim_{\lambda
\to \infty} F_u(\lambda)=\|u\|_H^2$ together with right continuity of $F_u$ 
imply $F_u(\lambda)=0$ for $\lambda <
\lambda_0$ and $F_u(\lambda)=\|u\|_H^2$ for $\lambda \geq \lambda_0$.
Hence, $\lambda_0$ is an eigenvalue of $A$ with eigenvector $u \in \Dom{A}$ 
if and only if $E(\lambda)u=0$ for $\lambda < \lambda_0$ and $E(\lambda)u=u$ for
$\lambda \geq \lambda_0$. 

Similarly, one can show that $\lambda_0 \in \sigma_c(A)$ is equivalent to
$\lim_{\varepsilon \to 0-}E(\lambda_0+\varepsilon)u=E(\lambda_0)u$ for all $u
\in H$ and $E(\lambda_1)\not=E(\lambda_2)$ for $\lambda_1 < \lambda_0 < \lambda_
2$. 
As already proved in sec. \ref{strucspec_sec}, the residual spectrum
$\sigma_r(A)$ is empty.

Thus, we have $\lambda_0 \in \rho(A)$ if and only if the associated spectral
family $E$ is constant on a
neighbourhood of $\lambda_0$.
This, together with the requirements 
$\lim_{\lambda \to -\infty}E(\lambda)u=0$ and
$\lim_{\lambda \to \infty}E(\lambda)u=u$ implies that the spectrum of a
self--adjoint operator is nonempty!

\section{Functions of Self--Adjoint Operators}
Let $A: \Dom{A} \subset H \to H$ be a self--adjoint operator on a Hilbert space
$H$. 
With the help of the spectral theorem we are now able to define functions of
$A$.
Again, all the material presented in this section can be found in standard
textbooks, e.g. \cite{Yosida1980}. 
We also mention the freely available book \cite{Teschl2007}.

\subsection{Definitions and Properties}
\paragraph{Definition of $f(A)$}
By the spectral theorem there exists a unique spectral family $\{E(\lambda):
\lambda \in \mathbb{R}\}$ such that $A$ has the representation
$A=\int_\mathbb{R} \lambda dE(\lambda)$ or, less symbolically,
$(Au|v)_H=\int \mathrm{id}_\mathbb{R} d\mu_{u,v}$ for all $u \in \Dom{A}$ and $v
\in H$ where $\mu_{u,v}$ is the complex--valued measure associated to
$\{E(\lambda)\}$ defined in sec.
\ref{specfam_sec}.
Let $f: \mathbb{R} \to \mathbb{C}$ be a measurable function.
Then, we define an operator $f(A): \Dom{f(A)} \subset H \to H$ by
$\Dom{f(A)}:=\Dom{A_f}$ and $f(A):=A_f$ where $A_f$ is the operator defined in
sec. \ref{opmeasure_sec}, i.e. we have
$$ \Dom{f(A)}=\left \{u \in H: \int |f|^2 d\mu_u < \infty \right \} $$
and 
$$ (f(A)u|v)_H=\int f d\mu_{u,v} $$
for all $u \in \Dom{f(A)}$ and $v \in H$ where, as before, $\mu_u:=\mu_{u,u}$.
As already mentioned, $f(A)$ is densely defined and linear.

\paragraph{Calculus for functions of self--adjoint operators}
One can show that $E(\lambda)f(A) \subset f(A)E(\lambda)$ for any $\lambda \in
\mathbb{R}$, i.e. $f(A)$ commutes with $E(\lambda)$.
Furthermore, we have the following functional calculus for $f(A)$ (see e.g.
\cite{Yosida1980} for a proof).
\begin{itemize}
\item $\Dom{f(A)}=\Dom{\overline{f}(A)}$ and
$(f(A)u|v)_H=(u|\overline{f}(A)v)_H$ for all $u,v \in \Dom{f(A)}$.
\item Let $g: \mathbb{R} \to \mathbb{C}$ be a measurable function.
Then, we have the representation 
$$ (f(A)u|g(A)v)_H=\int f\overline{g} d\mu_{u,v} $$
for all $u \in \Dom{f(A)}$ and $v \in \Dom{g(A)}$.
\item $(\alpha f)(A)u=\alpha f(A)u$ for all $u \in \Dom{f(A)}$, $\alpha
\in \mathbb{C}$ and $(f+g)(A)u=f(A)u+g(A)u$ for all $u \in \Dom{f(A)} \cap
\Dom{g(A)}$.
\item $g(A)u \in \Dom{f(A)}$ for $u \in \Dom{g(A)}$ is equivalent to 
$u \in \Dom{fg(A)}$ and we have
$fg(A)u=f(A)g(A)u$ for all $u \in \Dom{fg(A)}$.
\item If $f$ is finite everywhere on $\mathbb{R}$ then the adjoint 
$f(A)^*$ is given by $f(A)^*=\overline{f}(A)$ and hence, $f(A)f(A)^*=f(A)^*f(A)$
which shows that $f(A)$ is a \emph{normal operator}.
In particular, $f(A)$ is self--adjoint if $f$ is finite everywhere and
real--valued. 
\end{itemize}

\subsection{Unitary Groups}
Let $H$ be a Hilbert space.
A mapping $U: \mathbb{R} \to \mathcal{B}(H)$ is called a \emph{strongly
continuous one--parameter
group of linear operators on $H$} if $U$ satisfies
$U(0)=\mathrm{id}_H$, $U(t+s)=U(t)U(s)$ for all $t,s \in \mathbb{R}$
and the mapping $t \mapsto U(t)u$ is continuous for all $u \in H$.
$U$ is called \emph{unitary} if $\|U(t)u\|_H=\|u\|_H$ for all $u \in
H$ and any $t \in \mathbb{R}$.
In particular, every strongly continuous one--parameter group is a strongly
continuous one--parameter semigroup of linear operators and hence, the notion of
a generator of a semigroup carries over to groups.
The following proposition shows how to explicitly construct a strongly
continuous unitary one--parameter group with the help of the functional calculus for
a given self--adjoint generator.

\begin{proposition}
\label{funsa_prop_unit}
Let $A: \Dom{A} \subset H \to H$ be a self--adjoint operator on a Hilbert space
$H$. Define $U_+(t):=\exp(itA)$ and $U_-(t):=\exp(-itA)$ via the functional
calculus.
Then, $U_\pm$ have the following properties.
\begin{enumerate}
\item For any $t \in \mathbb{R}$, $U_\pm(t)$ is a bounded linear operator on 
$H$ which satisfies $\|U_\pm(t)u\|_H=\|u\|_H$ for all $u \in H$.
\item $U_\pm(0)=\mathrm{id}_H$ and $U_\pm(t+s)=U_\pm(t)U_\pm(s)$ for 
all $t,s \in \mathbb{R}$.
\item For any $u \in H$, the function $t \mapsto U_\pm(t)u: \mathbb{R} \to H$ 
is continuous.
\item The limit $\lim_{h \to 0}\frac{1}{h}(U_\pm(t+h)u-U_\pm(t))$ exists if and
only if $u \in \Dom{A}$ and in this case we have
$$ U_\pm'(t)u:=\lim_{h \to 0}\frac{U_\pm(t+h)u-U_\pm(t)u}{h}=\pm iU_\pm(t)Au
=\pm iAU_\pm(t)u. $$
Thus, $\pm iA$ is the generator of $U_\pm$.
\end{enumerate}
\end{proposition}

\begin{proof}
\begin{enumerate}
\item Let $t \in \mathbb{R}$. Application of the functional calculus yields  
$$ (U_\pm(t)u|U_\pm(t)u)_H=\int d\mu_u=\|u\|_H^2 $$
for any $u \in H$ where 
$\mu_u$ is the measure defined in sec. \ref{specfam_sec} for the spectral
family $\{E(\lambda): \lambda \in \mathbb{R}\}$ associated to the self--adjoint
operator $A$.
Hence, $U(t): H \to H$ is a unitary linear operator on $H$.
\item This follows immediately from the functional calculus.
\item Fix $u \in H$ and let $t_0 \in \mathbb{R}$. 
Applying the functional calculus we have
$$ \lim_{t \to t_0}\|U_\pm(t)u-U_\pm(t_0)u\|_H^2
=\lim_{t \to t_0}\int_\mathbb{R}|\exp(\pm
it\lambda)-\exp(\pm it_0\lambda)|^2 d\mu_u(\lambda)=0 $$
by Lebesgue's theorem on dominated convergence.
\item Let $u \in \Dom{A}$. Applying the functional calculus we obtain
\begin{multline*}
 \lim_{t \to 0} \|t^{-1}(U_\pm(t)u-u)\mp iAu\|_H^2=\lim_{t \to 0}
\int_\mathbb{R} |t^{-1}(\exp(\pm it\lambda)-1) \mp i\lambda|^2 
d\mu_u(\lambda) \\
=\int_\mathbb{R} \left |\left. \frac{d}{dt}\exp(\pm it\lambda) \right |_{t=0} \mp
i\lambda \right |^2 d\mu_u(\lambda)=0
\end{multline*}
by Lebesgue's theorem on dominated convergence.
Hence, we have shown $U_\pm'(0)u=\pm iAu$ for all $u \in \Dom{A}$.
However, using the group property and strong continuity of $U_\pm$ we infer
$$ \lim_{h \to 0}h^{-1} (U_\pm(t+h)u-U_\pm(t)u)=U_\pm(t)\lim_{h \to
0}h^{-1}(U_\pm(h)u-u)=\pm i U_\pm(t)Au $$
and similarly we obtain $U_\pm'(t)u=\pm iAU(t)u$.

Now let $\tilde{A}$ be the generator of $\mp i U_\pm$, i.e.
$$ \Dom{\tilde{A}}=\left \{u \in H: \lim_{t \to 0}\frac{1}{t}(U_\pm(t)u-u) 
\mbox{ exists} \right \} $$
and $\tilde{A}:=\mp i \lim_{t \to 0}\frac{1}{t}(U_\pm(t)u-u)$.
Then we have
\begin{multline*}
(\tilde{A}u|v)_H
=\mp i \lim_{t \to 0}t^{-1}(u|U_\pm(t)^*v-v)_H \\
=\left ( u|\pm i \lim_{t \to 0}t^{-1} (U_\pm(-t)v-v) \right )_H
=(u|\tilde{A}v)_H
\end{multline*}
for all $u,v \in \Dom{\tilde{A}}$ since 
$U_\pm(t)^*=U_\pm(-t)$ by the functional calculus.
This shows that the operator $\tilde{A}$ 
is a symmetric extension
of $A$.
However, since $A$ is self--adjoint there do not exist proper symmetric
extensions and therefore we have $\Dom{A}=\Dom{\tilde{A}}$. 

\end{enumerate}
\end{proof}

\begin{remark}
We have shown that $iA$ generates a strongly continuous unitary one--parameter
group if $A$ is self--adjoint.
It turns out that the converse is also true, i.e. that any strongly 
continuous unitary
one--parameter group $U$ on a Hilbert space $H$ is of the form $U(t)=\exp(itA)$ 
for a self--adjoint operator $A: \Dom{A} \subset H \to H$.
This result is known as \emph{Stone's theorem} (cf. \cite{Yosida1980}).
\end{remark}

\section{Well--Posedness of Wave Equations}
Applying the theory outlined in the previous sections we are now able to
consider abstract wave equations on Hilbert spaces.
This yields a well--posedness result for a certain class of
equations and moreover, it is possible to write down the solution in a rather
explicit form with the help of the functional calculus.
For more information we refer to \cite{Beyer2007}.

\subsection{Well--Posedness for Strictly Positive Generators}
The first result is the analogue of Theorem \ref{cauchy_thm_gen} in semigroup
theory.
Let $A: \Dom{A} \subset H \to H$ be a self--adjoint operator on a Hilbert space
$H$ satisfying $(Au|u)_H \geq 0$ for all $u \in \Dom{A}$.
Then there exists the square root $A^{1/2}$ of $A$ which is again self--adjoint
(Theorem \ref{growth_thm_sqrt} or directly via functional calculus).
We consider the unitary groups $t \mapsto \exp(\pm itA^{1/2})$.

\begin{lemma}
\label{weq_lem_exp}
Let $A: \Dom{A} \subset H \to H$ be a self--adjoint operator satisfying
$(Au|u)_H \geq 0$ for all $u \in \Dom{A}$.
For $u_0 \in \Dom{A}$ define $u(t):=\exp(itA^{1/2})u_0$ via the 
functional calculus.
Then, the function $u: \mathbb{R} \to H$ satisfies the wave equation
$u''(t)=-Au(t)$ for all $t \in \mathbb{R}$ where
$u'(t):=\lim_{h \to 0}\frac{1}{h}(u(t+h)-u(t))$.
Furthermore, the same holds true for $u(t):=\exp(-itA^{1/2})u_0$.
\end{lemma} 

\begin{proof}
Define $u(t):=\exp(itA^{1/2})u_0$ for $u_0 \in \Dom{A}$.
From Prop. \ref{funsa_prop_unit} we already know that $u'(t)=
iA^{1/2}u(t)$ for all $t \in \mathbb{R}$.
Furthermore, $\exp(itA^{1/2})$ commutes with $A^{1/2}$.
Thus, since $u_0 \in \Dom{A}$, we have $A^{1/2}u_0 \in \Dom{A^{1/2}}$ and hence,
applying Prop. \ref{funsa_prop_unit} again, we obtain
$$ u''(t)=\frac{d}{dt}iA^{1/2}\exp(itA^{1/2})u_0=
\frac{d}{dt}i\exp(itA^{1/2})A^{1/2}u_0=-Au(t). $$
The same reasoning goes through for $u(t)=\exp(-itA^{1/2})u_0$.
\end{proof}

Now we assume the slightly stronger condition $(Au|u)_H \geq \gamma (u|u)_H$ 
for all $u \in \Dom{A}$ and some $\gamma > 0$
and consider the operators $\sin(tA^{1/2})$ and $\cos(tA^{1/2})$.
According to the functional calculus they are given by
$$ \sin(tA^{1/2})=\frac{1}{2i} \left ( \exp(itA^{1/2})-\exp(-itA^{1/2}) \right ) $$ and
$$ \cos(tA^{1/2})=\frac{1}{2} \left (\exp(itA^{1/2})+\exp(-itA^{1/2}) \right). $$
Furthermore, since $(Au|u)_H>\gamma(u|u)_H$, it follows that $\sigma(A) \subset
[\gamma,\infty)$ (cf. Lemma \ref{cauchy_lem_specA}) 
and in particular, $A$ is bounded 
invertible.
Hence, the operator $A^{-1/2}:=A^{1/2}A^{-1}: H \to \Dom{A^{1/2}} \subset H$ 
exists and is inverse to
$A^{1/2}$.
Now we have collected all the necessary tools to prove the analogue of the
generation result Theorem \ref{cauchy_thm_gen}.

\begin{theorem}
\label{weq_thm_gen}
Let $A: \Dom{A} \subset H \to H$ be a self--ajoint operator on a Hilbert space
$H$ satisfying $(Au|u)_H \geq \gamma (u|u)_H$ for all $u \in \Dom{A}$ and a
$\gamma>0$.
Then, for given $u_0,u_1 \in \Dom{A}$, 
there exists a unique function 
$u: \mathbb{R} \to H$ such that
$u''(t)=-Au(t)$ for all $t \in \mathbb{R}$ and $u(0)=u_0$, $u'(0)=u_1$.
The unique solution $u$ can be given explicitly as
$$ u(t)=\cos(tA^{1/2})u_0+\sin(tA^{1/2})A^{-1/2}u_1. $$
Moreover, the real--valued function 
$t \mapsto (A^{1/2}u(t)|A^{1/2}u(t))_H+(u'(t)|u'(t))_H$
("the \emph{energy}") is constant for all $t \in \mathbb{R}$.
\end{theorem}

\begin{proof}
Let $u_0, u_1 \in \Dom{A}$.
Applying Lemma \ref{weq_lem_exp} it follows immediately that $u$ given by
$$ u(t)=\cos(tA^{1/2})u_0+\sin(tA^{1/2})A^{-1/2}u_1 $$
satisfies $u''(t)=-Au(t)$ for all $t \in \mathbb{R}$ and $u(0)=u_0$,
$u'(0)=u_1$.
Moreover, inserting for $u$ and using the functional calculus 
we directly compute
$$ (A^{1/2}u(t)|A^{1/2}u(t))_H+(u'(t)|u'(t))_H=\|A^{1/2}u_0\|_H^2+\|u_1\|_H^2 $$
which shows conservation of energy.

Let $\tilde{u}: \mathbb{R} \to H$ be another solution with initial data
$\tilde{u}(0)=u_0$ and $\tilde{u}'(0)=u_1$.
Then, $u-\tilde{u}$ is again a solution with zero initial data.
However, conservation of energy implies that
$\|A^{1/2}(u(t)-\tilde{u}(t))\|_H+\|u'(t)-\tilde{u}'(t)\|_H=0$.
Note that by definition we have
$\|u'(t)-\tilde{u}'(t)\|_H=\frac{d}{dt}\|u(t)-\tilde{u}(t)\|_H$ and hence,
$$ 0=\int_0^t \frac{d}{ds}\|u(s)-\tilde{u}(s)\|_H
ds=\|u(t)-u_0-\tilde{u}(t)+u_0\|_H $$
which shows $u(t)=\tilde{u}(t)$ for all $t \in \mathbb{R}$.
\end{proof}

\begin{remark}
Note that the solution operators $\cos(tA^{1/2})$ and $\sin(tA^{1/2})A^{-1/2}$ 
are
bounded and hence, they can be applied to general functions $u_0, u_1 \in H$ and
not only to $u_0, u_1 \in \Dom{A}$.
Analogous to semigroup theory this leads to the notion of generalized solutions
of abstract wave equations.
\end{remark}

\subsection{Well--Posedness for Nonnegative Generators}
The condition $(Au|u)_H \geq \gamma(u|u)_H$ for a $\gamma>0$ assumed in the
previous section was necessary to assure existence of the operator $A^{-1/2}$.
However, the operator $A^{-1/2}$ appears only as
$\sin(tA^{1/2})A^{-1/2}$ in the solution formula.
Thus, the question is whether $\sin(tA^{1/2})A^{-1/2}$ can be
defined reasonably even if $A^{-1/2}$ does not exist.
This problem is analogous to considering the real--valued function 
$f: (0,\infty) \to
\mathbb{R}$ defined by 
$f(x):=\frac{\sin \sqrt{x}}{\sqrt{x}}$.
Since $\lim_{x \to 0}f(x)=1$ by de l'Hospital, $f$ can be
continuously extended to $[0,\infty)$.
Thanks to the functional calculus it is exactly this construction which can be
used for the operator $\sin(tA^{1/2})A^{-1/2}$ as well.

Let $A: \Dom{A} \subset H \to H$ be a self--adjoint operator on a Hilbert space
$H$ which satisfies $(Au|u)_H \geq 0$ for all $u \in \Dom{A}$.
Then, the self--adjoint square root $A^{1/2}$ exists and it satisfies
$(A^{1/2}u|u)_H \geq 0$ for all $u \in \Dom{A^{1/2}}$ (Theorem
\ref{growth_thm_sqrt}).
For $t \in \mathbb{R}$ we define $f_t: \mathbb{R} \to \mathbb{R}$ by
$$ f_t(\lambda):=\left \{ \begin{array}{l}
\sin (t \lambda) \lambda^{-1} \mbox{ for }\lambda > 0 \\
t \mbox{ for } \lambda \leq 0 
\end{array} \right . $$ 
Then, $f_t$ is continuous and hence, $f_t(A^{1/2})$ is well--defined 
via the functional calculus.
Note that the inequality $(A^{1/2}u|u)_H \geq 0$ for all $u \in \Dom{A^{1/2}}$
implies $\sigma(A^{1/2}) \subset [0,\infty)$ (cf. Lemma \ref{cauchy_lem_specA}) 
and hence, the distribution
function $\lambda \mapsto (E(\lambda)u|u)_H$ for $u \in H$ is constant on
$(-\infty,0)$ where $\{E(\lambda)\}$ is the spectral family associated to $A^{1/2}$.
Thus, the values of $f_t$ on $(-\infty,0)$ do not contribute.

\begin{lemma}
\label{weq_lem_ft}
The function $u: \mathbb{R} \to H$ defined by $u(t):=f_t(A^{1/2})u_0$ for a given
$u_0 \in \Dom{A}$ satisfies $u''(t)=-Au(t)$ for all $t \in
\mathbb{R}$.
\end{lemma}

\begin{proof}
Let $\{E(\lambda): \lambda \in \mathbb{R}\}$ be the spectral family of 
$A^{1/2}$ (spectral theorem) and $d\mu_u$ the spectral measure associated to
$\{E(\lambda)\}$ defined in sec. \ref{specfam_sec}.
Observe that $\Dom{f_t(A^{1/2})}=H$ since $f_t$ is bounded and therefore, the
operator $f_t(A^{1/2}): H \to H$ is bounded for any $t \in \mathbb{R}$.
Applying the functional calculus we obtain
\begin{multline*}
\lim_{h \to 0} 
\|h^{-1}(f_{t+h}(A^{1/2})u_0-f_t(A^{1/2})u_0)-\cos(tA^{1/2})u_0\|_H^2 \\
=\lim_{h \to 0}
\int_0^\infty \left | h^{-1}(f_{t+h}(\lambda)-f_t(\lambda))-\cos(t
\lambda) \right |^2 d\mu_{u_0}(\lambda)=0 
\end{multline*}
for any $u_0 \in H$ by Lebesgue's theorem on dominated convergence 
and therefore, 
$u'(t)=\cos(tA^{1/2})u_0$.
Analogously, we have 
\begin{multline*}
\|u''(t)+Au(t)\|^2_H \\
=\lim_{h \to 0} \int_0^\infty \left |
h^{-1}(\cos((t+h)\lambda)-\cos(t\lambda))+
\lambda^2 f_t(\lambda) \right |^2 d\mu_{u_0}=0 
\end{multline*}
for any $u_0 \in \Dom{A}$ which is the claim.
\end{proof}

\begin{remark}
In particular it follows that $f_t(A^{1/2})u_0 \in \Dom{A}$ for all $t \in
\mathbb{R}$ if $u_0 \in \Dom{A}$.
\end{remark}

Now we are able to prove the well--posedness result for nonnegative generators.

\begin{theorem}
\label{weq_thm_gen2}
Let $A: \Dom{A} \subset H \to H$ be a self--adjoint operator on a Hilbert space
$H$ satisfying $(Au|u)_H \geq 0$ for all $u \in \Dom{A}$.
Then, for given $u_0, u_1 \in \Dom{A}$ there exists a unique function $u:
\mathbb{R} \to \Dom{A} \subset H$ such that $u''(t)=-Au(t)$ for all $t \in
\mathbb{R}$ and $u(0)=u_0$, $u'(0)=u_1$.
This unique $u$ can be given explicitly as
$$ u(t)=\cos(tA^{1/2})u_0+f_t(A^{1/2})u_1 $$
where $f_t: \mathbb{R} \to \mathbb{R}$ is defined by
$$ f_t(\lambda):=\left \{ \begin{array}{l} 
\sin(t \lambda)\lambda^{-1} \mbox{ for }\lambda > 0 \\
t \mbox{ for } \lambda \leq 0 \end{array} \right. $$
Furthermore, the \emph{energy} 
$$ t \mapsto (A^{1/2}u(t)|A^{1/2}u(t))_H+(u'(t)|u'(t))_H $$
is constant.
\end{theorem}

\begin{proof}
Let $u_0,u_1 \in \Dom{A}$. Applying Lemmas \ref{weq_lem_ft} and 
\ref{weq_lem_exp} we
conclude that $u$, given by $u(t):=\cos(tA^{1/2})u_0+f_t(A^{1/2})u_1$, satisfies
$u''(t)=-Au(t)$ for all $t \in \mathbb{R}$ and $u(0)=u_0$, $u'(0)=u_1$.
By inserting for $u(t)$ and direct computation (functional calculus) we obtain
$$ (A^{1/2}u(t)|A^{1/2}u(t))_H+(u'(t)|u'(t))_H=
(A^{1/2}u_0|A^{1/2}u_0)_H+(u_1|u_1)_H. $$
Uniqueness can be obtained analogously to Theorem \ref{weq_thm_gen}.
\end{proof}

\begin{remark}
Again, we also have generalized solutions since the solution operators
$\cos(tA^{1/2})$ and $f_t(A^{1/2})$ are bounded and hence, they can be applied
to any element in $H$.
\end{remark}

\subsection{Application to the Linearized Wave Map Problem}
Now we return to the equation
\begin{equation}
\label{weq_eq_phitilde}
\tilde{\phi}_{\sigma \sigma}-(1-\rho^2)^2
\tilde{\phi}_{\rho
\rho}-\frac{2(1-\rho^2)^2}{\rho}\tilde{\phi}_\rho+\frac{2(1-\rho^2)\cos(2
f_0)-\rho^2}{\rho^2}\tilde{\phi}=0 
\end{equation}
which describes the (rescaled) linearized flow around the Turok Spergel
solution in hyperbolic coordinates (cf. sec. \ref{wp_sec}).

\paragraph{Operator formulation}
As before we define $H:=L^2_w(0,1)$ for $w(\rho)=\frac{\rho^2}{(1-\rho^2)^2}$
and $A_0: \Dom{A_0} \subset H \to H$ denotes the
operator constructed in sec. \ref{opa_sec}.
Hence, the equation $u''(t)=-A_0u(t)$ for a function $u: \mathbb{R} \to H$ is an
operator formulation of eq. (\ref{weq_eq_phitilde}). 

\paragraph{Well--posedness}
According to sec. \ref{opa_sec}, the operator $A_0$ is self--adjoint and 
from Theorem
\ref{speca_thm_speca} we know that $\sigma(A_0)=[0,\infty)$.
Thus, $A_0$ satisfies $(A_0u|u)_H \geq 0$ for all $u \in \Dom{A_0}$ (Lemma 
\ref{propsa_prop_semibound}).
Applying Theorem \ref{weq_thm_gen2} we infer that the Cauchy problem
\begin{equation}
\label{weq_eq_opphitilde}
\left \{ \begin{array}{l} u''(\sigma)=-A_0u(\sigma) \mbox{ for } \sigma>0 \\
u(0)=u_0, u'(0)=u_1 
\end{array} \right. 
\end{equation}
for given $u_0,u_1 \in \Dom{A}$ and a function $u: [0,\infty) \to H$ is
well--posed.

Moreover, Theorem \ref{weq_thm_gen2} tells us that for 
any classical solution of eq. (\ref{weq_eq_opphitilde}) the energy $\sigma
\mapsto \|A_0^{1/2}u(\sigma)\|_H^2+\|u'(\sigma)\|_H^2$ is conserved.

\paragraph{Nonexistence of growing solutions}
As discussed previously (sec. \ref{growth_sec}), the mapping 
$u \mapsto \|A_0^{1/2}u\|_H$ is a norm on $\Dom{A_0}$
and hence, conservation of energy implies the nonexistence of growing solutions
with respect to $u \mapsto \|A_0^{1/2}u\|_H$.
Thus, there are no solutions of the equation
\begin{equation*}
\phi_{\sigma \sigma}-2 \phi_\sigma -(1-\rho^2)^2 \phi_{\rho \rho}-\frac{2
(1-\rho^2)^2}{\rho}\phi_\rho +\frac{2(1-\rho^2)\cos(2 f_0)}{\rho^2}\phi=0
\end{equation*}
that grow faster than the gauge instability which behaves as $e^{\sigma}$ for
$\sigma \to \infty$.
This shows that the Turok Spergel solution is as stable as it can be in the
hyperbolic coordinates.

\paragraph{Discussion}
We conclude that the functional calculus yields essentially the same result 
as the
semigroup approach but it is more explicit.
In either case the main effort lies in determining the spectrum of the operator
$A_0$. 

Thus, we have exploited the self--adjoint approach (hyperbolic coordinates)
to the linear stability problem and have obtained the best possible result.
To gain further insight one has to change coordinates.
However, in a different coordinate system the involved operators become much
less convenient since they are not self--adjoint anymore.

\chapter{The Spectra of $A_n$}
\thispagestyle{empty}
\label{specln_ch}

This chapter is devoted to the study of the spectra of the operators 
$A_n$ for $n \geq 1$.
We show that the operator $A_n$ has exactly $n$ negative eigenvalues and give a
rough lower bound for the smallest eigenvalue.
Furthermore, we investigate the spectral behaviour of $A_n$ for $n \to \infty$.  

Througout this chapter we adopt the previously used notation, i.e. 
$H:=L^2_w(0,1)$ with
$w(\rho):=\frac{\rho^2}{(1-\rho^2)^2}$, $A: \Dom{A} \subset H \to H$ 
is the self--adjoint operator 
generated by the Sturm--Liouville differential expression $a$ given by
$au:=\frac{1}{w}(-(pu')'+qu)$ where $p(\rho):=\rho^2$ and $q(\rho) \equiv 2$.
Moreover, the operator $A_n: \Dom{A_n} \subset H \to H$ is defined by
$\Dom{A_n}:=\Dom{A}$ and
$$ A_nu:=Au+g_nu $$
for $u \in \Dom{A_n}$ and 
$$ g_n(\rho):=\frac{2(1-\rho^2)\cos(2f_n(\rho))-
\rho^2-2(1-\rho^2)^2}{\rho^2} $$
where $f_n$ denotes the $n$--th self--similar wave map. 

\section{The Spectrum}

\subsection{The Continuous Spectrum}

As already mentioned (Theorem \ref{selfsimilar_thm_bizon}), Bizo\'n has shown
existence of the smooth self--similar wave maps $f_n$.
However, in fact 
the solutions $f_n$ are not only smooth but even analytic on $[0,1]$,
i.e. they can be expanded in a convergent power series around any point 
$\rho \in [0,1]$ (cf. \cite{Bizon2002}).
Furthermore, they satisfy $f_n(0)=0$ and $f_n(1)=\frac{\pi}{2}$.
Hence, all the results of sec. \ref{speca_sec} which do not depend on the
explicit form of $f_0$ but solely on the asymptotic behaviour for $\rho \to 0$
and $\rho \to 1$ carry over to $A_n$ without change.
In particular, the whole Frobenius analysis is equally valid for $A_n$.
Thus, we have the following proposition.

\begin{proposition}
The continuous spectrum $\sigma_c(A_n)$ of $A_n$ for any 
$n=0,1,2,\dots$ is given by
$\sigma_c(A_n)=[0,\infty)$.
\end{proposition} 

\subsection{The Point Spectrum}
As in the case $n=0$, the gauge instability (cf. sec. \ref{wp_sec}) is present.
Thus, the function $\theta_n$ defined by $\theta_n(\rho):=\rho
\sqrt{1-\rho^2}f_n'(\rho)$ is a formal solution of 
$A_n u=0$.
More explicitly, $(\lambda-A_n)u=0$ reads
\begin{equation}
\label{spAn_eq_ev}
u''+\frac{2}{\rho}u'+\frac{(1+\lambda)\rho^2-2(1-\rho^2)
\cos(2f_n)}{\rho^2(1-\rho^2)^2} u=0
\end{equation}
According to theorem \ref{selfsimilar_thm_bizon}, the self--similar wave map 
$f_n$ has exactly $n$ intersections with the line $\frac{\pi}{2}$ on $[0,1)$.
Thus, since $f_n(1)=\frac{\pi}{2}$, it follows that $\theta_n$ has exactly $n$
zeros on $(0,1)$.
Hence, an oscillation argument 
similar to Lemma \ref{speca_lem_Sturm} implies
that there are exactly $n+1$ numbers 
$0 =\lambda_0 > \lambda_1 > \dots > \lambda_n$ such that eq.
(\ref{spAn_eq_ev}) with $\lambda=\lambda_j$ has a nontrivial solution $u_j$
satisfying $u_j(0)=u_j(1)=0$ ($j=0,1,2 \dots, n$).
According to the Frobenius analysis in sec. \ref{speca_sec}, $u_j$
has the asymptotic behaviour $u_j(\rho) \sim \rho$ for $\rho \to 0$ and 
$u_j(\rho) \sim (1-\rho)^{\alpha_j}$ for $\rho \to 1$ where
$\alpha_j:=\frac{1+\sqrt{-\lambda_j}}{2}$.
Thus, we immediately observe that 
$u_j \in \Dom{A_n}$ for
$j>0$ and hence, $u_j$ is an eigenfunction of $A_n$ if $j>0$.
This shows that the operator $A_n$ has exactly $n$ negative eigenvalues
$\lambda_1, \lambda_2, \dots, \lambda_n$.
Analogous to sec. \ref{speca_sec} it follows that $\lambda-A$ is invertible for
negative $\lambda$ which are not in the point spectrum and hence, we arrive at
the following theorem.

\begin{theorem}
\label{spAn_thm_specAn}
The spectrum of the operator $A_n$ is given by 
$$\sigma_p(A_n)=\{\lambda_1,
\dots, \lambda_n \in \mathbb{R}: 0>\lambda_1>\lambda_2 > \dots > \lambda_n\},$$
$$ \sigma_c(A_n)=[0,\infty) \mbox{ and } \sigma_r(A_n)=\emptyset $$
for $n=0,1,2,\dots$.
\end{theorem}

\subsection{A Simple Estimate}
We apply an elementary argument to obtain a rough lower bound for the smallest
eigenvalue of $A_n$.

\begin{lemma}
The smallest eigenvalue $\lambda_n$ of $A_n$ satisfies the estimate 
$$\lambda_n \geq \inf_{\rho \in (0,1)} 
\left \{\frac{2(1-\rho^2)\cos(2f_n(\rho))}{\rho^2}-1 \right \} > -\infty $$
for any $n \in \mathbb{N}$.
\end{lemma} 

\begin{proof}
First of all we show that the infimum exists.
We abbreviate 
$$ h_n(\rho):=\frac{2(1-\rho^2)\cos(2f_n(\rho))}{\rho^2}-1. $$
Since $f_n(0)=0$ for all $n \in \mathbb{N}$ it follows that
$\lim_{\rho \to 0+}h_n(\rho)=\infty$ and $\lim_{\rho \to 1-}h_n(\rho)=-1$.
However, $h_n \in C(0,1)$ and therefore it must have a minimum on $[0,1]$.

Now suppose $u \in \Dom{A_n}$ is an eigenfunction of $A_n$ with eigenvalue
$\lambda<\inf_{\rho \in (0,1)} h_n(\rho)$.
By Frobenius it follows that $u(0)=u(1)=0$ and $u'(0) \not=0$.
Without loss of generality we assume $u'(0)>0$.
In order to satisfy the boundary condition $u(1)=0$, $u$ has to have a maximum.
Let the first maximum be located at $\rho_0 \in (0,1)$.
Thus, we have $u(\rho_0)>0$, $u'(\rho_0)=0$ and $u''(\rho_0)\leq 0$.
Inserting in eq. (\ref{spAn_eq_ev}) we obtain
$$ u''(\rho_0)=-\frac{\lambda-h_n(\rho_0)}{(1-\rho_0^2)^2}u(\rho_0)>0 $$
which is a contradiction.
\end{proof}

\section{Numerics}
We intend to numerically calculate the point spectrum of the operator $A_n$.
To this end it is necessary to construct the self--similar wave map $f_n$.
As already mentioned, this has first been done in \cite{Aminneborg1995} with a
shooting and matching procedure and we will reproduce these results.
The point spectrum of $A_n$ can be obtained by the same technique and this
has been done in \cite{Bizon2000}.

\subsection{Construction of Self--Similar Wave Maps}
\begin{figure}[h]
\centering
\includegraphics[totalheight=7cm,angle=-90]{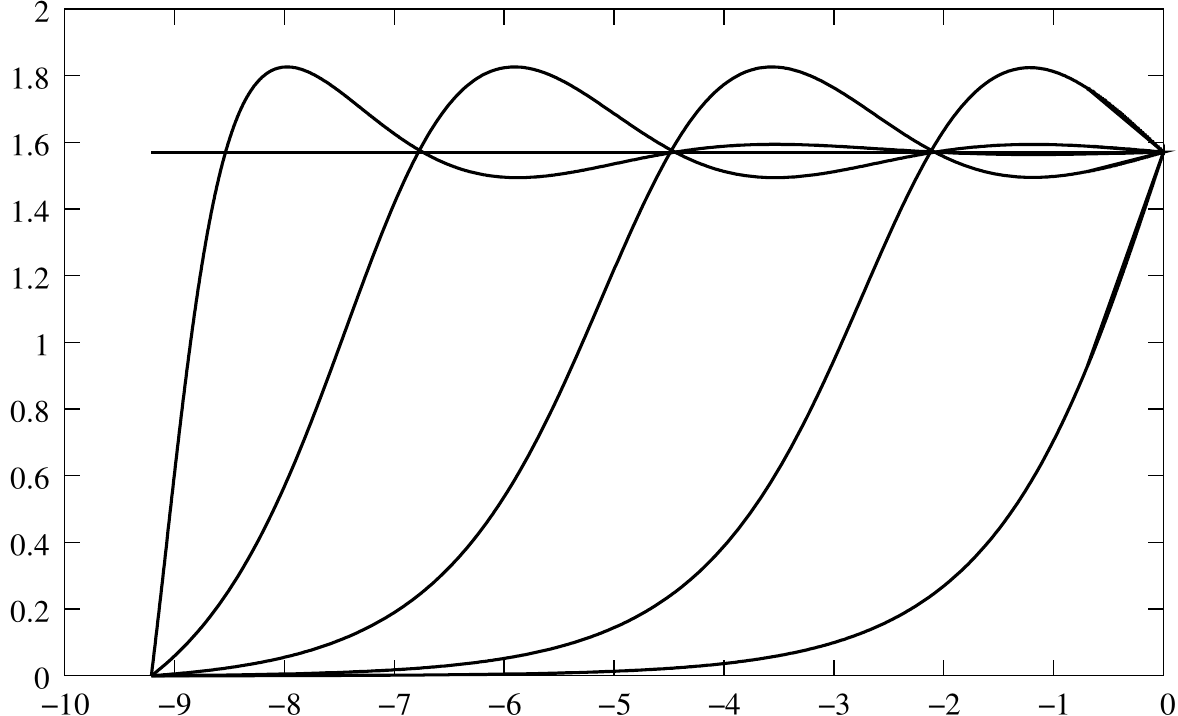}
\caption{The first five self--similar profiles calculated with $\alpha=10^{-4}$.}
\label{num_fig_css}
\end{figure}

We construct the self--similar solutions $f_n$ with a standard shooting and
matching technique.
It turns out that the derivative $f_n'(0)$ increases very quickly with $n$
becoming larger and hence, it is advantageous to use a logarithmic coordinate.
Thus, we define $x:=\log(\alpha+\rho)$ for a small $\alpha>0$.
Eq. (\ref{selfsimilar_eq_css}) transforms into
\begin{equation}
\label{num_eq_csslog}
f''+\frac{e^x+\alpha}{e^x-\alpha}f'-\frac{e^{2x}\sin(2f)}{(1-(e^x-\alpha)^2)
(e^x-\alpha)^2}=0
\end{equation}
where $x \in [\log \alpha, \log (\alpha+1)]$.
We have the regularity conditions $f(\log \alpha)=0$ and
$f(\log(\alpha+1))=\frac{\pi}{2}$.
For integrating eq. (\ref{num_eq_csslog}) we use the ODE solver provided by the
GNU Scientific Library \cite{Gough2003}.
Fig. \ref{num_fig_css} shows the first five solutions calculated with
$\alpha=10^{-4}$.

\subsection{Calculation of the Point Spectrum}
We numerically calculate solutions of the eigenvalue equation $(\lambda-A_n)u=0$
which is given explicitly by
\begin{equation}
\label{num_eq_eveq}
u''+\frac{2}{\rho}u'+\frac{(1+\lambda)\rho^2-2(1-\rho^2) \cos(2
f_n)}{\rho^2(1-\rho^2)^2}u=0
\end{equation}
for $\rho \in [0,1]$.
We have $\sigma_p(A_n) \subset (-\infty,0)$ by Theorem \ref{spAn_thm_specAn} and
hence, we restrict ourselves to $\lambda < 0$.
The boundary conditions at the singular points $\rho=0$ and $\rho=1$ are
dictated by the requirement $u \in \Dom{A_n}$.
According to the Frobenius analysis of sec. \ref{speca_sec}, $u \in \Dom{A_n}$
implies
$u(\rho) \sim \rho$ for $\rho \to 0$ and $u(\rho) \sim (1-\rho)^\beta$ for
$\rho \to 1$ where
$\beta=\frac{1+\sqrt{-\lambda}}{2}$.
However, since $\beta$ increases as $\lambda$ decreases we encounter a technical
diffculty:
In order to "shoot away" from the singular point $\rho=1$ we have to calculate
more and more derivatives as $\lambda$ decreases.
To go around this problem we define a new unknown 
$v(\rho):=(1-\rho)^{-\beta}u(\rho)$.
It follows that $v(\rho) \sim 1$ as $\rho \to 1$ if 
$u \in \Dom{A_n}$ and hence, $v$ is
analytic if $u \in \Dom{A_n}$.
Moreover, the "bad" solution around $\rho=1$ behaves as 
$(1-\rho)^{-\sqrt{-\lambda}}$ and thus,
it can easily be distinguished numerically from the "good" one since it is 
singular at $\rho=1$.
This is essential in order to obtain a well--behaved numerical approximation.
Eq. (\ref{num_eq_eveq}) transforms into the generalized eigenvalue problem
\begin{equation}
\label{num_eq_eveqv}
v''+\left (\frac{2}{\rho}-\frac{2 (1+\mu)}{1-\rho^2} \right )v'-\left ( 
\frac{2\cos(2f_n)}{\rho^2(1-\rho^2)}+\frac{(1+\mu)(2+\mu)}{1-\rho^2} \right
)v=0
\end{equation}
where we have set $\mu:=\sqrt{-\lambda}$.

We solve the eigenvalue problem with the
shooting and matching method similar to the construction of self--similar wave
maps in the previous section: 
Let $v_l$ and $v_r$ denote the solutions obtained 
by integrating eq. (\ref{num_eq_eveqv}) from $\rho=0$ to $\rho=\frac{1}{2}$ and
$\rho=1$ to $\rho=\frac{1}{2}$, respectively.
As a matching condition we require the Wronskian $W(v_l,v_r)(\frac{1}{2})$ to
vanish.
This has the advantage that we can fix $v_l'(0)=1$ as well as $v_r(1)=1$ and 
the only free parameter is the eigenvalue $\mu$.
Hence, eigenvalues can be found by a simple bisection and one is not forced to
use a multidimensional root finder which can be tricky.  
The numerical results are summarized in Table \ref{num_tbl_evmu}.

\begin{table}
\begin{center}
\begin{tabular}{|c|c|c|c|c|}
\hline
 $\mu_n$ & $f_1$ & $f_2$ & $f_3$ & $f_4$ \\
 \hline 
 $\mu_1$ & 5.333625 & 5.304 & 5.30 & 5.3 \\ 
 $\mu_2$ & & 58.0701 & 57.68 & 57.6 \\
 $\mu_3$ & & & 625 & 620 \\
\hline
\end{tabular}
\end{center}
\caption{Approximate eigenvalues}
\label{num_tbl_evmu}
\end{table}

We avoid a detailed error analysis since we do not need these numbers in the
sequel but merely note the interesting observation that the horizontal rows seem
to converge.
This numerical convergence raises the question whether there exists a 
relation like
"$\lim_{n \to \infty}\sigma(A_n)=\sigma(A_\infty)$" where $A_\infty$ is the
perturbation operator around the limiting solution $f_\infty \equiv
\frac{\pi}{2}$ (cf. sec. \ref{selfsimilar_sec}).
We will discuss this issue in the following section.

\section{The Operator $A_\infty$}

We study the operator $A_\infty$ which describes linear perturbations around the
limiting solution $f_\infty$ (cf. sec. \ref{selfsimilar_sec}).

The (rescaled) flow of linear perturbations around the limiting solution 
$f_\infty=\frac{\pi}{2}$ is
governed by eq. (\ref{wp_eq_linwmtilde}) with $f_n$ substituted by the constant
$\frac{\pi}{2}$, i.e.
\begin{equation}
\label{Ainf_eq_linpert}
\tilde{\phi}_{\sigma \sigma}-(1-\rho^2)^2
\tilde{\phi}_{\rho
\rho}-\frac{2(1-\rho^2)^2}{\rho}\tilde{\phi}_\rho
+\frac{-2+\rho^2}{\rho^2} \tilde{\phi}=0.
\end{equation}
Thus, the operator $A_\infty$ is generated by the formal differential 
expression $a_\infty$ defined by
\begin{equation} 
\label{Ainf_eq_ainf}
a_\infty:=\frac{1}{w}\left ( -\frac{d}{d\rho}p \frac{d}{d\rho}+q \right ).
\end{equation}
where
$w(\rho):=\frac{\rho^2}{(1-\rho^2)^2}$, $p(\rho):=\rho^2$ and 
$q(\rho):=\frac{-2+\rho^2}{(1-\rho^2)^2}$.
Thus, we define the underlying
Hilbert space $H$ as $H:=L^2_w(0,1)$
and apply the method of Frobenius in order to obtain
asymptotic estimates for solutions of $a_\infty u=0$.
Around $\rho=1$ the indices of the Frobenius analysis are
both equal to $\frac{1}{2}$ and thus, the
situation is completely analogous to $A_n$ and $\rho=1$ is in the limit--point
case.
However, around $\rho=0$ the indices are $-\frac{1 \pm i\sqrt{7}}{2}$
which means that all nontrivial solutions of $a_\infty u=0$
belong to $H$ near $\rho=0$.
Thus, $\rho=0$ is in the limit--circle case and in order to define a 
self--adjoint operator we have to specify a boundary condition at $\rho=0$.
But which boundary condition is the "correct" one?
Bizo\'n \cite{Bizon2002a} has studied the analogous problem for Yang--Mills
equations and he proposed to choose the boundary condition in such a way that
there exists a nontrivial solution of $a_\infty u=0$ similar to the gauge 
instability for $A_n$. 

\subsection{The Boundary Condition}
For $\lambda \leq 0$ let $u(\cdot, \lambda)$ be the solution of 
$(\lambda-a_\infty)u(\cdot,\lambda)=0$ with asymptotic behaviour
$u(\rho,\lambda) \sim (1-\rho)^\alpha$ for $\rho \to 1$ where
$\alpha=\frac{1+\sqrt{-\lambda}}{2}$ (which exists by Frobenius' method).
We are interested in the asymptotic behaviour of $u(\cdot, \lambda)$ for $\rho
\to 0$.
According to Frobenius' method we have 
$$ u(\rho,\lambda) \propto \rho^{-1/2} \left
(\rho^{i\sqrt{7}/2}\widetilde{\varphi}(\rho,\lambda)
+m(\lambda)\rho^{-i\sqrt{7}/2}
\widetilde{\psi}(\rho,\lambda) \right )$$
where $\widetilde{\varphi}(\cdot,\lambda)$ as well as 
$\widetilde{\psi}(\cdot, \lambda)$ are
holomorphic around $\rho=0$ and satisfy
$\widetilde{\varphi}(0,\lambda)=\widetilde{\psi}(0,\lambda)=1$.
The complex number $m(\lambda)$ is called the \emph{connection coefficient}.
The asymptotic behaviour of $u(\cdot,\lambda)$ can be written in a more
convenient way.

\begin{lemma}
\label{Ainf_lem_usin}
For $\lambda \leq 0$ let $u(\cdot,\lambda)$ be a real solution of
$(\lambda-a_\infty)u=0$ with asymptotic behaviour $u(\rho,\lambda) \sim
(1-\rho)^\alpha$ for $\rho \to 1$ where $\alpha=\frac{1+\sqrt{-\lambda}}{2}$.
Then, $u(\cdot,\lambda)$ can be written as
$$ u(\rho,\lambda)=c \rho^{-1/2} \left ( \sin \left ( \frac{\sqrt{7}}{2} \log \rho+\delta(\lambda)
\right ) + h(\rho,\lambda) \right )$$
where $c$ is a real constant, $|\delta(\lambda)| \leq \frac{\pi}{2}$, 
$h(\cdot,\lambda) \in C^1(0,1)$ and 
$$ \lim_{\rho \to 0}h(\rho,\lambda)=\lim_{\rho \to 0}\rho
h'(\rho,\lambda)=0. $$
\end{lemma}

\begin{proof}
As already mentioned above, we have
$$ u(\rho, \lambda) \propto \rho^{-1/2} \left
(\rho^{i\sqrt{7}/2}\widetilde{\varphi}(\rho,\lambda)
+m(\lambda)\rho^{-i\sqrt{7}/2}
\widetilde{\psi}(\rho,\lambda) \right ). $$ 
Since $u(\cdot,\lambda)$ is real by assumption, we have $u(\cdot,\lambda)=\Re
u(\cdot, \lambda)$ and hence, we infer
$$ \rho^{1/2} u(\rho, \lambda)=a(\lambda)\cos 
\left( \frac{\sqrt{7}}{2}\log \rho \right )+b(\lambda) \sin \left ( 
\frac{\sqrt{7}}{2}
\log \rho \right ) 
+h(\rho,\lambda)
$$
where $a(\lambda)$ and $b(\lambda)$ are real--valued and defined by
$m(\lambda)$.
Moreover, $h(\cdot,\lambda) \in C^1(0,1)$, $\lim_{\rho \to 0}h(\rho,\lambda)=0$ 
and $\lim_{\rho \to 0}\rho h'(\rho,\lambda)=0$ which follows immediately from
the properties of $\widetilde{\varphi}(\cdot,\lambda)$ and
$\widetilde{\psi}(\cdot, \lambda)$.
Thus, the identity $a \cos x + b \sin x = \sqrt{a^2+b^2} \sin(x+d)$ where
$d=\arctan{\frac{b}{a}}$ finishes the proof.
\end{proof}

According to Bizo\'n's proposal, eigenfunctions should satisfy
$\delta(\lambda)=\delta(0)$.
The following lemma shows how this requirement can be translated into a boundary
condition.

\begin{lemma}
For $\lambda \leq 0$ let $u(\cdot, \lambda)$ be a real solution of 
$(\lambda-a_\infty)u=0$ with asymptotic behaviour $u(\rho,\lambda) \sim
(1-\rho)^\alpha$ where $\alpha=\frac{1+\sqrt{-\lambda}}{2}$.
Assume further that $|\delta(0)|<\frac{\pi}{2}$ where $\delta$ is defined in
Lemma \ref{Ainf_lem_usin}.
Then, $\delta(\lambda)=\delta(0)$ is equivalent to
$$ \lim_{\rho \to 0}\rho^2 \left (
u(\rho,\lambda)u'(\rho,0)-u'(\rho,\lambda)u(\rho,0) \right )=0 $$
where $':=\frac{d}{d\rho}$.
\end{lemma}

\begin{proof}
Applying Lemma \ref{Ainf_lem_usin} and using the addition theorem for
trigonometric functions we readily obtain 
$$ \lim_{\rho \to 0}\rho^2
(u(\rho,\lambda)u'(\rho,0)-u'(\rho,\lambda)u(\rho,0))=
\sin(\delta(\lambda)-\delta(0)) $$
which yields the claim since $|\delta(\lambda)-\delta(0)| < \pi$.
\end{proof}

\subsection{Construction of the Operator}
We construct the self--adjoint operator $A_\infty$ by imposing the boundary
condition discussed in the previous section.

To this end recall the definition of the maximal operator 
$A_\mathrm{max}: \Dom{A_\mathrm{max}} \subset H \to H$ generated by $a_\infty$ which is given by 
$$\Dom{A_\mathrm{max}}:=\{u \in H: u,pu' \in
AC_\mathrm{loc}(0,1), a_\infty u \in H\}$$
and $A_\mathrm{max}u:=a_\infty u$ for $u \in \Dom{A_\mathrm{max}}$.

We denote
by $\tilde{\chi}$ a real--valued nontrivial function satisfying 
$a_\infty \tilde{\chi}=0$ with asymptotic
behaviour $\tilde{\chi}(\rho) \sim (1-\rho)^{1/2}$ for $\rho \to 1$.
Such a function exists by Frobenius' method and it is unique up to 
constant multiples
since the other linearly independent solution of $a_\infty u=0$ around 
$\rho=1$ contains 
a logarithmic term.
Note, however, that $\tilde{\chi} \notin H$ due to its asymptotic behaviour for
$\rho \to 1$.
Using a smooth cut--off function we construct a $\chi \in \Dom{A_\mathrm{max}}$ satisfying
$\chi(\rho)=\tilde{\chi}(\rho)$ for $\rho \in (0,\frac{1}{2})$.
Obviously, there exists a $v \in \Dom{A_\mathrm{max}}$ with $[\chi,v]_p(0)
\not=0$.
We define the operator $A_\infty$ by
$$
\Dom{A_\infty}:=\{u \in \Dom{A_\mathrm{max}}: [u,\chi]_p(0)=0\}
$$
and $A_\infty u:=a_\infty u$ for $u \in \Dom{A_\infty}$.
Invoking Lemma \ref{opth_lem_Aselfadj} we immediately infer that $A_\infty$ is
self--adjoint.

\subsection{Calculation of the Point Spectrum}
Consider the formal eigenvalue equation $(\lambda-a_\infty)u=0$.
We make a coordinate transformation $\rho \mapsto z:=\rho^2$
and define a new unknown 
$$v(z):=z^\alpha(1-z)^\beta u(\sqrt{z})$$ 
where
$\alpha:=-\frac{1}{4}(-1+i\sqrt{7})$ and 
$\beta:=-\frac{1}{2}(1+\sqrt{-\lambda})$.
The equation $(\lambda-a_\infty)u=0$ transforms into
\begin{equation}
\label{Ainf_eq_hypgeom}
z(1-z) v''+ [c-(a+b+1)z]v'-ab v=0
\end{equation}
where $a:=\frac{1}{4}(1+2\sqrt{-\lambda}+i\sqrt{7})$, 
$b:=\frac{1}{4}(3+2\sqrt{-\lambda}+i\sqrt{7})$ and 
$c:=1+\frac{i\sqrt{7}}{2}$.
Eq. (\ref{Ainf_eq_hypgeom}) is the \emph{hypergeometric differential equation}
(cf. \cite{Abramowitz1964}).
Around $z=1$ there exist two linearly independent solutions 
$v_1(\cdot, \lambda)$ and
$\tilde{v}_1(\cdot, \lambda)$ given by
$v_1(z,\lambda):={}_2F_1(a, b; a+b+1-c; 1-z)$ and  
$\tilde{v}_1(z,\lambda):=(1-z)^{c-a-b}
{}_2F_1(c-a, c-b; c+1-a-b; 1-z)$ where 
$_2F_1$ denotes
the \emph{hypergeometric function} (cf. \cite{Abramowitz1964}).
We are interested in eigenfunctions and hence, the condition $u \in H$ rules out
the solution $\tilde{v}_1(\cdot, \lambda)$ since 
$_2F_1(c-a, c-b; c+1-a-b; 1-z) \sim 1$ for $z
\to 1$.
Around $z=0$ the two linearly independent solutions $v_0(\cdot, \lambda)$ and 
$\tilde{v}_0(\cdot, \lambda)$
are given by $v_0(z,\lambda)={}_2F_1(a,b; c; z)$ and
$\tilde{v}_0(z,\lambda)
=z^{1-c}{}_2F_1(a+1-c,b+1-c; 2-c; z)$.
By the general theory of linear second order ordinary differential equations we
infer that for $z \in (0,1)$ we can write 
$$ v_1(z,\lambda)=v_0(z,\lambda)+m(\lambda)\tilde{v}_0(z,\lambda). $$
The connection coefficient $m(\lambda)$ can be given explicitly in terms
of the $\Gamma$--function \cite{Erdelyi1953} and reads
$$ m(\lambda)=\frac{\Gamma(a+1-c)\Gamma(b+1-c)\Gamma(c-1)}{\Gamma(a)
\Gamma(b)\Gamma(1-c)}. $$ 
The boundary condition specified in the definition of $\Dom{A_\infty}$ 
translates into the requirement $m(\lambda)=m(0)$.
This transcendental equation can be solved numerically using bisection and by
this we obtain the point spectrum $\sigma_p(A_\infty)$ of $A_\infty$.
The results are given in Table \ref{Ainf_tbl_ev} where we have set
$\mu:=\sqrt{-\lambda}$.
Furthermore, we have duplicated Table \ref{num_tbl_evmu} for comparison.

\begin{table}
\begin{center}
\begin{tabular}{|c|c|c|c|c|c|}
\hline
 $\mu_n$ & $f_1$ & $f_2$ & $f_3$ & $f_4$ & $f_\infty$ \\
 \hline 
 $\mu_1$ & 5.333625 & 5.304 & 5.30 & 5.3 & 5.3009 \\ 
 $\mu_2$ & & 58.0701 & 57.68 & 57.6 & 57.637 \\
 $\mu_3$ & & & 625 & 620 & 619.61 \\
\hline
\end{tabular}
\end{center}
\caption{Approximate eigenvalues}
\label{Ainf_tbl_ev}
\end{table}

Hence, the numerical results suggest that the point spectrum of $A_n$ converges to 
$\sigma_p(A_\infty)$ for $n \to \infty$.

\chapter{Further Results and Outlook}
\label{outlook_ch}

So far we have studied linear stability of the Turok Spergel solution $f_0$ in
hyperbolic coordinates.
We have identified two features which spoil the analysis to a certain degree.
First, the time translation symmetry of the original problem leads to an
exponentially growing solution of the perturbation equation.
This fact corresponds to an isolated point in the continuous spectrum of the
operator $A_0$.
However, one could easily go around this problem by defining a projection
operator which removes this point from the spectrum of $A_0$.
By this, one would immediately gain linear stability of $f_0$ as conjectured.
The reason why this does not work is the nature of the coordinate system which
induces an unbounded part in the continuous spectrum of $A_0$.
Thus, the only possibility is to introduce a new time coordinate $\tau$
different from $\sigma$.
However, this destroys the self--adjoint character of the problem which makes it
much more difficult.
To illustrate the problems one has to deal with we consider the (possibly simplest) choice $\tau:=-\log(T-t)$.
The resulting linarized equation around the self--similar solution $f_n$ reads
\begin{equation}
\label{outlook_eq_linwm}
\phi_{\tau \tau}-(1-\rho^2)\phi_{\rho \rho}+2 \rho \phi_{\tau
\rho}+\phi_\tau-\frac{2(1-\rho^2)}{\rho^2}\phi_\rho+\frac{2
\cos(2f_n)}{\rho^2}\phi=0.
\end{equation}
Thus, one obtains a mixed derivative since the coordinate lines are not
orthogonal anymore.
Such a term turns out to be very inconvenient.
We formally rewrite this evolution equation as a first order system of the form
$$ \frac{d}{d\tau}\mathbf{u}(\tau)=L_n \mathbf{u}(\tau) $$
for a matrix differential operator $L_n$. 
Now we consider the eigenvalue equation $(\lambda-L_0)\mathbf{u}=0$
which reduces to
\begin{equation}
\label{outlook_eq_genev}
 u''+\left ( \frac{2}{\rho}-\frac{2\lambda \rho}{1-\rho^2} \right )u'- \left (
\frac{2\cos(2f_0)}{\rho^2(1-\rho^2)}+\frac{\lambda(1+\lambda)}{1-\rho^2} \right
)u=0. 
\end{equation}
Since $f_0(\rho)=\frac{1-4 \rho^2+\rho^4}{(1+\rho^2)^2}$ this equation has six
regular singular points $\rho=0,\pm 1, \pm i, \infty$.
Using the transformation $\rho \mapsto z:=\rho^2$ one is left with
the four singularities $z=-1,0,1,\infty$.
Thus, solutions are given in terms of Heun's functions (cf.
\cite{Ronveaux1995}).
However, most of the knowledge concerning Heun's functions is based on numerical
techniques and thus, it is very difficult to obtain the "spectrum" of this
generalized eigenvalue problem.
Furthermore, it is by no means clear what boundary conditions one should
specify at the singular points $\rho=0$ and $\rho=1$.
In the self--adjoint formulation the choice of the function space is dictated 
by the
operator itself but in this non--self--adjoint setting we do not have such an
information.
Thus, as a first step one would have to find a function space such that $L_0$ is
properly defined and the
linear evolution problem eq. (\ref{outlook_eq_linwm}) is well--posed which is
probably not so easy.
Secondly, one has to analyse the spectrum of $L_0$, i.e. study eq.
(\ref{outlook_eq_genev}) with appropriate boundary conditions.
There exist partial results addressing this issue by the author
\cite{Donninger2007} which state that eq. (\ref{outlook_eq_genev}) has no
analytic solutions for real $\lambda$ unless $\lambda=1$.
Furthermore, this problem has been investigated numerically (cf.
\cite{Bizon2005}, \cite{Donninger2006}) and these studies strongly suggest that
$\lambda=1$ is indeed the only "eigenvalue" with positive real part.
We note that the origin of this unstable mode is well understood 
since it stems from the
time translation symmetry of the wave map equation similar to the gauge
instability we have encountered in the self--adjoint formulation.
Thus, if one could make these ideas rigorous it would be possible to use a 
spectral projection
which removes the eigenvalue $1$ from the spectrum of $L_0$ and prove linear
stability of $f_0$.

Of course, the ultimate goal would be proving \emph{nonlinear} stability of this
solution.
We do not try to make this precise but merely note that quite recently
\cite{Krieger2007}
rigorous results have been obtained for a different (easier) problem where
some of the difficulties arising there seem to be similar to our wave map model.
Concerning self--similar blow up for semilinear wave equations we also 
mention \cite{Galaktionov2004} and \cite{Merle2006}, \cite{Merle2005} although
the latter deal with the energy (sub)critical case.
There is also a notion of \emph{orbital stability} pioneered by Weinstein
(cf. e.g. \cite{Weinstein1986}) which might be able to deal with the 
difficulties arising 
from the fact that
$f_0$ is a one--parameter family of functions rather than a single solution. 
However, this approach has mainly been worked out for the Schr\"odinger equation
which is quite a different problem.
To conclude we have to admit that a rigorous proof of nonlinear stability 
of $f_0$ seems to be beyond 
the scope of present techniques.

Finally, we emphasize that the results obtained in this thesis are not
confined to the particular wave map model which has been considered.
In principle, they are equally applicable to any evolution equation of the form
$$ \psi_{tt}-\psi_{rr}-\frac{2}{r}\psi_r+\frac{f(\psi)}{r^2}=0 $$
if it shows similar behaviour (existence of self--similar solutions).
A prominent example is the Yang--Mills field in $5+1$ dimensions which is in the
same criticality class as our wave maps model and analogous phenomena have been
observed for this system.
Bizo\'n has studied this model \cite{Bizon2002a} and he has proved existence of
a countable family of self--similar solutions.
The analysis presented in this thesis carries over to the Yang--Mills model
with minor changes.

\begin{appendix}
\chapter{Symbols}

\begin{tabular}{ll}
$\mathbb{N}$ & The natural numbers $1, 2, \ldots$ \\
$\mathbb{N}_0$ & The natural numbers including zero $0, 1, 2, \ldots$ \\
$\mathbb{Z}$ & The integer numbers $0, -1, 1, -2, 2, -3, 3, \ldots$ \\
$\mathbb{R}$ & The real numbers \\
$\mathbb{R}^k$ & $\{(x_1, \ldots, x_k): x_j \in \mathbb{R} \: \forall 1 \leq j \leq
k\}$ \\ 
$(a, b)$ & $\{x \in \mathbb{R}: a<x<b\}$ \\
$[a, b)$ & $\{x \in \mathbb{R}: a\leq x < b\}$ \\
$(a, b]$ & $\{x \in \mathbb{R}: a<x \leq b\}$ \\
$[a, b]$ & $\{x \in \mathbb{R}: a \leq x \leq b\}$ \\
$\mathbb{C}$ & The complex numbers \\
$\mathbb{C}^k$ & $\{(x_1, \ldots, x_k): x_j \in \mathbb{C} \: \forall 1 \leq j \leq
k\}$ \\ 
$\overline{x}$ & The complex conjugate of $x \in \mathbb{C}$ \\
$\mathrm{Re} x$ & The real part of $x \in \mathbb{C}$ \\
$\mathrm{Im} x$ & The imaginary part of $x \in \mathbb{C}$ \\
$u_x, \partial_x u, \frac{\partial u}{\partial x}, \frac{
\partial}{\partial x}u$ & Partial derivative of $u$ with respect to $x$ \\
$\mathrm{id}_M$ & The identity on a set $M$, i.e. $\mathrm{id}_M(x):=x$ for $x
\in M$ \\
$\ker(A)$ & The null space or kernel of a linear mapping $A: V \to W$ \\
& between two vector spaces $V$ and $W$, \\
& i.e. $\ker(A):=\{u \in V: Au=0\}$ \\
$\im(A)$ & The image or range of a mapping $A: X \to Y$ between \\
& two sets $X$,$Y$, i.e. $\im(A):=\{Ax: x \in X\}$
\end{tabular}


\end{appendix}

\bibliography{bib/bib}{}

\begin{thebibliography}{10}

\bibitem{Abramowitz1964}
Milton Abramowitz and Irene~A. Stegun.
\newblock {\em Handbook of mathematical functions with formulas, graphs, and
  mathematical tables}, volume~55 of {\em National Bureau of Standards Applied
  Mathematics Series}.
\newblock For sale by the Superintendent of Documents, U.S. Government Printing
  Office, Washington, D.C., 1964.

\bibitem{Aichelburg2005}
Peter~C. Aichelburg, Piotr Bizo{\'n}, and Zbislaw Tabor.
\newblock Bifurcation and fine structure phenomena in critical collapse of a
  self-gravitating {$\sigma$}-field.
\newblock {\em Classical Quantum Gravity}, 23(16):S299--S306, 2006.

\bibitem{Aminneborg1995}
Stefan {\AA}minneborg and Lars Bergstr{\"o}m.
\newblock On self-similar global textures in an expanding universe.
\newblock {\em Phys. Lett. B}, 362(1-4):39--45, 1995.

\bibitem{Bauer2001}
Heinz Bauer.
\newblock {\em Measure and integration theory}, volume~26 of {\em de Gruyter
  Studies in Mathematics}.
\newblock Walter de Gruyter \& Co., Berlin, 2001.
\newblock Translated from the German by Robert B.\ Burckel.

\bibitem{Beyer2007}
Horst~Reinhard Beyer.
\newblock {\em Beyond partial differential equations}, volume 1898 of {\em
  Lecture Notes in Mathematics}.
\newblock Springer, Berlin, 2007.
\newblock On linear and quasi-linear abstract hyperbolic evolution equations.

\bibitem{Bizon2000a}
Piotr Bizo{\'n}.
\newblock Equivariant self-similar wave maps from {M}inkowski spacetime into
  3-sphere.
\newblock {\em Comm. Math. Phys.}, 215(1):45--56, 2000.

\bibitem{Bizon2002a}
Piotr Bizo{\'n}.
\newblock Formation of singularities in {Y}ang-{M}ills equations.
\newblock {\em Acta Phys. Polon. B}, 33(7):1893--1922, 2002.

\bibitem{Bizon2005}
Piotr Bizo{\'n}.
\newblock An unusual eigenvalue problem.
\newblock {\em Acta Phys. Polon. B}, 36(1):5--15, 2005.

\bibitem{Bizon2007}
Piotr Bizo\'n, Tadeusz Chmaj, and Andrzej Rostworowski.
\newblock On asymptotic stability of the skyrmion.
\newblock Preprint math-ph/0701037, 2007.

\bibitem{Bizon2000}
Piotr Bizo{\'n}, Tadeusz Chmaj, and Zbis{\l}aw Tabor.
\newblock Dispersion and collapse of wave maps.
\newblock {\em Nonlinearity}, 13(4):1411--1423, 2000.

\bibitem{Bizon2002}
Piotr Bizo{\'n} and Arthur Wasserman.
\newblock On the existence of self-similar spherically symmetric wave maps
  coupled to gravity.
\newblock {\em Classical Quantum Gravity}, 19(12):3309--3321, 2002.

\bibitem{Cap2004}
Andreas Cap.
\newblock Differentialgeometrie 1.
\newblock Lecture notes University of Vienna, 2004.

\bibitem{Cazenave1998}
Thierry Cazenave, Jalal Shatah, and A.~Shadi Tahvildar-Zadeh.
\newblock Harmonic maps of the hyperbolic space and development of
  singularities in wave maps and {Y}ang-{M}ills fields.
\newblock {\em Ann. Inst. H. Poincar\'e Phys. Th\'eor.}, 68(3):315--349, 1998.

\bibitem{Georgiev2004}
Piero D'Ancona and Vladimir Georgiev.
\newblock On the continuity of the solution operator to the wave map system.
\newblock {\em Comm. Pure Appl. Math.}, 57(3):357--383, 2004.

\bibitem{Donninger2006}
Roland Donninger and Peter~C. Aichelburg.
\newblock A note on the eigenvalues for equivariant maps of the {SU}(2)
  sigma--model.
\newblock Preprint math-ph/0601019, 2006.

\bibitem{Donninger2007}
Roland Donninger and Peter~C. Aichelburg.
\newblock On the mode stability of a self--similar wave map.
\newblock Preprint math-ph/0702025, 2007.

\bibitem{Engel2000}
Klaus-Jochen Engel and Rainer Nagel.
\newblock {\em One-parameter semigroups for linear evolution equations}, volume
  194 of {\em Graduate Texts in Mathematics}.
\newblock Springer-Verlag, New York, 2000.
\newblock With contributions by S. Brendle, M. Campiti, T. Hahn, G. Metafune,
  G. Nickel, D. Pallara, C. Perazzoli, A. Rhandi, S. Romanelli and R.
  Schnaubelt.

\bibitem{Erdelyi1953}
Arthur Erd{\'e}lyi, Wilhelm Magnus, Fritz Oberhettinger, and Francesco~G.
  Tricomi.
\newblock {\em Higher transcendental functions. {V}ols. {I}, {II}}.
\newblock McGraw-Hill Book Company, Inc., New York-Toronto-London, 1953.
\newblock Based, in part, on notes left by Harry Bateman.

\bibitem{Evans1998}
Lawrence~C. Evans.
\newblock {\em Partial differential equations}, volume~19 of {\em Graduate
  Studies in Mathematics}.
\newblock American Mathematical Society, Providence, RI, 1998.

\bibitem{Fattorini1985}
H.~O. Fattorini.
\newblock {\em Second order linear differential equations in {B}anach spaces},
  volume 108 of {\em North-Holland Mathematics Studies}.
\newblock North-Holland Publishing Co., Amsterdam, 1985.
\newblock , Notas de Matem\'atica [Mathematical Notes], 99.

\bibitem{Fuller1954}
F.~B. Fuller.
\newblock Harmonic mappings.
\newblock {\em Proc. Nat. Acad. Sci. U. S. A.}, 40:987--991, 1954.

\bibitem{Galaktionov2004}
V.~A. Galaktionov and S.~I. Pohozaev.
\newblock On similarity solutions and blow-up spectra for a semilinear wave
  equation.
\newblock {\em Quart. Appl. Math.}, 61(3):583--600, 2003.

\bibitem{Gell-Mann1960}
M.~Gell-Mann and M.~L{\'e}vy.
\newblock The axial vector current in beta decay.
\newblock {\em Nuovo Cimento (10)}, 16:705--726, 1960.

\bibitem{Gough2003}
Brian Gough, editor.
\newblock {\em GNU Scientific Library Reference Manual}.
\newblock Network Theory Limited, 2003.

\bibitem{Gustafsson1995}
Bertil Gustafsson, Heinz-Otto Kreiss, and Joseph Oliger.
\newblock {\em Time dependent problems and difference methods}.
\newblock Pure and Applied Mathematics (New York). John Wiley \& Sons Inc., New
  York, 1995.
\newblock , A Wiley-Interscience Publication.

\bibitem{Kato1980}
Tosio Kato.
\newblock {\em Perturbation theory for linear operators}.
\newblock Classics in Mathematics. Springer-Verlag, Berlin, 1995.
\newblock Reprint of the 1980 edition.

\bibitem{Klainerman1986}
S.~Klainerman.
\newblock The null condition and global existence to nonlinear wave equations.
\newblock In {\em Nonlinear systems of partial differential equations in
  applied mathematics, Part 1 (Santa Fe, N.M., 1984)}, volume~23 of {\em
  Lectures in Appl. Math.}, pages 293--326. Amer. Math. Soc., Providence, RI,
  1986.

\bibitem{Klainerman2002a}
Sergiu Klainerman and Igor Rodnianski.
\newblock On the global regularity of wave maps in the critical {S}obolev norm.
\newblock {\em Internat. Math. Res. Notices}, (13):655--677, 2001.

\bibitem{Klainerman2002}
Sergiu Klainerman and Sigmund Selberg.
\newblock Bilinear estimates and applications to nonlinear wave equations.
\newblock {\em Commun. Contemp. Math.}, 4(2):223--295, 2002.

\bibitem{Kovalyov1987}
Mikhail Kovalyov.
\newblock Long-time behaviour of solutions of a system of nonlinear wave
  equations.
\newblock {\em Comm. Partial Differential Equations}, 12(5):471--501, 1987.

\bibitem{Krieger2007}
J.~Krieger and W.~Schlag.
\newblock On the focusing critical semi-linear wave equation.
\newblock {\em Amer. J. Math.}, 129(3):843--913, 2007.

\bibitem{Krieger2005}
Joachim Krieger.
\newblock Global regularity of wave maps from {$\bold R\sp {2+1}$} to {$H\sp
  2$}. {S}mall energy.
\newblock {\em Comm. Math. Phys.}, 250(3):507--580, 2004.

\bibitem{Liebling2000}
Steven~L. Liebling, Eric~W. Hirschmann, and James Isenberg.
\newblock Critical phenomena in nonlinear sigma models.
\newblock {\em J. Math. Phys.}, 41(8):5691--5700, 2000.

\bibitem{Locker1986}
John Locker.
\newblock {\em Functional analysis and two-point differential operators},
  volume 144 of {\em Pitman Research Notes in Mathematics Series}.
\newblock Longman Scientific \& Technical, Harlow, 1986.

\bibitem{Merle2005}
Frank Merle and Hatem Zaag.
\newblock Determination of the blow-up rate for a critical semilinear wave
  equation.
\newblock {\em Math. Ann.}, 331(2):395--416, 2005.

\bibitem{Merle2006}
Frank Merle and Hatem Zaag.
\newblock On growth rate near the blowup surface for semilinear wave equations.
\newblock {\em Int. Math. Res. Not.}, (19):1127--1155, 2005.

\bibitem{Misner1978}
Charles~W. Misner.
\newblock Harmonic maps as models for physical theories.
\newblock {\em Phys. Rev. D (3)}, 18(12):4510--4524, 1978.

\bibitem{Misner1982}
Charles~W. Misner.
\newblock Nonlinear model field theories based on harmonic mappings.
\newblock In {\em Spacetime and Geometry: The Alfred Schild Lectures}, pages
  82--101. University of Texas Press, Austin (Texas), 1982.

\bibitem{Naimark1968}
M.~A. Na{\u\i}mark.
\newblock {\em Linear differential operators. {P}art {II}: {L}inear
  differential operators in {H}ilbert space}.
\newblock With additional material by the author, and a supplement by V. \`E.
  Ljance. Translated from the Russian by E. R. Dawson. English translation
  edited by W. N. Everitt. Frederick Ungar Publishing Co., New York, 1968.

\bibitem{Pazy1983}
A.~Pazy.
\newblock {\em Semigroups of linear operators and applications to partial
  differential equations}, volume~44 of {\em Applied Mathematical Sciences}.
\newblock Springer-Verlag, New York, 1983.

\bibitem{Reed1975}
Michael Reed and Barry Simon.
\newblock {\em Methods of modern mathematical physics. {II}. {F}ourier
  analysis, self-adjointness}.
\newblock Academic Press [Harcourt Brace Jovanovich Publishers], New York,
  1975.

\bibitem{Reed1978}
Michael Reed and Barry Simon.
\newblock {\em Methods of modern mathematical physics. {IV}. {A}nalysis of
  operators}.
\newblock Academic Press [Harcourt Brace Jovanovich Publishers], New York,
  1978.

\bibitem{Ronveaux1995}
A.~Ronveaux, editor.
\newblock {\em Heun's differential equations}.
\newblock Oxford Science Publications. The Clarendon Press Oxford University
  Press, New York, 1995.
\newblock With contributions by F. M. Arscott, S. Yu.\ Slavyanov, D. Schmidt,
  G. Wolf, P. Maroni and A. Duval.

\bibitem{Sell2002}
George~R. Sell and Yuncheng You.
\newblock {\em Dynamics of evolutionary equations}, volume 143 of {\em Applied
  Mathematical Sciences}.
\newblock Springer-Verlag, New York, 2002.

\bibitem{Shatah1988}
Jalal Shatah.
\newblock Weak solutions and development of singularities of the {${\rm
  SU}(2)$} {$\sigma$}-model.
\newblock {\em Comm. Pure Appl. Math.}, 41(4):459--469, 1988.

\bibitem{Shatah1998}
Jalal Shatah and Michael Struwe.
\newblock {\em Geometric wave equations}, volume~2 of {\em Courant Lecture
  Notes in Mathematics}.
\newblock New York University Courant Institute of Mathematical Sciences, New
  York, 1998.

\bibitem{Shatah1994}
Jalal Shatah and A.~Shadi Tahvildar-Zadeh.
\newblock On the {C}auchy problem for equivariant wave maps.
\newblock {\em Comm. Pure Appl. Math.}, 47(5):719--754, 1994.

\bibitem{Sideris1989}
Thomas~C. Sideris.
\newblock Global existence of harmonic maps in {M}inkowski space.
\newblock {\em Comm. Pure Appl. Math.}, 42(1):1--13, 1989.

\bibitem{Spivak1999}
Michael Spivak.
\newblock {\em A comprehensive introduction to differential geometry. {V}ol.
  {I}}.
\newblock Publish or Perish Inc., Wilmington, Del., second edition, 1979.

\bibitem{Szybka2004}
Sebastian~J. Szybka.
\newblock Chaotic self-similar wave maps coupled to gravity.
\newblock {\em Phys. Rev. D (3)}, 69(8):084014, 7, 2004.

\bibitem{Tao2001}
Terence Tao.
\newblock Global regularity of wave maps. {II}. {S}mall energy in two
  dimensions.
\newblock {\em Comm. Math. Phys.}, 224(2):443--544, 2001.

\bibitem{Tataru2004}
Daniel Tataru.
\newblock The wave maps equation.
\newblock {\em Bull. Amer. Math. Soc. (N.S.)}, 41(2):185--204 (electronic),
  2004.

\bibitem{Tataru2005}
Daniel Tataru.
\newblock Rough solutions for the wave maps equation.
\newblock {\em Amer. J. Math.}, 127(2):293--377, 2005.

\bibitem{Teschl2007}
Gerald Teschl.
\newblock Mathematical methods in quantum mechanics.
\newblock Freely available on--line at {\tt http://www.mat.univie.ac.at/\~{
  }gerald}, 2007.

\bibitem{Triebel1995}
Hans Triebel.
\newblock {\em Interpolation theory, function spaces, differential operators}.
\newblock Johann Ambrosius Barth, Heidelberg, second edition, 1995.

\bibitem{Turok1990}
N.~Turok and D.~Spergel.
\newblock Global texture and the microwave background.
\newblock {\em Phys. Rev. Lett.}, 64:2736--2739, 1990.

\bibitem{Weinstein1986}
M.~I. Weinstein.
\newblock Remarks on orbital stability of ground states for subcritical and
  critical nonlinearities.
\newblock In {\em Semigroups, theory and applications, Vol.\ I (Trieste,
  1984)}, volume 141 of {\em Pitman Res. Notes Math. Ser.}, pages 249--252.
  Longman Sci. Tech., Harlow, 1986.

\bibitem{Xiao1998}
Ti-Jun Xiao and Jin Liang.
\newblock {\em The {C}auchy problem for higher-order abstract differential
  equations}, volume 1701 of {\em Lecture Notes in Mathematics}.
\newblock Springer-Verlag, Berlin, 1998.

\bibitem{Yosida1980}
K{\=o}saku Yosida.
\newblock {\em Functional analysis}.
\newblock Classics in Mathematics. Springer-Verlag, Berlin, 1995.
\newblock Reprint of the sixth (1980) edition.

\end{thebibliography}
\bibliographystyle{plain}

\newpage

\thispagestyle{empty}
\begin{center}
{\large \bf Curriculum Vitae} \\
Roland Donninger
\end{center}

\begin{tabular}{ll}
27. Nov. 1977 & Geboren in Ried im Innkreis \\
& Eltern Otmar und Barbara Donninger \\
1984 -- 1988 & Volksschule in Ried im Innkreis \\
1988 -- 1996 & Realgymnasium in Ried im Innkreis \\
Juni 1996 & Matura mit gutem Erfolg \\
Okt. 1996 -- Mai 1997 & Pr\"asenzdienst in H\"orsching und Wels \\
Okt. 1997 -- Juni 2004 & Studium der Physik / Mathematik Lehramt \\
& an der Universit\"at Wien, \\
& Abschluss mit ausgezeichnetem Erfolg \\
Ab Sept. 2004 & Doktoratsstudium der Physik an \\
& der Universit\"at Wien \\
& sowie wissenschaftlicher \\
& Mitarbeiter in den FWF Projekten \\
& P15738 und P19126 an der \\
& Universit\"at Wien unter Leitung von \\
& Prof. Peter C. Aichelburg
\end{tabular}

\begin{center}
{\large \bf Teilnahme an Konferenzen und Auszeichnungen} \\
\end{center}

\begin{tabular}{ll}
20. Sept. bis & Summer School "Structure and dynamics \\
25. Sept. 2004 & of compact objects", Albert Einstein Institut \\
& f\"ur Gravitationsphysik, Golm, Deutschland \\
24. Juli bis & Workshop "Spectral Theory and its Applications", \\
28. Juli 2006 & Isaac Newton Institute for Mathematical Sciences,  \\
& Cambridge, UK \\
12. Sept. bis & Workshop "Evolution equations and self--gravitating \\
14. Sept. 2007 & systems", Albert Einstein Institut f\"ur \\
& Gravitationsphysik, Golm, Deutschland
\end{tabular}

\vspace{1cm}

\begin{tabular}{ll}
Mai 2005 & Verleihung des Alfred Wehrl Preises f\"ur Mathematische \\
& Physik f\"ur die Diplomarbeit "Perturbation Analysis of \\
& Self--Similar Solutions of the SU(2) $\sigma$--Model \\
& on Minkowski Spacetime"
\end{tabular}

\end{document}